\documentclass[cleveref, autoref]{paper}

\usepackage{fullpage,float,amsthm,amsfonts,amsmath,amssymb,keytheorems,url,tikz,circuitikz,tabularx,multirow,multicol,hhline,tablefootnote,xcolor,etoolbox,clipboard,authblk,titlesec,setspace,titletoc,tocloft,enumitem}
\usepackage[sortcites=true,maxnames=1000]{biblatex}
\usepackage[toc,page]{appendix}
\usepackage[margin=0pt,font={small},labelfont={bf},labelsep=colon,format=plain]{caption}
\usepackage[left]{lineno}
\usepackage[mathscr]{eucal}
\usepackage[ruled,vlined]{algorithm2e}
\usepackage[hidelinks]{hyperref}
\makeatletter 
\let\original@algocf@latexcaption\algocf@latexcaption
\long\def\algocf@latexcaption#1[#2]{%
  \@ifundefined{NR@gettitle}{%
    \def\@currentlabelname{#2}%
  }{%
    \NR@gettitle{#2}%
  }%
  \original@algocf@latexcaption{#1}[{#2}]%
}
\makeatother 

\usetikzlibrary{fadings}
\usetikzlibrary{patterns}
\usetikzlibrary{shadows.blur}
\usetikzlibrary{shapes}
\usetikzlibrary{positioning}

\def\titlename{}
\def\authorsnames{}

\def\illustrationsPath{Illustrations/}

\def\apaspImagesPath{\illustrationsPath APASP/}

\def\squiggylinescale{0.08}
\def\squiggylinedist{0.6}
\def\squiggylinerange{3.2}

\definecolor{turquoise}{RGB}{64,224,208}
\definecolor{salmon}{RGB}{250,128,114}

\newcommand{\ceil}[1]{\left\lceil #1 \right\rceil}
\newcommand{\paren}[1]{\left( #1 \right)}
\newcommand{\bracke}[1]{\left[ #1 \right]}
\newcommand{\bracce}[1]{\left\{ #1 \right\}}
\newcommand{\pair}[1]{\left< #1 \right>}
\newcommand{\abs}[1]{\left| #1 \right|}

\newcommand{\tabref}[1]{\autoref{tab:#1}}
\newcommand{\secref}[1]{\hyperref[#1]{Section~\ref*{#1}}}
\newcommand{\appref}[1]{\hyperref[#1]{Appendix~\ref*{#1}}}
\newcommand{\thmref}[1]{~\protect\autoref{thm:#1}}
\newcommand{\defref}[1]{~\protect\hyperref[deg:#1]{Definition~\ref*{def:#1}}}
\newcommand{\lemref}[1]{\hyperref[lem:#1]{Lemma~\ref*{lem:#1}}}
\newcommand{\algref}[1]{\hyperref[alg:#1]{Algorithm~\ref*{alg:#1}}}

\newcommand{\cncref}[1]{~\protect\hyperref[cnc:#1]{Conclusion~\ref*{cnc:#1}}}
\newcommand{\corref}[1]{~\protect\hyperref[cor:#1]{Corollary~\ref*{cor:#1}}}
\newcommand{\figref}[1]{\autoref{fig:#1}}

\newcommand{\obvref}[1]{\hyperref[obv:#1]{Observation~\ref*{obv:#1}}}
\newcommand{\clmref}[1]{\hyperref[clm:#1]{Claim~\ref*{clm:#1}}}
\newcommand{\stpref}[3]{\hyperref[alg:#1_#2]{\textit{#3}}}

\newtheorem{question}{Question}
\newcommand{\prbref}[1]{\hyperref[prb:#1]{Problem~\ref*{prb:#1}}}
\newcommand{\queref}[1]{\hyperref[que:#1]{Question~\ref*{que:#1}}}

\NewDocumentCommand\mmin{mgg}{\displaystyle \min\IfNoValueTF{#2}{}{_{#2}}\IfNoValueTF{#3}{}{^{#3}}\bracce{#1}}
\NewDocumentCommand\mmax{mgg}{\displaystyle \max\IfNoValueTF{#2}{}{_{#2}}\IfNoValueTF{#3}{}{^{#3}}\bracce{#1}}
\NewDocumentCommand\mprod{mgg}{\displaystyle \prod\IfNoValueTF{#2}{}{_{#2}}\IfNoValueTF{#3}{}{^{#3}}{#1}}
\NewDocumentCommand\msum{mgg}{\displaystyle \sum\IfNoValueTF{#2}{}{_{#2}}\IfNoValueTF{#3}{}{^{#3}}{#1}}
\NewDocumentCommand\mcup{mgg}{\displaystyle \bigcup\IfNoValueTF{#2}{}{_{#2}}\IfNoValueTF{#3}{}{^{#3}}{#1}}
\NewDocumentCommand\mcap{mgg}{\displaystyle \bigcap\IfNoValueTF{#2}{}{_{#2}}\IfNoValueTF{#3}{}{^{#3}}{#1}}
\newcommand{\textoverline}[1]{\={#1}}
\NewDocumentCommand\case{mg}
    {
    \ensuremath{\textbf{Case (#1\IfNoValueTF{#2}{}{$_#2$})}}
    }
\NewDocumentCommand\notcase{mg}
    {
    \ensuremath{\textbf{Case (\textoverline{#1}\IfNoValueTF{#2}{}{$_#2$})}}
    }
\newcommand{\midline}{\,\middle| \,\,}

\newcommand{\qoute}[1]{``#1''}
\newcommand{\squiggly}{
\begin{tikzpicture} 
    {
    \draw[domain=-\squiggylinedist:0,smooth,variable=\x,scale=\squiggylinescale] plot({\x},{0.75+sin(4*0 r)});
    \draw[domain=0:\squiggylinerange,smooth,variable=\x,scale=\squiggylinescale] plot({\x},{0.75+sin(4*\x r)});
    \draw[domain=\squiggylinerange:\squiggylinerange+\squiggylinedist,smooth,variable=\x,scale=\squiggylinescale] plot({\x},{0.75+sin(4*\squiggylinerange r)});
    }
\end{tikzpicture}
}
\newcommand{\straight}{
\begin{tikzpicture} 
    {
    \draw[domain=-\squiggylinedist:-\squiggylinedist+0.0001,smooth,variable=\x,scale=\squiggylinescale] plot({\x},{-0.25+sin(4*\squiggylinerange r)});
    \draw[domain=-\squiggylinedist+0.0001:\squiggylinerange+\squiggylinedist,smooth,variable=\x,scale=\squiggylinescale] plot({\x},{0.75+sin(4*\squiggylinerange r)});
    }
\end{tikzpicture}
}
\newcommand{\fatrightarrow}
{
    \begin{tikzpicture}
        \draw[-{Triangle[width=6pt,length=6pt]}, line width=2pt](0,0) -- (0.3, 0);
    \end{tikzpicture}
}
\newcommand{\hollowrightarrow}
{
    \begin{tikzpicture}
        \node[draw, single arrow,
              minimum height=1pt, minimum width=1pt,
              single arrow head extend=4pt,
              anchor=west, rotate=0] at (0,-2) {};
    \end{tikzpicture}
}
\newcommand{\ubrace}[2]{{\underbrace{#1}_{#2}}}
\newcommand{\textubrace}[2]{%
  \ubrace{#1}{%
    \mbox{\scriptsize\begin{socgtabular}{@{}c@{}}#2\end{socgtabular}}%
  }%
}
\ifdefined\socgtabular\else
  \let\socgtabular\tabular
\fi

\newcommand{\dashedlinesep}[0]{
\\[-0.5em]
    \text{\,--\,--\,--\,--\,--\,--\,--\,--\,--\,--\,--\,--\,--\,--\,--\,--\,--\,--\,--\,--\,--\,} &  \text{\,--\,--\,--\,--\,--\,--\,--\,--\,--\,--\,--\,--\,--\,--\,--\,--\,--\,--\,--\,--\,--\,--\,--\,--\,--\,} \\[-0.5em]
}
\newcommand{\longspace}[0]{\,\,\,\,\,\,\,\,\,\,\,\,\,\,\,\,\,\,\,\,\,\,\,\,\,\,\,\,\,\,\,\,\,\,}
\newcommand\cond[1]{Condition (#1)}
\newcommand\lemcond[2]{\cond{#1} of \lemref{#2}}
\newcommand*{\settitlename}[1]{\title{#1}\def\titlename{#1}}
\newcommand*{\addauthor}[2]{\author[#2]{#1} 
\IfNoValueTF{\authorsnames}{\def\authorsnames{#1}}{\expandafter\def\expandafter\authorsnames\expandafter{\authorsnames, #1}}}
\renewcommand{\keywords}[1]{\def\keywordnames{#1}}

\ExplSyntaxOn
\cs_new:Npn 
\ballformatvar:n #1
 {
  \tl_clear:N \l_tmpa_tl
  \tl_map_inline:nn {#1}
   {
    \str_if_in:nnTF {ijkl\ell} {##1} 
      { \tl_put_right:Nn \l_tmpa_tl {\textit{##1}} }
      { \tl_put_right:Nn \l_tmpa_tl {##1} } 
   }
  \l_tmpa_tl
}
\cs_new:Npn \ballformatarg:n #1
 {
  \regex_match:nnTF { \A (?:[0-9]|i|j|k|\\ell)(?:[0-9ij k\\ell\+\-]*?) \Z } { #1 }
       {S\textsubscript{\ballformatvar:n{#1}}} 
       { #1 }
 }

\NewDocumentCommand \ball {mmg}
 {
    ball\paren{#1 ,
               \ballformatarg:n {#2}
               \IfNoValueTF{#3}
                           {}
                           {, \ballformatarg:n {#3}}
              }
 }
\ExplSyntaxOff

\newtheorem{theorem}{Theorem}[section]
\newtheorem{lemma}[theorem]{Lemma}
\newtheorem{claim2}[theorem]{Claim}

\newtheorem{observation}[theorem]{Observation}

\settitlename{Additive, Near-Additive, and Multiplicative Approximations for APSP in Weighted Undirected Graphs: Trade-offs and Algorithms}

\addauthor{Liam Roditty}{1,$\S$}
\addauthor{Ariel Sapir}{1,$\propto$}

\affil[1]{Department of \href{http://cs.biu.ac.il/}{Computer Science}, Bar-Ilan University}
\affil[$\S$]{liam.roditty@biu.ac.il}
\affil[$\propto$]{sapirar@biu.ac.il}

\date{\today}


\makeatletter
\def\@maketitle{%
  \newpage
  \null
  \vskip 2em%
  \begin{center}%
  \let \footnote \thanks
    {\huge \textsf{\textbf{\@title}} \par}%
    \vskip 1.5em%
    {\large
      \lineskip .5em%
      \begin{tabular}[t]{c}%
        \@author
      \end{tabular}\par}%
    \vskip 1em
    {\large \@date}
  \end{center}
  \par
  \vskip 1.5em}
\makeatother

\makeatletter
\renewenvironment{abstract}{
    \begin{center}%
    {\bfseries\sffamily \Large\abstractname\vspace{\z@}}
      \end{center}%
      \quotation
    }
    \endquotation
\makeatother

\titleformat*{\section}{\Large\bfseries\sffamily}
\titleformat*{\subsection}{\large\bfseries\sffamily}

\dottedcontents{section}[1.5em]{}{1.3em}{.6em}

\setlist[itemize,1]{label=\scalebox{1.6}[1.0]{$\blacklozenge$}}
\setlist[itemize,2]{label=\scalebox{0.9}[1.3]{$\blacktriangleright$}}
\setlist[itemize,3]{label=\textbf{\scalebox{0.9}[1.2]{\fatrightarrow}}}
\setlist[itemize,4]{label=\textbf{\scalebox{0.666}[0.47]{\hollowrightarrow}}}
\setlist[itemize,5]{label=\textbf{\scalebox{0.9}[1.2]{$>$}}}

\begin{document}

\let\originalleft\left
\let\originalright\right
\renewcommand{\left}{\mathopen{}\mathclose\bgroup\originalleft}
\renewcommand{\right}{\aftergroup\egroup\originalright}

\setlength{\textfloatsep}{8pt}


\maketitle

\begin{abstract}
\small We present a $+2\sum_{i=1}^{k+1} {W_i}$-APASP algorithm for dense weighted graphs with a runtime of $\tilde O\left(n^{2+\frac{1}{3k+2}}\right)$, where $W_{i}$ is the weight of an $i^\textnormal{th}$ heaviest edge on a shortest path between two vertices. Dor, Halperin and Zwick [FOCS'96 and SICOMP'00] introduced two algorithms for the commensurate unweighted $+2\cdot \left( k+1\right)$-APASP problem: one for sparse graphs with a runtime of $\tilde O \left(n^{2-\frac{1}{k+2}} m^{\frac{1}{k+2}}\right)$ and one for dense graphs with a runtime of $\tilde O\left(n^{2+\frac{1}{3k+2}}\right)$. Subsequently, Cohen and Zwick [SODA'97 and JALG'01] adapted the algorithm for sparse graphs to the weighted setting, namely a $+2\sum_{i=1}^{k+1} {W_i}$-APASP algorithm with the same $\tilde O \left(n^{2-\frac{1}{k+2}} m^{\frac{1}{k+2}}\right)$ runtime. We fill the nearly three decades old gap by providing an algorithm for dense weighted graphs, matching the runtime for the unweighted setting. 

In addition, we explore \emph{nearly additive} APASP, where the multiplicative stretch is $1+\varepsilon$. We present a $\left(1+\varepsilon, \min{\left\{2W_1,4W_{2}\right\}}\right)$-APASP algorithm with a runtime of $\tilde O\left(\left(\frac{1}{\varepsilon}\right)^{O\left(1\right)} \cdot n^{2.15135313} \cdot \log W\right)$. This improves upon Saha and Ye [SODA'24], which had the same runtime, yet computed a  $\left(1+\varepsilon, 2W_1\right)$-APASP. 

For pure multiplicative APASP, we begin by presenting a $\left(\frac{7}{3}+\varepsilon\right)$-APASP algorithm with a runtime of $\tilde O \left( \left(\frac{1}{\varepsilon}\right)^{O\left(1\right)} \cdot n^{2.15135313} \cdot \log W\right)$. This improves, for dense graphs, the $\tilde O \left( nm^{\frac{2}{3}}+n^2\right)$ runtime of the $\frac{7}{3}$-APASP algorithm by Baswana and Kavitha [FOCS'06 and SICOMP'10], while introducing an additional $\varepsilon$ to the multiplicative stretch.

We further view this result in a broader framework of $\left(\frac{3\ell + 4}{\ell + 2} + \varepsilon\right)$-APASP algorithms, similarly to the framework of  $\frac{3\ell + 4}{\ell + 2}$-APASP algorithms by Akav and Roditty [ESA'21]. This also  generalizes the $\left(2+\varepsilon\right)$-APASP algorithm by Dory, Forster, Kirkpatrick, Nazari, Vassilevska Williams, and de Vos [SODA'24].
 
Finally, we show that it is possible to \qoute{bypass} an $\tilde \Omega \left(n^\omega\right)$ conditional lower bound by Dor, Halperin, and Zwick for  $\alpha$-APASP with $\alpha < 2$, by allowing an additive component to the approximation (e.g. a $\paren{\frac{6k+3}{3k+2},\sum_{i=1}^{k+1}{W_{i}}}$-APASP with  $\tilde O\left(n^{2+\frac{1}{3k+2}}\right)$ runtime.). 
\end{abstract}

\setstretch{1.3}

\newpage

\tableofcontents
\newpage

\section{Introduction}~\label{intro}
We study the classic \textbf{A}ll-\textbf{P}airs \textbf{S}hortest \textbf{P}aths (\emph{APSP}) problem in a weighted undirected graph $G=\paren{V , E,w}$ with non-negative weights $w:E\rightarrow \mathbb{R^{+}}$. Floyd-Warshall's algorithm \cite{Floyd1962, Roy1959, Warshall1962} solves APSP in $O\left(n^3\right)$ time. Johnson's algorithm \cite{Johnson1977} improves it to $O\left(n^2 \log n + nm\right)$, which is still $O\paren{n^3}$ for dense graphs. Despite advances (e.g. \cite{Fredman1976, Takaoka1992, RWilliams2018}), no truly sub-cubic APSP algorithm currently exists. The fastest known one runs in $O\paren{ \frac{n^3}{2^{\sqrt{\Omega\left(\log n\right)}}} }$ time \cite{RWilliams2018}. 

Facing this, an immediate question arises: {\it Can we approximate APSP in less than $\tilde O \paren{n^3}$ time?} To explore this, we properly define the problem of \textbf{A}ll-\textbf{P}airs \textbf{A}pproximated \textbf{S}hortest \textbf{P}aths (\emph{APASP}). For parameters $\alpha,\beta$, the goal of an $\paren{\alpha,\beta}$-APASP is to compute, for all $u,v\in V$, an approximate distance $d\bracke{u,v}$ s.t. $\delta\paren{u,v} \leq d\bracke{u,v} \leq \alpha\cdot \delta\paren{\alpha,\beta} + \beta$. If $\beta=0$, the approximation is purely multiplicative, referred to as an $\alpha$-APASP, where $\alpha$ is the \textit{multiplicative stretch}; if $\alpha=1$, the approximation is purely additive, known as $+\beta$-APASP, where $\beta$ is the \textit{additive stretch}; otherwise, the approximation is \qoute{mixed}. 

In weighted graphs, defining additive approximation in a meaningful way is a bit tricky. By scaling edge weights, we can show that any constant $+\beta$-APASP is equivalent to APSP. To overcome this, we explore additive stretches that depend on $w$. One approach is a $+\beta W$-APASP, where $W=\mmax{w\paren{e}}{e\in E}$. However, it might hold that most shortest paths use edges whose weight is much smaller than $W$, hence a $+\beta W$-APASP yields a very \qoute{weak} guarantee.

Cohen and Zwick \cite{CohZwi1997} addressed this issue by defining $W_{i}=W_{i}\paren{u\squiggly v}$ to be the weight of an $i^{\textnormal{th}}$ heaviest edge on a shortest path $u\squiggly v$. This approach has since influenced many approximation results for various graphs distance problems (e.g., \cite{SahYe2023, LaLe2024, AhmBodSahKobSpe2021, ElkGitNei2021}). 

We consider the  approach of Cohen and Zwick under a broader spectrum which generalizes the concept of equivalence between weighted and unweighted  approximations. Consider an $\paren{\alpha,\beta}$-APASP in an unweighted graph for arbitrary $\alpha,\beta$ and let $f\paren{\beta,w}$ be a function. The problem of an $\paren{\alpha,f\paren{\beta,w}}$-APASP in weighted graphs is a \textit{commensurate version} of the  $\paren{\alpha,\beta}$-APASP problem in unweighted graphs if, when the weight function $w\paren{e}=1$ for all $e\in E$, the additive stretch of the two problems is the same, that is, $f\paren{\beta,w}=\beta$.

Two algorithms $\mathscr{A}_1$ and $\mathscr{A}_2$, with runtimes $T_{1}\paren{n}$ and $T_{2}\paren{n}$, respectively,  are \textit{runtime equivalent} if $T_{1}\paren{n} \in \tilde \Theta_{W} \paren{T_{2}\paren{n}}$, where $\tilde \Theta_{W}\paren{\cdot}$ hides poly-logarithmic factors of $n$ or $W$. An $\paren{\alpha,f\paren{\beta,w}}$-APASP algorithm $\mathscr{A}_1$ for weighted graphs is a \textit{strongly commensurate version} of an $\paren{\alpha,\beta}$-APASP algorithm $\mathscr{A}_2$ in unweighted graphs if $\paren{\alpha,f\paren{\beta,w}}$-APASP is a commensurate version of $\paren{\alpha,\beta}$-APASP and $\mathscr{A}_1$ and $\mathscr{A}_2$ are runtime equivalent. Thus, as long as $\mathscr{A}_2$ is not improved, $\mathscr{A}_1$ is optimal. We consider the broad question:

\begin{question}~\label{que:equivWnoW}
For an $\paren{\alpha,\beta}$-APASP algorithm in unweighted graphs, what are its strongly commensurate versions $\paren{\alpha,f\paren{\beta,w}}$-APASP algorithms in weighted graphs?
\end{question}

Partial answers are already known.  Cohen and Zwick \cite{CohZwi1997} had two $+2W_{1}$-APASP algorithm with runtimes $\tilde O \paren{n^{\frac{3}{2}}m^{\frac{1}{2}}}$ and $\tilde O \paren{n^{\frac{7}{3}}}$, matching the runtimes of the two $+2$-APASP algorithms in unweighted graphs by Dor, Halperin, and Zwick \cite{DorHalZwi2000}. Since $+2W_{1}=+2$ when $w\paren{e}=1$ for all $e\in E$, the two pairs are, respectively, strongly commensurate.

\begin{table}[t]
  \begin{center}
\small\addtolength{\tabcolsep}{-3pt}
    \renewcommand\arraystretch{1.5}
    \begin{tabular}{|c|c|c|c|} 
      \hline
      \textbf{\small Approximation} & \textbf{\small Combinatorial}  & \textbf{\small Algebraic} & \textbf{\small Reference}  \\
      \hhline{|=|=|=|=|}
      $+2\msum{W_{i}}{i=1}{k+1} $& $n^{2+\frac{1}{3k+2}}$  &  --  & \secref{+2Wi}\\
      \hline
      
      \multirow{2}{*}{$\frac{7}{3}+\varepsilon$}& \multirow{2}{*}{--} &  $mn^{\beta} + n^{2+\gamma} +\left(\frac{1}{\varepsilon} \right)^{O\left(1\right)} \cdot n^{\omega\left(1-\beta-\gamma\right)} \cdot  \log W$  & \multirow{2}{*}{\secref{7over3+eps}} \\
      && $\approx \left(\frac{1}{\varepsilon}\right)^{O\left(1\right)} \cdot n^{2.15135313} \cdot \log W$ & \\
      \hline
      $\frac{3\ell+4}{\ell+2}+\varepsilon$& -- & $mn^{\beta}+n^{2+\gamma}+\paren{\frac{1}{\varepsilon} }^{O\paren{1}} \cdot n^{\omega\paren{1-\beta-\ell\cdot \gamma}} \cdot  \log W $ & \secref{3ellplus4overellplus2+eps} \\
      \hline
      \multirow{2}{*}{$\left(1+\varepsilon,\mmin{2W_{1},4W_{2}}\right)$}& \multirow{2}{*}{--}  &  $mn^{\beta} + n^{2+\gamma} +\left(\frac{1}{\varepsilon} \right)^{O\left(1\right)} \cdot n^{\omega\left(1-\beta-\gamma\right)} \cdot  \log W$  & \multirow{2}{*}{\secref{1+eps,4W2}} \\
      &  & $\approx \left(\frac{1}{\varepsilon}\right)^{O\left(1\right)} \cdot n^{2.15135313} \cdot \log W$ & \\
      \hline
    \end{tabular}
    \renewcommand\arraystretch{1}
    \caption{Our results for combinatorial and algebraic APASP for weighted graphs.}~\label{tab:our_weighted}
  \end{center}

  \vspace{-6mm}
\end{table}

Cohen and Zwick also generalized one of their $+2W_{1}$-APASP algorithms to a $+2\msum{W_{i}}{i=1}{k+1}$-APASP algorithm for weighted sparse graphs with $\tilde O \paren{n^{2-\frac{1}{k+2}}m^{\frac{1}{k+2}}}$ runtime, hence finding a strongly commensurate version to the unweighted $+2\cdot\paren{k+1}$-APASP algorithm of Dor, Halperin, and Zwick for sparse graphs. Yet, the $+2\cdot\paren{k+1}$-APASP algorithm for unweighted dense graphs with $\tilde O \paren{n^{2+\frac{1}{3k+2}}}$ runtime of Dor, Halperin and Zwick was left unaddressed, raising the following question:

\begin{question}~\label{que:+2Wiand+2k}
Is there a $+2\msum{W_{i}}{i=1}{k+1}$-APASP algorithm which is a strongly commensurate version of the $+2\cdot\paren{k+1}$-APASP algorithm for dense unweighted graphs of  \cite{DorHalZwi2000}?
\end{question}

\queref{+2Wiand+2k} remained open for nearly three decades since the work of Cohen and Zwick \cite{CohZwi1997}, despite extensive progress on $\paren{\alpha,\beta}$-APASP algorithms for weighted graphs (e.g., \cite{BasKav2010, Kavitha2012, BerKas2007, AkaRod2021, SahYe2023, DorForKirNazVasVos2023}). In this paper, we answer it in the affirmative by presenting a $+2\msum{W_{i}}{i=1}{k+1}$-APASP algorithm for dense weighted graphs, running in $\tilde O \paren{n^{2+\frac{1}{3k+2}}}$ time -- 
 establishing a strongly commensurate version of Dor, Halperin, and Zwick’s algorithm.

In our approach, we utilized only hitting sets and \textbf{S}ingle \textbf{S}ource \textbf{S}hortest \textbf{P}aths (\emph{SSSP}) invocations. However, other powerful tools exist. \textbf{M}atrix \textbf{M}ultiplication (\emph{MM}), particularly \textbf{M}in \textbf{P}lus \textbf{M}atrix \textbf{M}ultiplication (\emph{MPMM}), is a key tool for reducing the runtime of shortest path algorithms. Over the years, there have been many improvements (e.g. \cite{AmbFilLeg2014, DuaWuZho2022, VasXuXuZho2024, AlmDuaVasXuXuZho2024}) on the exponent of $n$ for the runtime of fast MM. However, these fast MM algorithms \qoute{hide} large constants in their runtime, which make them inefficient for practical use. 

Motivated by this, Aingworth, Chekuri, Indyk and Motwani 
 \cite{AinCheIndMot1999} initiated the study of algorithms for graph approximation problems (not necessarily APASP) that avoid fast MM. Since then,  APASP algorithms have been broadly categorized into either: an \emph{algebraic approach} -- which allows the use of fast MM  (e.g., \cite{BriKunWun2019, DorForKirNazVasVos2023, Kavitha2012, SahYe2023, Zwick2002, ElkNei2020}), and a \emph{combinatorial approach} -- where such algorithms are not used (e.g., \cite{AkaRod2021, BasKav2010, DorForKirNazVasVos2023, Kavitha2012, Roditty2023, CohZwi1997}). We summarize\footnote{Some results depend on certain parameters, such as $\omega\paren{\alpha,\beta,\gamma}$ or $MPCRBD\left(n,m\right)$, which is the runtime for a MPMM on column and row bounded difference matrices of dimensions $n\times m$ and $m\times n$. We provide the mathematical expression and the numerical exponent with currently known values.}  results for unweighted graphs in \tabref{known_unweighted} and for weighted graphs in \tabref{known_weighted}.

\begin{table}[t]
  \begin{center}
\small\addtolength{\tabcolsep}{-3pt}
    \renewcommand\arraystretch{1.5}
    \begin{tabular}{|c|c|c|} 
      \hline
      \textbf{\small Approximation} & \textbf{\small Combinatorial}  & \textbf{\small Algebraic} \\
      \hhline{|=|=|=|}
      \multirow{2}{*}{$+2$}& \multirow{2}{*}{$\mmin{n^{\frac{7}{3}},n^{\frac{3}{2}}m^{\frac{1}{2}}}$ \cite{DorHalZwi2000}}  & $n^{2} t^{\frac{1}{2}}+MPCRBD(n,\frac{n}{t})$ \multirow{2}{*}{\qquad\cite{DenKirVicVasZho2022}}\\
      &&$\approx n^{2.2867}$\\
      \hline
      \multirow{2}{*}{$\left(1+\varepsilon,2\right)$}& \multirow{2}{*}{--} & $n^{2+\frac{1-r}{2}}+\frac{1}{\varepsilon}n^{\omega\left(r\right)}$ \multirow{2}{*}{\qquad\cite{DorForKirNazVasVos2023}}\\ 
      & & $\approx \frac{1}{\varepsilon}\cdot n^{2.15135313}$ \\
      \hline
      $+2\cdot\left(k+1\right)$& $\mmin{n^{2+\frac{1}{3k+2}},n^{2-\frac{1}{k+2}}m^{\frac{1}{k+2}}}$ \cite{DorHalZwi2000} & --\\
      \hline
    \end{tabular}
    \renewcommand\arraystretch{1}
    \caption{Known additive (and nearly) combinatorial and algebraic APASP for unweighted graphs. }~\label{tab:known_unweighted}
  \end{center}

    \vspace{-6mm}
\end{table} 

One algebraic approach focuses on \emph{nearly additive} approximations. Specifically, we show an algorithm for $\paren{1+\varepsilon,\mmin{2W_{1},4W_{2}}}$-APASP with a runtime of $\tilde O \paren{\frac{1}{\varepsilon}\cdot n^{2.15135313} \cdot \log W}$. This improves upon Saha and Ye \cite{SahYe2023}, which had the same runtime, yet computed a $\paren{1+\varepsilon,2W_{1}}$-APASP. 

Another approach considers purely multiplicative stretches. We present a general framework of  $\paren{\frac{3\ell+4}{\ell+2}+\varepsilon}$-APASP algorithms with $\tilde O \paren{mn^{\beta}+n^{2+\gamma}+\paren{\frac{1}{\varepsilon}}^{O\paren{1}}\cdot n^{\omega\paren{1-\beta-\ell\cdot \gamma}} \cdot  \log W}$ runtime, where $\beta $ and $\gamma $ are parameters which depend on the graph's density and $\ell$ is a parameter that indicates the quality of the stretch. When the graph is dense, the runtime becomes $\tilde O \paren{n^{2+\beta}+\paren{\frac{1}{\varepsilon}}^{O\paren{1}}\cdot n^{\omega\paren{1-\paren{\ell+1}\cdot\beta}} \cdot \log W}$, where $\beta$ is chosen such that $2+\beta=\omega\paren{1-\paren{\ell+1}\cdot\beta}$. 

To properly understand the results that lie within this framework, we must first examine purely multiplicative stretches. In this setting, the goal is to compute an $\alpha$-APASP as efficiently as possible. To do so, we first consider the known upper and lower bounds.

Dor, Halperin and Zwick \cite{DorHalZwi2000} show that computing a $\paren{2-\varepsilon}$-APASP, for any $\varepsilon>0$, is as hard as \textbf{B}oolean \textbf{M}atrix \textbf{M}ultiplication (\emph{BMM}),  which is only known to be solved in $\tilde O \paren{n^{\omega}}$\footnote{$\omega\paren{\alpha,\beta,\gamma}$ is the infimum of the exponent of the runtime for all algorithms which compute the multiplication of an $n^{\alpha}\times n ^{\beta}$ and $n^{\beta}\times n^{\gamma}$ matrices. $\omega\paren{\beta} = \omega\paren{1,\beta,1}$ and $\omega=\omega\paren{1}$.}. This conditional lower bound means that $2$-APASP is the smallest multiplicative stretch s.t. an  $\alpha$-APASP algorithm may require less than $\tilde O \paren{n^{\omega}}$ time.  

\begin{table}[t]
  \begin{center}
    
\small\addtolength{\tabcolsep}{-3pt}
    \renewcommand\arraystretch{1.5}
    \begin{tabular}{|c|c|c|} 
      \hline
      \textbf{\small Approximation} & \textbf{\small Combinatorial}  & \textbf{\small Algebraic} \\
      \hhline{|=|=|=|}
      $+2W_{2}$& $n^{\frac{3}{2}} m ^{\frac{1}{2}} $ \cite{CohZwi1997} &  - \\
      \hline
      $+\mmin{2W_{1},4W_{2}}$& $n^{\frac{7}{3}} $ \cite{CohZwi1997} &  - \\
      \hline
      \multirow{2}{*}{$\left(1+\varepsilon,2W_{1}\right)$}& \multirow{2}{*}{--}  &  $\frac{1}{\varepsilon}\cdot n^{\omega\paren{1,1-x,1}}\cdot \log W + n^{2+\frac{x}{2}}$ \multirow{2}{*}{\qquad\cite{SahYe2023}} \\
      &  & $\approx \frac{1}{\varepsilon}\cdot n^{2.15135313}\cdot \log W$\\
      \hline
      $+2\msum{W_{i}}{i=1}{k+1} $& $n^{2-\frac{1}{k+2}}m^{\frac{1}{k+2}}$ \cite{CohZwi1997} &  --  \\
      \hline
      $\left(1+\varepsilon,2\msum{W_{i}}{i=1}{k+1}\right)$& --  &  $\frac{1}{\varepsilon}\cdot n^{\omega\paren{1-\frac{k}{k+2}\cdot x,1-x,1-\frac{k-1}{k+2}\cdot x}}\cdot \log W + n^{2+\frac{x}{k+2}}$ \cite{SahYe2023} \\
      \hline
      $2$& $n^{\frac{1}{2}}m+n^2$ \cite{BasKav2010,AkaRod2021} &  --  \\
      \hline
      \multirow{2}{*}{$2+\varepsilon$}& \multirow{2}{*}{--} &  $n^{1-r}m + \left(\frac{1}{\varepsilon}\right)^{O\left(1\right)}n^{\omega\left(r\right)}\cdot \log W $ \multirow{2}{*}{\qquad\cite{DorForKirNazVasVos2023}}  \\
      && $\approx \left(\frac{1}{\varepsilon}\right)^{O\left(1\right)} \cdot n^{2.21235201}\cdot \log W$\\
      \hline
      $\left(2,W_{1}\right)$& $n^2$ \cite{BasKav2010,BerKas2007} &  -- \\
      \hline
      $\frac{7}{3}$& $nm^{\frac{2}{3}}+n^2$ \cite{BasKav2010,AkaRod2021} &  -- \\
      \hline
      $\frac{3\ell+4}{\ell+2}$& $n^{2-\frac{3}{\ell+2}}m^{\frac{2}{\ell+2}}+n^2$ \cite{AkaRod2021} & --  \\
      \hline
      $3$& $n^{2}$ \cite{CohZwi1997,BasKav2010}  & --  \\
      \hline
    \end{tabular}
    \renewcommand\arraystretch{1}
    \caption{Known combinatorial and algebraic APASP runtimes for weighted graphs.}~\label{tab:known_weighted}
    \vspace{-4mm}
  \end{center}
\end{table}    

Since APASP algorithms output an $n \times n$ matrix, $\Omega(n^2)$ time is unavoidable. For weighted graphs, the only known purely multiplicative stretch that can be computed in quadratic time is $3$-APASP  \cite{CohZwi1997,BasKav2010}. This stands in contrast to the conditional lower bound for $\paren{2-\varepsilon}$-APASP, implying the existence of a  threshold $\alpha \in \bracke{2,3}$, for which there is an $\alpha$-APASP algorithm with $\tilde O\paren{n^{2}}$ runtime, while any  $\paren{\alpha-\varepsilon}$-APASP algorithm would require more\footnote{Assuming BMM cannot be solved in quadratic time.} than quadratic time. 

\begin{question}~\label{que:quadAPASP}
What is the smallest $\alpha\in\left[2,3\right]$, for which there is an $\alpha$-APASP algorithm that runs in $\tilde O\left(n^{2}\right)$ time, while any  $\paren{\alpha-\varepsilon}$-APASP algorithm necessitates  $\tilde \Omega \left(n^{2+\gamma}\right)$ time?
\end{question}

While exact quadratic-time algorithms are rare, there is notable progress for \qoute{near quadratic} ones, as summarized in \tabref{known_weighted}. This paper contributes in this direction by presenting a $\paren{\frac{7}{3}+\varepsilon}$-APASP algorithm with $\tilde O \paren{\paren{\frac{1}{\varepsilon}}^{O\paren{1}}\cdot n^{2.15135313} \cdot \log W}$ runtime. For dense graphs, we improve the $\tilde O \left( m^{\frac{2}{3}}n+n^{2}\right)$ runtime of the  $\frac{7}{3}$-APASP algorithm  by Baswana and Kavitha \cite{BasKav2010}, at the cost of adding an $\varepsilon$ to the multiplicative stretch. 

Baswana and Kavitha also presented a $2$-APASP algorithm with $\tilde O \paren{n^{\frac{3}{2}} m^{\frac{1}{2}}+n^{2}}$ runtime. 
Recently, Dory, Forster, Kirkpatrick, Nazari, Vassilevska Williams and de Vos \cite{DorForKirNazVasVos2023} presented a $\paren{2+\varepsilon}$-APASP algorithm with a $\tilde O \paren{\paren{\frac{1}{\varepsilon}}^{O\paren{1}} \cdot n^{2.21235201}\cdot \log W}$ runtime, improving the runtime for dense graphs at the cost of adding an $\varepsilon$. This algorithm and our own $\paren{\frac{7}{3}+\varepsilon}$-APASP algorithm belong to a the previously discussed unified framework of  $\paren{\frac{3\ell+4}{\ell+2}+\varepsilon}$-APASP algorithms with $\tilde O \paren{mn^{\beta}+n^{2+\gamma}+\paren{\frac{1}{\varepsilon}}^{O\paren{1}}\cdot n^{\omega\paren{1-\beta-\ell\cdot \gamma}} \cdot  \log W}$ runtime.

\begin{table}[t]
  \begin{center}
    
\small\addtolength{\tabcolsep}{-3pt}
    \renewcommand\arraystretch{1.5}
    \begin{tabular}{|c|c|c|} 
      \hline
      \textbf{\small Approximation} & \textbf{\small Combinatorial} & \textbf{\small Reference} \\
      \hhline{|=|=|=|}
      $+2\msum{W_{i}}{i=1}{k+1} $& $n^{2+\frac{1}{3k+2}}$    & \secref{+2Wi} \\
      \hline
      $\paren{\frac{6k+3}{3k+2},\msum{W_{i}}{i=1}{k+1}} $& $n^{2+\frac{1}{3k+2}}$ & \secref{tradeoffs}  \\
      \hline     $\paren{\frac{\paren{9k+4}\cdot\alpha+\paren{3k+2}\cdot\beta}{\paren{\alpha+\beta}\cdot\paren{3k+2}},\frac{2\beta}{\alpha+\beta}\msum{W_{i}}{i=1}{k+1}}$& $n^{2+\frac{1}{3k+2}}$ & \secref{tradeoffs} \\
      \hline
      $\frac{9k+4}{3k+2}$& $\leq n^{2+\frac{1}{3k+2}}$  & \cite{AkaRod2021}  \\
      \hline
    \end{tabular}
    \renewcommand\arraystretch{1}
    \caption{Runtime equivalence for $\tilde O\paren{n^{2+\frac{1}{3k+2}}}$.}~\label{tab:our_res_equiv_n1over3k+2}
    \vspace{-4mm}
  \end{center}
\end{table}

While these results primarily explore trade-offs between the runtime and the multiplicative stretch, we would also like to fix the runtime and provide trade-offs between the multiplicative and additive stretches, proceeding in alignment with the direction of Saha and Ye \cite{SahYe2023}.

We begin by focusing on cases where the multiplicative stretch is below $2$, offering a way to \qoute{bypass} the conditional lower bound of $\Omega\paren{n^{\omega}}$ of Dor, Halperin and Zwick \cite{DorHalZwi2000} for $\paren{\alpha,\beta}$-APASP algorithms where $\alpha<2$. 

\begin{question}~\label{que:under2}
For which values of $\alpha,\beta$ such that $\alpha<2$ is it possible to compute an $\paren{\alpha,\beta}$-APASP in less than $\tilde O \paren{n^{\omega}}$ runtime?
\end{question}

We partially answer this by presenting a family of $\paren{\alpha,\beta}$-APASP algorithms with $\tilde O \paren{n^{2+\frac{1}{3k+2}}}$ runtime, such as $\paren{\frac{6k+3}{3k+2},\msum{W_{i}}{i=1}{k+1}}$-APASP. Furthermore, our approach helps bridge a known gap in the literature: no combinatorial algorithms currently exists for purely multiplicative $\alpha$-APASP in weighted graphs where $\alpha\in\paren{2,\frac{7}{3}}$. The lack of such can be partially accommodated by an $\paren{\alpha,\beta}$-APASP s.t.  $\alpha\in \paren{2,\frac{7}{3}}$ and $\beta$ is small. For instance, we obtain a $\paren{\frac{11}{5},\frac{W_{1}+W_{2}}{2}}$-APASP in $\tilde O\paren{n^{\frac{11}{5}}}$ time and a $\paren{\frac{13}{6},\frac{2\cdot\paren{W_{1}+W_{2}+W_{3}}}{3}}$-APASP in $\tilde O\paren{n^{\frac{17}{8}}}$ time (See: \secref{tradeoffs}). These belong to a wider family of approximations computable in $\tilde O(n^{2+\frac{1}{3k+2}})$ time (see \tabref{our_res_equiv_n1over3k+2}). We broaden the question to any runtime function $T\paren{n}$:

\begin{question}~\label{que:tradeoffs}
Given an $\paren{\alpha_{1},\beta_{1}}$-APASP with $T\paren{n}$ runtime, what other $\paren{\alpha_{2},\beta_{2}}$-APASP can be computed within $\tilde O \paren{T\paren{n}}$ runtime?\end{question}

\subsection{Related Work}\label{related}
 Cohen and Zwick \cite{CohZwi1997} had a $2$-APASP algorithm with $\tilde O \paren{n^{\frac{3}{2}} m^{\frac{1}{2}}+n^{2}}$ runtime and a $\frac{7}{3}$-APASP in $\tilde O \paren{n^{\frac{7}{3}}}$ time. Both runtimes were improved by Baswana and Kavitha \cite{BasKav2010} to $\tilde O \left(m\sqrt{n} +n^{2}\right)$ and $\tilde O \left( m^{\frac{2}{3}}n+n^{2}\right)$, respectively. Kavitha \cite{Kavitha2012} showed a $\frac{5}{2}$-APASP algorithm with $\tilde O \paren{n^{\frac{9}{4}}}$ runtime, which was improved by Akav and Roditty \cite{AkaRod2021} to $\tilde O\left(n^{\frac{5}{4}} m ^{\frac{1}{2}} +n^{2}\right)$ time. 

These multiplicative stretches  were initially computed in an ad-hoc manner, each requiring its own tailored algorithm. The approaches for these APASP algorithms were developed without a unifying framework or a recognition of a broader underlying structure. Only later did Akav and Roditty \cite{AkaRod2021} observe that these seemingly distinct stretches are, in fact, a part of a family of problems characterized by the form of $\frac{3\ell+4}{\ell+2}$-APASP that can be computed in $\tilde O \left(n^{2-\frac{3}{\ell+2}}m^{\frac{2}{\ell+2}}+n^{2}\right)$ time. For example, $\ell=0$ indicates a $2$-APASP algorithm that runs in $\tilde O\paren{n^{\frac{1}{2}}m+n^{2}}$ and $\ell = 3$ indicates a $\frac{13}{5}$-APASP algorithm that runs in $\tilde O\paren{n^{\frac{7}{5}}m^{\frac{2}{5}}+n^{2}}$ time. 

A related problem is \textbf{M}ulti-\textbf{S}ource \textbf{S}hortest 
 \textbf{P}aths (\emph{MSSP}), where instead of computing all pairs of distances, we are given a set $S\subseteq V$ of sources and we need to compute the distances from this set to any $v\in V$. We define an $\paren{\alpha,\beta}$-MSASP similarly to an $\paren{\alpha,\beta}$-APASP. Let $\varepsilon>0$, $\abs{S} \in O\paren{n^{r}}$ and assume edge weights are from $\bracce{0,1,\ldots,W}\cup\bracce{\infty}$. Recently, Elkin and Neiman \cite{ElkNei2020} a $\paren{1+\varepsilon}$-MSASP algorithm with $\tilde O \paren{m^{1+o\paren{1}} +  \paren{ \frac{1}{\varepsilon} }^{O\paren{1}} \cdot n^{\omega\paren{r}} \cdot \log W }$ runtime. 

Dory, Forster, Kirkpatrick, Nazari, Vassilevska Williams and de Vos \cite{DorForKirNazVasVos2023} utilized the result of Elkin and Neiman \cite{ElkNei2020} and the $2$-APASP of Baswana and Kavitha \cite{BasKav2010}. They achieved a  $\paren{2+\varepsilon}$-APASP algorithm with a $\tilde O \paren{\paren{\frac{1}{\varepsilon}}^{O\paren{1}} \cdot n^{2.21235201}\cdot \log W}$ runtime, which improves the runtime of the $2$-APASP of Baswana and Kavitha \cite{BasKav2010} for dense graphs. We generalize this  $\paren{2+\varepsilon}$-APASP algorithm \cite{DorForKirNazVasVos2023} and our own $\paren{\frac{7}{3}+\varepsilon}$-APASP algorithm, similarly to Akav and Roditty \cite{AkaRod2021}, and present a unified framework of $\paren{\frac{3\ell+4}{\ell+2}+\varepsilon}$-APASP algorithms.

An equivalent approach was taken by Saha and Ye \cite{SahYe2023}. They  computed a $(1+\varepsilon,2\msum{W_{i}}{i=1}{k+1})$-APASP. Their base case was $\paren{1+\varepsilon,2W_{1}}$-APASP with $\tilde O \paren{\frac{1}{\varepsilon}\cdot n^{2.15135313} \cdot \log W}$ runtime. We improve the base case and compute a $\paren{1+\varepsilon,\mmin{2W_{1},4W_{2}}}$-APASP, in the same runtime.

\section{Toolkit}\label{toolkit}
\subsection{Basic graph notions}\label{notions}
Unless stated otherwise, we consider an undirected graph $G=\left(V,E,w\right)$ with a non-negative weight function $w:E\rightarrow \mathbb{R}^{\geq 0}$, where the weight of an edge $\paren{u,v}\in E$ is $w\paren{u,v}$. Denote $\abs{V} = n$ and $\abs{E}= m$. For $k\in \mathbb{N}$, let $\bracke{k}=\bracce{1,2,\ldots,k}$.  Each vertex $u\in V$ has its set of neighbours $N\paren{u}$ and the set $\Gamma\paren{u,k}$ of $k$-nearest neighbours. A \emph{path} from a vertex $u$ to a vertex $v$ is an ordered list $u=y_{0},y_{1},y_{2},\ldots,y_{k} , y_{k+1}=v$ s.t. for any $i\in\bracke{k+1}$: $\paren{y_{i-1},y_{i}}\in E$. The weight of a path is $w\paren{P} = \sum_{e\in P}{w\paren{e}}$. A path with the smallest weight is a \textit{shortest path} and its weight is the \emph{distance}, denoted by $\delta\left(u,v\right)$. The distance between a vertex $u$ to a set of vertices $S\subseteq V$ is the distance from $u$ to the nearest $s\in S$. That is, $\delta\left(u,S\right) = \mmin{\delta\left(u,s\right)}{s\in S} $. The \emph{pivot} of $u$ is a\footnote{If there are several candidates we can break ties by a consistent criteria.} vertex $p_{S}\paren{u}$ such that $\delta\paren{u,p_{S}\paren{u}}=\delta\paren{u,S}$.

Denote a\footnote{Several shortest paths may exist. If nothing is specifically mentioned, we consider an arbitrary one.} shortest path between $u$ and $v$ by $u \squiggly v$: $w\paren{u\squiggly v } = \delta\paren{u,v}$. If $u \squiggly v$ is a single edge, we write $u\straight v$, hence $w\paren{u,v}=\delta\paren{u,v}$. Let  $W_{i}\paren{u\squiggly v}$ be the weight of the $i^{\textnormal{th}}$ heaviest edge. That is, $W_{1}\paren{u\squiggly v}$ for the  heaviest, $W_{2}\paren{u\squiggly  v}$ for the second heaviest, etc. The notion of a $W_{i}\paren{u\squiggly v}$ applies to the fixed path $u\squiggly v$, and so does $\msum{W_{i}}{i\in I}$. Yet, if no specific $u\squiggly v$ is stated, we consider \textbf{any}\footnote{Or, the smallest sum.} shortest path, but all edges must be \textbf{from the same path}. 

We denote by $d\bracke{u,v}$ the distance estimate computed by our algorithms. For an $\paren{\alpha,\beta}$-APASP algorithm, we prove that $\delta\paren{u,v} \leq d\bracke{u,v} \leq \alpha\cdot \delta\paren{u,v} + \beta$. Since $G$ is undirected, we assume that any update to $d\bracke{u,v}$ also updates $d\bracke{v,u}$. This adds only $O\paren{1}$ per update and does not affect the overall runtime.

\subsection{Nearest vertices and pivots}~\label{nearest}
\begin{figure}[H]
\centering

\begin{tikzpicture}[x=0.75pt,y=0.75pt,yscale=-1,xscale=1]

\draw  [color={rgb, 255:red, 74; green, 144; blue, 226 }  ,draw opacity=0.55 ][fill={rgb, 255:red, 74; green, 144; blue, 226 }  ,fill opacity=0.29 ] (301.45,234.55) .. controls (272.97,209.21) and (270.06,166) .. (294.94,138.04) .. controls (319.82,110.09) and (363.07,107.97) .. (391.55,133.32) .. controls (420.03,158.67) and (422.94,201.87) .. (398.06,229.83) .. controls (373.18,257.78) and (329.93,259.9) .. (301.45,234.55) -- cycle ;
\draw    (314.73,189.53) -- (361.23,162.53) ;
\draw    (314.73,189.53) -- (358.73,226.2) ;
\draw    (358.73,226.2) -- (434.75,226.5) ;
\draw    (359.73,225.2) -- (434.25,164) ;
\draw    (434.25,164) -- (434.75,222.5) ;
\draw    (198.43,200.43) -- (248.43,190.43) ;
\draw    (198.43,200.43) -- (255,153.67) ;
\draw    (252.43,186.43) -- (259,153.67) ;
\draw    (263.43,153.43) -- (357.23,162.53) ;
\draw    (365.23,162.53) -- (434.25,163) ;
\draw  [color={rgb, 255:red, 74; green, 144; blue, 226 }  ,draw opacity=1 ][fill={rgb, 255:red, 74; green, 144; blue, 226 }  ,fill opacity=1 ] (357.23,162.53) .. controls (357.23,160.32) and (359.02,158.53) .. (361.23,158.53) .. controls (363.44,158.53) and (365.23,160.32) .. (365.23,162.53) .. controls (365.23,164.74) and (363.44,166.53) .. (361.23,166.53) .. controls (359.02,166.53) and (357.23,164.74) .. (357.23,162.53) -- cycle ;
\draw  [color={rgb, 255:red, 0; green, 0; blue, 0 }  ,draw opacity=1 ][fill={rgb, 255:red, 0; green, 0; blue, 0 }  ,fill opacity=1 ] (354.73,226.2) .. controls (354.73,223.99) and (356.52,222.2) .. (358.73,222.2) .. controls (360.94,222.2) and (362.73,223.99) .. (362.73,226.2) .. controls (362.73,228.41) and (360.94,230.2) .. (358.73,230.2) .. controls (356.52,230.2) and (354.73,228.41) .. (354.73,226.2) -- cycle ;
\draw  [color={rgb, 255:red, 0; green, 0; blue, 0 }  ,draw opacity=1 ][fill={rgb, 255:red, 0; green, 0; blue, 0 }  ,fill opacity=1 ] (310.73,189.53) .. controls (310.73,187.32) and (312.52,185.53) .. (314.73,185.53) .. controls (316.94,185.53) and (318.73,187.32) .. (318.73,189.53) .. controls (318.73,191.74) and (316.94,193.53) .. (314.73,193.53) .. controls (312.52,193.53) and (310.73,191.74) .. (310.73,189.53) -- cycle ;
\draw  [color={rgb, 255:red, 0; green, 0; blue, 0 }  ,draw opacity=1 ][fill={rgb, 255:red, 0; green, 0; blue, 0 }  ,fill opacity=1 ] (430.25,163) .. controls (430.25,160.79) and (432.04,159) .. (434.25,159) .. controls (436.46,159) and (438.25,160.79) .. (438.25,163) .. controls (438.25,165.21) and (436.46,167) .. (434.25,167) .. controls (432.04,167) and (430.25,165.21) .. (430.25,163) -- cycle ;
\draw  [color={rgb, 255:red, 0; green, 0; blue, 0 }  ,draw opacity=1 ][fill={rgb, 255:red, 0; green, 0; blue, 0 }  ,fill opacity=1 ] (255,153.67) .. controls (255,151.46) and (256.79,149.67) .. (259,149.67) .. controls (261.21,149.67) and (263,151.46) .. (263,153.67) .. controls (263,155.88) and (261.21,157.67) .. (259,157.67) .. controls (256.79,157.67) and (255,155.88) .. (255,153.67) -- cycle ;
\draw  [color={rgb, 255:red, 0; green, 0; blue, 0 }  ,draw opacity=1 ][fill={rgb, 255:red, 0; green, 0; blue, 0 }  ,fill opacity=1 ] (248.43,190.43) .. controls (248.43,188.22) and (250.22,186.43) .. (252.43,186.43) .. controls (254.64,186.43) and (256.43,188.22) .. (256.43,190.43) .. controls (256.43,192.64) and (254.64,194.43) .. (252.43,194.43) .. controls (250.22,194.43) and (248.43,192.64) .. (248.43,190.43) -- cycle ;
\draw  [color={rgb, 255:red, 0; green, 0; blue, 0 }  ,draw opacity=1 ][fill={rgb, 255:red, 0; green, 0; blue, 0 }  ,fill opacity=1 ] (194.43,200.43) .. controls (194.43,198.22) and (196.22,196.43) .. (198.43,196.43) .. controls (200.64,196.43) and (202.43,198.22) .. (202.43,200.43) .. controls (202.43,202.64) and (200.64,204.43) .. (198.43,204.43) .. controls (196.22,204.43) and (194.43,202.64) .. (194.43,200.43) -- cycle ;
\draw  [color={rgb, 255:red, 0; green, 0; blue, 0 }  ,draw opacity=1 ][fill={rgb, 255:red, 0; green, 0; blue, 0 }  ,fill opacity=1 ] (430.75,226.5) .. controls (430.75,224.29) and (432.54,222.5) .. (434.75,222.5) .. controls (436.96,222.5) and (438.75,224.29) .. (438.75,226.5) .. controls (438.75,228.71) and (436.96,230.5) .. (434.75,230.5) .. controls (432.54,230.5) and (430.75,228.71) .. (430.75,226.5) -- cycle ;
\draw    (248.43,190.43) -- (314.73,189.53) ;
\draw[shorten <=4pt, shorten >=4pt] (361.23,161.53) -- (358.73,226.2);

\draw (365,194.4) node
[anchor=west,inner sep=0.75pt,font=\footnotesize]
{$\displaystyle 1$};

\draw (367.5,170.5) node [anchor=north west][inner sep=0.75pt]
[font=\footnotesize,color={rgb,255:red,74; green,144; blue,226},opacity=1,xslant=-0.03]
{$\displaystyle u$};

\draw (339.92,226.5) node [anchor=north west][inner sep=0.75pt]  [font=\footnotesize,xslant=-0.03] [align=left] {$\displaystyle a_{2}$};
\draw (443.83,219.17) node [anchor=north west][inner sep=0.75pt]  [font=\footnotesize,xslant=-0.03] [align=left] {$\displaystyle a_{7}$};
\draw (443,155) node [anchor=north west][inner sep=0.75pt]  [font=\footnotesize,xslant=-0.03] [align=left] {$\displaystyle a_{6}$};
\draw (192.08,205.5) node [anchor=north west][inner sep=0.75pt]  [font=\footnotesize,color={rgb, 255:red, 0; green, 0; blue, 0 }  ,opacity=1 ,xslant=-0.03] [align=left] {$\displaystyle a_{4}$};
\draw (254.08,132.17) node [anchor=north west][inner sep=0.75pt]  [font=\footnotesize,color={rgb, 255:red, 0; green, 0; blue, 0 }  ,opacity=1 ,xslant=-0.03] [align=left] {$\displaystyle a_{5}$};
\draw (247.51,196.43) node [anchor=north west][inner sep=0.75pt]  [font=\footnotesize,color={rgb, 255:red, 0; green, 0; blue, 0 }  ,opacity=1 ,xslant=-0.03] [align=left] {$\displaystyle a_{3}$};
\draw (397,213) node [anchor=north west][inner sep=0.75pt]  [font=\footnotesize] [align=left] {$\displaystyle 1$};
\draw (275.83,193.33) node [anchor=north west][inner sep=0.75pt]  [font=\footnotesize] [align=left] {$\displaystyle 2$};
\draw (389.5,163.5) node [anchor=north west][inner sep=0.75pt]  [font=\footnotesize] [align=left] {$\displaystyle 4$};
\draw (327,207.17) node [anchor=north west][inner sep=0.75pt]  [font=\footnotesize] [align=left] {$\displaystyle 1$};
\draw (398,193.33) node [anchor=north west][inner sep=0.75pt]  [font=\footnotesize] [align=left] {$\displaystyle 1$};
\draw (328.5,163.67) node [anchor=north west][inner sep=0.75pt]  [font=\footnotesize] [align=left] {$\displaystyle 2$};
\draw (424.33,191) node [anchor=north west][inner sep=0.75pt]  [font=\footnotesize] [align=left] {$\displaystyle 9$};
\draw (221,159.33) node [anchor=north west][inner sep=0.75pt]  [font=\footnotesize] [align=left] {$\displaystyle 5$};
\draw (223.17,198.67) node [anchor=north west][inner sep=0.75pt]  [font=\footnotesize] [align=left] {$\displaystyle 6$};
\draw (295.75,190.17) node [anchor=north west][inner sep=0.75pt]  [font=\footnotesize,xslant=-0.03] [align=left] {$\displaystyle a_{1}$};
\draw (307.47,127.01) node [anchor=north west][inner sep=0.75pt]  [font=\footnotesize,color={rgb, 255:red, 45; green, 127; blue, 226 }  ,opacity=1 ] [align=left] {$\displaystyle \Gamma ( u,2)$};
\draw (295.94,142.04) node [anchor=north west][inner sep=0.75pt]  [font=\footnotesize] [align=left] {$\displaystyle 3$};
\draw (260.17,165.67) node [anchor=north west][inner sep=0.75pt]  [font=\footnotesize] [align=left] {$\displaystyle 1$};

\end{tikzpicture}

\caption{\label{fig:nearestexample} The set $\Gamma \paren{u,2} = \bracce{a_{1},a_{2}}$.}
\end{figure}\vspace{-2mm}
We provide two equivalent definitions, one for random algorithms and one for determinisitic algorithms. Starting with deterministic,  
provided several sets $T_{1},T_{2},\ldots,T_{\ell}$ of elements from a universe $U=\bracce{u_{1},u_{2},\ldots,u_{n}}$, a \emph{hitting set} is a set $S\subseteq U$ such that for any $i\in \bracke{\ell}$: $S\cap T_{i} \neq \varnothing$. In our context, $U=V$ and $T_{u}=\Gamma\paren{u,k}$ for $u\in V$ and some $k$ that will be later specified. A hitting set, in this sense, is a set $S\subseteq V$  of small size which contains at least one  of the $k$-nearest vertices of $u$, for all $u\in V$. Hence its pivot $p_{S}\paren{u}$ is the nearest vertex between $u$ and $S$. That is, $\delta\paren{u,p_{S}\paren{u}} \leq \delta\paren{u,s}$ for all $s\in S$. These definitions are useful for APASP by distinguishing between pairs of vertices $u,v\in V$ that are \qoute{far} from each other -- and those that are \qoute{near} -- that is, one lies within $\Gamma\paren{\cdot,k}$ of the other. For near vertices, usually the exact distance is computed, by an algorithm \cite{BasKav2010, ThoZwi2005} similar to Dijkstra, while for \qoute{far} vertices, we use an approximation on paths that pass through the pivots.

\begin{observation}[\cite{AinCheIndMot1999}]~\label{obv:hs} 
Let $r\in \mathbb{N}$ such that $\abs{T_{i}}\geq r$ for any $i\in\bracke{\ell}$. A hitting set $S\subseteq U$ of size $\tilde O \paren{\frac{n}{r}}$ can be deterministically computed in $O\paren{nr}$ time.
\end{observation}

\begin{observation}~\label{obv:ball} 
Let $k\in \bracke{n}$. 
The sets $\Gamma\paren{u,k}$ can be deterministically computed, for all
$u\in V$, in
$\tilde O\paren{m+nk}$ time.
\end{observation}

\begin{proof}
Sort the neighbours of all vertices $u\in V$ by the weight of their edge in $\tilde O \left( m \right)$ time. Construct the set $\Gamma\paren{u,k}$ by adding $k$ vertices per $u\in V$. As each vertex has $k$  such neighbours, this requires $ O \paren{nk}$ runtime.
\end{proof}

For our random algorithms, we define these sets slightly differently.  For a parameter $q\in\paren{0,1}$ we now consider a set $S$ of vertices $s\in V$ sampled randomly and independently with probability $q$ each. For a vertex $u\in V$, we define its \emph{pivot} with respect to $S$ as a vertex $p_{S}\paren{u}\in S$ satisfying $\delta\paren{u,p_{S}\paren{u}}=\delta\paren{u,S}$. That is, the pivot of $u$ is the closest vertex to $u$ among members of  $S$. If there are several candidates we can break ties by a consistent criteria. We also associate $u$ with its \emph{ball} with respect to $S$, defined as: $\ball{u}{S} = \bracce{ v \in V \midline \delta\paren{u,v}<\delta\paren{u,p_{S}\paren{u}}}$.

The above can be extended for an $A\subseteq V$. Formally, we define
$\ball{u}{A}{S} = \bracce{ a\in A \midline \delta\paren{u,a} < \delta\paren{u,p_{S}\paren{u}}}$. Let $q'\in \paren{0,1}$ be the probability for a vertex from $A$ to belong to $S$.

\begin{observation}[\cite{ThoZwi2005}]~\label{obv:bunchsize} 
$E\bracke{\abs{\ball{u}{A}{S}}} \in \tilde O \paren{\frac{1}{q'}}$.
\end{observation}

 Regardless of $\ball{u}{A}{S}$ or $\Gamma\paren{u,k}$, we define $E_{S}\paren{u} = \left\{ \left(u,v\right)  \midline w\left(u,v\right) < \delta\left(u,S\right) \right\}$ as the set of edges near $u$ whose weight is smaller than $\delta\paren{u,S}$, and $E_{S} = \mcup{E_{S}\left(u\right)}{u\in V} $. We remark that if a vertex $u$ has less than $k$ neighbours, we include all its edges in $E_{S}\paren{u}$. 
 
 Denote the edges $H_{S} = \bracce{\paren{u,p_{S}\paren{u}} \midline u\in V}$. If there are multiple hitting sets, we set $H = \mcup{H_{S}}{S} $ as their union. For our deterministic algorithms, when concerning a hierarchy of hitting-sets $S_{1}\supseteq S_{2} \supseteq \ldots$, we assume\footnote{Otherwise, simply add $S_{i+1}$ to $S_{i}$. The asymptotic size will not change.} that $S_{i+1}\subseteq S_{i}$, and that each $S_{i}$ has its set of edges $E_{S_{i}}$ and pivots $p_{i}\paren{u}\in S_{i}$ for each $u\in V$. 

 \begin{observation}[\cite{BasKav2010}]~\label{obv:pathinball} 
Let $u\in V$, $S\subseteq V$ and  $v\in \ball{u}{S}$. The edges of $u\squiggly v$ are contained within $E_{S}$.
\end{observation}

\begin{observation}[\cite{ThoZwi2005}]~\label{obv:edgessize}
Let $u\in V$. For our deterministic algorithms $\abs{E_{S}\paren{u}} \leq k$ and $\abs{ E_{S} }\leq nk$. For our random algorithms, $E\bracke{\abs{E_{S}\paren{u}}} \in \tilde O \paren{\frac{1}{q}}$ and $E\bracke{\abs{E_{S}}} \in \tilde O \paren{\frac{n}{q}}$.
\end{observation}

We note that finding the pivots themselves and the exact distances from each vertex to its pivot, can be accomplished by a single SSSP invocation with an auxiliary vertex connected to $S$ with edges of weight $0$.

\begin{observation}~\label{obv:pivotsdistance} 
Let $k\in \bracke{n}$. The pivot $p_{S}\paren{u}$ of each vertex $u\in V$ and the exact distance $d\bracke{u,p_{S}\paren{u}} = \delta\paren{u,p_{S}\paren{u}}$ can be computed in $\tilde O \paren{m}$ time.
\end{observation}

\begin{observation}~\label{obv:edgesruntime}
The set $ E_{S} $ can be computed in $\tilde O \paren{nk}$ time deterministically or in an expected $\tilde O \paren{\frac{1}{q}}$ time randomly.
\end{observation}

\begin{observation}~\label{obv:ESiwithinESi+1}
If $S_{2}\subseteq S_{1}$ then  $ E_{S_{1}} \subseteq E_{S_{2}}$.
\end{observation}

\subsection{Min-plus matrix multiplication}~\label{minplus}
\textbf{M}in \textbf{P}lus \textbf{M}atrix \textbf{M}ultiplication\footnote{Also known as Distance Product (\emph{DP}).} (\emph{MPMM}) is a  problem similar to \textbf{M}atrix \textbf{M}ultiplication (\emph{MM}), yet uses a different multiplication operator: given two matrices $A$ and $B$, their product $C = A \star B$, is defined to as: 

$$C\left[i,j\right]  = \min_{k=1}^{n} \left\{A\left[i,k\right]+ B\left[k,j\right]\right\}$$

MPMM is runtime-equivalent to APSP \cite{VaiHopUll1974} and is often used as a tool to reduce the runtime of shortest paths problems. However, MM and MPMM are conjectured not to be runtime-equivalent\footnote{If they would have been, as we already know $\omega<3$, we could refute the APSP conjecture.}. Nonetheless, faster MM algorithms can be utilized for MPMM by redefining the matrices (e.g. \cite{ShoZwi1999, Zwick2002, BriKunWun2019}). Since Strassen's algorithm \cite{Strassen1969}, there was a series of works, many of whom in recent years (e.g., \cite{AmbFilLeg2014, DuaWuZho2022, VasXuXuZho2024, AlmDuaVasXuXuZho2024}), which improved the runtime of fast MM. However, these algorithms \qoute{hide} large constants in their runtime.
  
Zwick \cite{Zwick2002} introduced \textbf{A}pproximated \textbf{M}in-\textbf{P}lus \textbf{M}atrix \textbf{M}ultiplication (\emph{AMPMM}) in the same sense as APASP and presented a $\paren{1+\varepsilon}$-AMPMM algorithm with $\tilde O \left(\frac{1}{\varepsilon} \cdot n^\omega \cdot \log W  \right)$ runtime. That is, this algorithm computes a matrix $\tilde C$, where $C\bracke{i,j} \leq \tilde C\bracke{i,j} \leq \paren{1+\varepsilon}\cdot C\bracke{i,j}$. Bringmann, K\"{u}nnemann and Wegrzycki \cite{BriKunWun2019} later utilized for a $\paren{1+\varepsilon}$-APASP   in $\tilde O \left(\frac{1}{\varepsilon} \cdot n^{\omega} \cdot \log\frac{n}{\varepsilon} \right)$ time. 


\section{Overview}\label{overview}
Common to all our algorithms is a hierarchy of  hitting sets $V = S_{0} \supseteq S_{1} \supseteq \ldots \supseteq \varnothing$, where any $S_{i} \subseteq S_{i+1}$. Each level $S_i$ has its own pivots $p_{i}\paren{u}$ for every $u \in V$ and an associated edge set $E_{S_{i}}$ (see: \secref{nearest}). The number and sizes of the hitting sets differs between the algorithms.

From the combinatorial perspective, our methods consist of selecting pivots through hitting sets and invoking multiple SSSP on graphs with sets of original and auxiliary sets of edges, tailored for the desired approximation. Some of our algorithms are order-sensitive considering the various SSSP invocations, as explicitly stated if so. We also consider distance updates through selected vertices, i.e. we have $d\bracke{u,x}$ and $d\bracke{x,v}$ and we update $d\bracke{u,v}$ to be $\mmin{d\bracke{u,v},d\bracke{u,x}+d\bracke{x,v}}$.

Our main contribution in the combinatorial setting is our $+2\msum{W_{i}}{i=1}{k+1}$-APASP algorithm for dense weighted graphs with a runtime of $\tilde O\paren{n^{2+\frac{1}{3k+2}}}$. Extending the $+2\cdot\paren{k+1}$-APASP algorithm of Dor, Halperin, and Zwick \cite{DorHalZwi2000} to the weighted setting gives rise to several challenges. First, we use $\Gamma\paren{u,k}$ instead of $N\paren{u}$. This change is required to ensure that if there exists an edge $\paren{x,y} \in u\squiggly v$ s.t. $\paren{x,y}\notin E_{S_{i}}$, then $\delta\paren{x,p_{i}\paren{x}}\leq w\paren{x,y}$, hence a path  $u\squiggly  p_{i}\paren{x}  \squiggly v$ would not \qoute{cost} us more than $2\cdot w\paren{x,y}$, which is at most $2W_{1}\paren{u\squiggly v}$. Repeating this approach $k+1$ times results in an additive error of at most $2\cdot\paren{k+1}\cdot W_{1}\paren{u\squiggly v}$. However, our goal is to avoid reusing the weight $W_1$ of the heaviest edge  repeatedly and to use the lighter edges $W_{2}, W_{3},...,W_{k+1}$. To obtain this, we must fix a certain shortest path and carefully make sure we do not use any edge more than once in the upper-bound for our additive error. We achieve that by recursively concatenating auxiliary paths through pivots, ensuring that each additive term corresponds to a distinct edge from the original $u\squiggly v$. We emphasize that this issue $\,\,$ does not exist in the unweighted setting, where the upper-bound is simply $+2\cdot\paren{k+1}$.

From the algebraic perspective, our methods consist of invoking $\paren{1+\varepsilon}$-MSASP instead of several SSSP invocations, and $\paren{1+\varepsilon}$-AMPMM for designated matrices. Note that rectangular MPMM can be used as a tool to replace  distance updates through selected vertices. 

Our main contribution in the algebraic setting is our unified framework of $\paren{\frac{3\ell+4}{\ell+2}+\varepsilon}$-APASP algorithms with $\tilde O \paren{mn^{\beta}+n^{2+\gamma}+\paren{\frac{1}{\varepsilon}}^{O\paren{1}}\cdot n^{\omega\paren{1-\beta-\ell\cdot \gamma}} \cdot  \log W}$ runtime. For dense graphs $m \in \Theta\paren{n^{2}}$, setting $\ell=0$ results in the $\paren{2+\varepsilon}$-APASP algorithm of Dory, Forster, Kirkpatrick, Nazari, Vassilevska Williams and de Vos \cite{DorForKirNazVasVos2023}, setting $\ell=1$ results in our own $\paren{\frac{7}{3}+\varepsilon}$-APASP algorithm.  For $\ell=2$ we obtain a $\paren{\frac{5}{2}+\varepsilon}$-APASP algorithm with $\tilde O \paren{\paren{\frac{1}{\varepsilon}}^{O\paren{1}}\cdot n^{2.1185119}\cdot \log W}$ runtime. These results improve upon the  framework of Akav and Roditty \cite{AkaRod2021} for dense graphs, at the cost of an additional $\varepsilon$ to the multiplicative stretch.

\section{Warm-up: additive $+2W_{1}$-APASP}\label{+2W1}

\begin{algorithm}[H]
\caption{$+2W_{1}$-APASP$\left(G=\left(V,E,w\right)\right)$}\label{alg:+2W1}

\textbf{Input:} An undirected non-negative weighted graph $G=\left(V,E,w\right)$.

\textbf{Output:} A $+2W_{1}$-APASP. 

\medskip

Let $\beta,\gamma \in \left( 0 , 1 \right)$ to be fixed later

Initialize $d$ as the edges' weights and $\infty$ otherwise

Compute $\Gamma\paren{u,n^{\beta}}$ (respectively, $\Gamma\paren{u,n^{\beta+\gamma}}$) for all $u\in V$

Compute $S_{1}$ (respectively, $S_{2}$) to be a hitting set to $\bracce{\Gamma\paren{u,n^{\beta}} \mid u\in V}$ (respectively, $\Gamma\paren{u,n^{\beta+\gamma}}$)
 

Compute the distances from any $u\in V$ to its pivot $p_{1}\paren{u}$ (respectively, $p_{2}\paren{u}$)

Construct the set of edges $E_{S_{1}}$ (respectively, $E_{S_{2}}$)

\For{$i \in \pair{2,1,0}$}
{
    \For{$s\in  S_{i}$}
    {
        Invoke SSSP from $s$ on $G_{i} \paren{s} = \left( V,E_{ 
 S_{i+1}} \cup \paren{\bracce{s}\times V} \cup H \cup \paren{S_{2} \times S_{0}} , d \right)$ and update $d$ accordingly
    }
}

\Return $d$

\end{algorithm}

We begin by presenting \algref{+2W1}, which is a deterministic algorithm that computes a $+2W_{1}$-APASP and has a runtime of  $\tilde O \paren{n^{\frac{7}{3}}}$. \algref{+2W1} can be seen as a  a simplified version of a $\mmin{+2W_{1},+4W_{2}}$-APASP algorithm by Cohen and Zwick \cite{CohZwi1997}, which had the same runtime. However, we omit the \qoute{$+4W_{2}$} part in-order to present a simplified case-analysis. \algref{+2W1} will later serve as a foundation for the generalized $+2\msum{W_{i}}{i=1}{k+1}$-APASP algorithm, namely \algref{+2Wi}, that will be discussed in \secref{+2Wi}.

\algref{+2W1} works as follows: we initialize a matrix $d:V\times V \rightarrow \mathbb{R}$ to be the weight of the edge for two vertices with an edge or $\infty$ else-wise. We compute a hitting set $S_{1}$ (resp. $S_{2}$) to the sets $\Gamma\paren{u,n^{\beta}}$ (resp. $\Gamma\paren{u,n^{\beta+\gamma}}$) for all $u\in V$. For  $S_{1}$ (resp. $S_{2}$) we also compute its corresponding $E_{S_{1}}$ (resp. $E_{S_{2}}$). We set $S_{0}=V$ and $S_{3}=\varnothing$, hence $E_{S_{0}}=\varnothing$ and $E_{S_{3}}=E$. For each $u\in V$ we find its pivot $p_{1}\paren{u}$ (resp. $p_{2}\paren{u}$)  and the distance $d\bracke{u,p_{1}\paren{u}} \leftarrow \delta\paren{u,p_{1}\paren{u}}$ (resp. $d\bracke{u,p_{2}\paren{u}}$) by a single SSSP invocation (See: \obvref{pivotsdistance}). 

Let $i\in \pair{2,1,0}$ be an index\footnote{The order of $i\leftarrow 2,1,0$ is of importance here, as will be argued further on.}. \algref{+2W1} invokes SSSP from all $s\in S_{i}$ over the weighted graph $G_{i}\paren{s} = \paren{V,E_{S_{i+1}} \cup \paren{\bracce{s}\times V} \cup H \cup \paren{S_{2}\times S_{0}},d}$, where:

\begin{itemize}
    \item $E_{S_{i+1}}$ - The union, over \textbf{all} $u\in V$, of edges $\paren{u,v}$ such that $w\paren{u,v}<\delta\paren{u,p_{i+1}\paren{u}}$,
    \item $\bracce{s} \times V$ - Auxiliary edges from $s$ to  $V$, whose weight represents the distances that were computed in previous iterations,
    \item $H =  \bracce{\paren{u,p_{j}\paren{u}} \midline u \in V}$ - Auxiliary edges from \textbf{any} vertex $u$ to its pivots $p_{j}\paren{u}$ for $j\in\bracce{1,2}$,
    \item $S_{2} \times S_{0}$ - Auxiliary edges, , whose weight represents a distance that was previously computed, from \textbf{any} vertex $s_{2}\in S_{2}$ to \textbf{any} vertex $u\in V = S_{0}$.
\end{itemize}

The weight of each such edge, whether auxiliary or original, is given by the matrix $d:V\times V \rightarrow \mathbb{R}^{\geq 0}$. In other words, for any $i \in \pair{2,1,0}$ and $s\in S_{i}$, the matrix $d$ is a weight function over $G_{i}\paren{s}$. 

We emphasize that these SSSP invocations are performed in descending order from $i\leftarrow 2$ to $0$, as auxiliary edges will have a weight given by $d$, which is a distance estimate computed in a previous iteration. To see that, recall that whenever $d\bracke{p_{2}\paren{u},v}$ is assigned a smaller value than it previously held, the value $d\bracke{v,p_{2}\paren{u}}$ is updated as well. Therefore, when an SSSP invocation from $v\in V=S_{0}$ will occur, the auxiliary edge $\paren{v,p_{2}\paren{u}}$ will be in use as well, and its weight is the previously computed value $d\bracke{v,p_{2}\paren{u}}$. 

This concludes the description of \algref{+2W1}. We now turn to prove that \algref{+2W1} computes a $+2W_{1}$-APASP. To do so, we consider an arbitrary pair $u,v\in V$ and a shortest path $u\squiggly v$. Let $\paren{a,b}\in u \squiggly v$ be a heaviest edge on this path. Therefore, our goal is to prove that $d\bracke{u,v}\leq \delta\paren{u,v}+2w\paren{a,b}$. To do so, we consider the following cases and sub-cases: 
\begin{itemize}
    \item \notcase{a}: There exists a vertex $x\in u\squiggly v$ such that $\delta\paren{x,p_{2}\paren{x}}\leq w\paren{a,b}$,
    \item \case{a}: For any $x\in u\squiggly v$: $w\paren{a,b}<\delta\paren{x,p_{2}\paren{x}}$:
    \begin{itemize}
        \item \notcase{b}: There exists a vertex $z\in u\squiggly v$ such that $\delta\paren{z,p_{1}\paren{z}}\leq w\paren{a,b}$,
        \item \case{b}: For any $z\in u\squiggly v$: $w\paren{a,b}<\delta\paren{z,p_{1}\paren{z}}$.
    \end{itemize}
\end{itemize}

We prove an upper bound on $d$ for \notcase{a}, \notcase{b} and \case{b} in \lemref{+2W1_case_nota}, \lemref{+2W1_case_notb} and \lemref{+2W1_case_b}, respectively, hence covering  all possibilities. We refer to the collection of cases as the \textit{case tree} and provide a brief summary in \figref{+2W1_casetree}. We now begin with the proofs.

\begin{figure}[t]
\centering

\begin{tikzpicture}[x=0.75pt,y=0.75pt,yscale=-1,xscale=1]

\draw   (423.36,51.5) -- (488.77,51.5) .. controls (497.28,51.5) and (504.17,60.38) .. (504.17,71.33) .. controls (504.17,82.29) and (497.28,91.17) .. (488.77,91.17) -- (423.36,91.17) .. controls (414.86,91.17) and (407.97,82.29) .. (407.97,71.33) .. controls (407.97,60.38) and (414.86,51.5) .. (423.36,51.5) -- cycle ;
\draw    (409,62.83) -- (502.33,62.5) ;
\draw    (417.76,26.31) -- (435.47,47.99) ;
\draw [shift={(437.37,50.31)}, rotate = 230.76] [fill={rgb, 255:red, 0; green, 0; blue, 0 }  ][line width=0.08]  [draw opacity=0] (6.25,-3) -- (0,0) -- (6.25,3) -- cycle    ;
\draw    (370.57,26.31) -- (357.92,47.24) ;
\draw [shift={(356.37,49.81)}, rotate = 301.15] [fill={rgb, 255:red, 0; green, 0; blue, 0 }  ][line width=0.08]  [draw opacity=0] (6.25,-3) -- (0,0) -- (6.25,3) -- cycle    ;
\draw   (307.5,52.5) -- (379.04,52.5) .. controls (388.33,52.5) and (395.87,59.29) .. (395.87,67.67) .. controls (395.87,76.04) and (388.33,82.83) .. (379.04,82.83) -- (307.5,82.83) .. controls (298.2,82.83) and (290.67,76.04) .. (290.67,67.67) .. controls (290.67,59.29) and (298.2,52.5) .. (307.5,52.5) -- cycle ;
\draw    (292.17,63.17) -- (394.83,63.17) ;
\draw    (361.76,82.31) -- (379.47,103.99) ;
\draw [shift={(381.37,106.31)}, rotate = 230.76] [fill={rgb, 255:red, 0; green, 0; blue, 0 }  ][line width=0.08]  [draw opacity=0] (6.25,-3) -- (0,0) -- (6.25,3) -- cycle    ;
\draw    (314.57,82.31) -- (301.92,103.24) ;
\draw [shift={(300.37,105.81)}, rotate = 301.15] [fill={rgb, 255:red, 0; green, 0; blue, 0 }  ][line width=0.08]  [draw opacity=0] (6.25,-3) -- (0,0) -- (6.25,3) -- cycle    ;
\draw   (369.65,106.17) -- (433.48,106.17) .. controls (441.78,106.17) and (448.5,114.9) .. (448.5,125.67) .. controls (448.5,136.44) and (441.78,145.17) .. (433.48,145.17) -- (369.65,145.17) .. controls (361.36,145.17) and (354.63,136.44) .. (354.63,125.67) .. controls (354.63,114.9) and (361.36,106.17) .. (369.65,106.17) -- cycle ;
\draw    (356.67,117.17) -- (446.67,116.67) ;
\draw   (258.51,106.17) -- (326.53,106.17) .. controls (335.37,106.17) and (342.54,114.97) .. (342.54,125.83) .. controls (342.54,136.69) and (335.37,145.5) .. (326.53,145.5) -- (258.51,145.5) .. controls (249.67,145.5) and (242.5,136.69) .. (242.5,125.83) .. controls (242.5,114.97) and (249.67,106.17) .. (258.51,106.17) -- cycle ;
\draw    (244.5,117.17) -- (340.5,117.17) ;
\draw    (370.57,26.31) -- (396.67,14.67) ;
\draw    (417.76,26.31) -- (396.67,14.67) ;

\draw (441.57,53) node [anchor=north west][inner sep=0.75pt]  [font=\tiny] [align=left] {$\displaystyle {\displaystyle Case} \ (\overline{a})$};
\draw (403.97,63.5) node [anchor=north west][inner sep=0.75pt]  [font=\tiny] [align=left] {$\displaystyle  \begin{array}{{>{\displaystyle}l}}
\ \ \ \ \ \ \ \exists x\in u\rightsquigarrow v:\\
\ \delta ( x,p_{2}( x)) \leqslant w( a,b)
\end{array}$};
\draw (438.47,82) node [anchor=north west][inner sep=0.75pt]  [font=\tiny] [align=left] {\lemref{+2W1_case_nota}};
\draw (325.74,53.67) node [anchor=north west][inner sep=0.75pt]  [font=\tiny] [align=left] {$\displaystyle {\displaystyle Case} \ ( a)$};
\draw (294.8,63.17) node [anchor=north west][inner sep=0.75pt]  [font=\tiny] [align=left] {$\displaystyle  \begin{array}{{>{\displaystyle}l}}
\ \ \ \ \ \ \forall x\in u\rightsquigarrow v:\ \\
w( a,b) < \delta ( x,p_{2}( x))
\end{array}$};
\draw (382.24,106.67) node [anchor=north west][inner sep=0.75pt]  [font=\tiny] [align=left] {$\displaystyle {\displaystyle Case} \ (\overline{b})$};
\draw (352.54,117.64) node [anchor=north west][inner sep=0.75pt]  [font=\tiny] [align=left] {$\displaystyle  \begin{array}{{>{\displaystyle}l}}
\ \ \ \ \ \exists z\in u\rightsquigarrow v:\ \\
\delta ( z,p_{1}( z)) \leqslant w( a,b)
\end{array}$};
\draw (383.47,136.33) node [anchor=north west][inner sep=0.75pt]  [font=\tiny] [align=left] {\lemref{+2W1_case_notb}};
\draw (273.4,107.33) node [anchor=north west][inner sep=0.75pt]  [font=\tiny] [align=left] {$\displaystyle {\displaystyle Case} \ ( b)$};
\draw (240.8,118.5) node [anchor=north west][inner sep=0.75pt]  [font=\tiny] [align=left] {$\displaystyle  \begin{array}{{>{\displaystyle}l}}
\ \ \ \ \ \ \forall z\in u\rightsquigarrow v:\\
\ w( a,b) < \delta ( z,p_{1}( z))
\end{array}$};
\draw (269.13,136.67) node [anchor=north west][inner sep=0.75pt]  [font=\tiny] [align=left] {\lemref{+2W1_case_b}};

\end{tikzpicture}

\caption{\label{fig:+2W1_casetree} The case tree for \algref{+2W1}.}
\end{figure}

\begin{lemma}~\label{lem:+2W1_case_nota} 
   If \notcase{a} holds then $d\bracke{u,v} \leq \delta\paren{u,v}+2w\paren{a,b}$.
\end{lemma}

\begin{proof} 
Recall that we invoke SSSP from any member of $S_{2}$ over $E_{S_{3}} = E$. Hence, $d\bracke{p_{2}\paren{x},v} = \delta\paren{p_2\paren{x},v}$ and $d\bracke{u,p_{2}\paren{x}} = \delta\paren{u,p_2\paren{x}}$. Therefore, when we invoke SSSP from $u\in V = S_{0}$:

    \begin{alignat*}{3}
        d\bracke{u,v} & \leq \textubrace{d\bracke{u,p_{2}\paren{x}}+  d\bracke{p_{2}\paren{x},v}}{SSSP from $u\in V=S_{0}$}
        && = \textubrace{ \delta\paren{u,p_{2}\paren{x}} + \delta\paren{p_{2}\paren{x},v}}{Our preceding discussion} && \leq \textubrace{ \delta\paren{u,x } + 2 \cdot \delta\paren{x,p_{2}\paren{x}} + \delta\paren{x,v}}{Triangle inequality}\\
        &= \textubrace{\delta\paren{u,v } + 2 \cdot \delta\paren{x,p_{2}\paren{x}}}{$u\squiggly x \squiggly v$ is a shortest path}
        && \leq \textubrace{ \delta\paren{u,v } + 2 \cdot w\paren{a,b}}{By \notcase{a}}
    \end{alignat*}
\end{proof}

\begin{lemma}~\label{lem:+2W1_case_notb} 
  If \notcase{b} holds then $d\bracke{u,v} \leq \delta\paren{u,v}+2w\paren{a,b}$.
\end{lemma}
\begin{proof}
Let $\paren{x,y}\in u\squiggly v$. As \case{a} holds: $w\paren{x,y}\leq w\paren{a,b} < \delta\paren{x,p_{2}\paren{x}}$, which means that $\paren{x,y}\in E_{S_{2}}$. Therefore, $u\squiggly v \subseteq E_{S_{2}}$. When invoking SSSP from members of $S_{1}$, the edges of $u\squiggly v$ will be considered as well, therefore $d\bracke{s,v} = \delta\paren{s,v}$ for any $s\in S_{1}$. As \notcase{b} holds, there exists a vertex $z\in u\squiggly v$ such that $\delta\paren{z,p_{1}\paren{z}} \leq w \paren{a,b}$. Assume $z$ is the nearest vertex to $u$ that fulfills this inequality. Therefore, for any vertex $x\in u\squiggly z$, excluding $z$ itself: $w\paren{a,b} < \delta\paren{x,p_{1}\paren{x}}$, which means that $u\squiggly z \subseteq E_{S_{1}}$. Thus:

     \begin{alignat*}{3}
        d\bracke{u,v} & \leq \textubrace{d\bracke{u,p_{2}\paren{x}}+  d\bracke{p_{2}\paren{x},v}}{SSSP from $u\in V=S_{0}$}
        && = \textubrace{ \delta\paren{u,p_{1}\paren{z}} + \delta\paren{p_{1}\paren{z},v}}{Our preceding discussion} && \leq \textubrace{ \delta\paren{u,z } + 2 \cdot \delta\paren{z,p_{1}\paren{z}} + \delta\paren{z,v}}{Triangle inequality}\\
        &= \textubrace{\delta\paren{u,v } + 2 \cdot \delta\paren{z,p_{1}\paren{z}}}{$u\squiggly z \squiggly v$ is a shortest path}
        && \leq \textubrace{ \delta\paren{u,v } + 2 \cdot w\paren{a,b}}{By \notcase{a}}
        \end{alignat*}
\end{proof}

\begin{lemma}~\label{lem:+2W1_case_b} 
  If \case{b} holds then $d\bracke{u,v} = \delta\paren{u,v}$.
\end{lemma}
\begin{proof}
Let $\paren{z,w}\in u\squiggly v$. Therefore: $w\paren{z,w} \leq w\paren{a,b} < \delta\paren{z,p_{1}\paren{z}}$, which means that $\paren{z,w}\in E_{S_{1}}$. Hence, $u\squiggly v \subseteq E_{S_{1}}$. Recall that we invoke SSSP from $u\in V= S_{0}$, including the edge set $E_{S_{1}}$. Ergo, the exact distance $d\bracke{u,v}=\delta\paren{u,v}$ will be computed.

\end{proof}

We now turn to prove the main theorem of this section:

\begin{theorem}~\label{thm:+2W1} 
   \algref{+2W1} computes a $+2W_{1}$-APASP in $\tilde O\paren{n^\frac{7}{3}}$ time.
\end{theorem}
\begin{proof}
The correctness follows by the case analysis and \lemref{+2W1_case_nota} to \lemref{+2W1_case_b}. We consider the runtime. Initializing $d$ takes $O\paren{n^{2}}$ time. Computing $\Gamma\paren{u,n^{\beta}}$ (respectively, $\Gamma\paren{u,n^{\beta+\gamma}}$) for all vertices requires $\tilde O\paren{m+n^{1+\beta}}$ (respectively, $\tilde O\paren{m+n^{1+\beta+\gamma}}$) by \obvref{ball}. Computing the hitting set $S_{1}$ (respectively, $S_{2}$) can be done deterministically in $O\paren{n^{1+\beta}}$ (respectively, $O\paren{n^{1+\beta+\gamma}}$) by \obvref{hs}. Finding both pivots of each vertex and computing the distances to both pivots requires $\tilde O \paren{m}$ by \obvref{pivotsdistance}. Finding the edge set $E_{S_{1}}$ (respectively, $E_{S_{2}}$) requires $\tilde O \paren{n^{1+\beta}}$ (respectively, $\tilde O \paren{n^{1+\beta+\gamma}}$) by \obvref{edgesruntime}.

     By \obvref{hs} the size of $S_{1}$ is $\tilde O \paren{n^{1-\beta}}$ and the size of $S_{2}$ is $\tilde O \paren{n^{1-\beta-\gamma}}$. By \obvref{edgessize} the size of $E_{S_{1}}$ is $\tilde O \paren{n^{1+\beta}}$  and the size of $E_{S_{2}}$ is $\tilde O \paren{n^{1+\beta+\gamma}}$. As there are $n$ vertices, the size of $\paren{\bracce{s}\times V }\cup H$ is $\tilde O \paren{n}$, which is smaller than any $E_{S_{i+1}}$. Additionally, $\abs{S_{0}}\cdot \abs{S_{2}}$ is $\tilde O \paren{n^{2-\beta-\gamma}}$. Implementing the SSSP invocation by Dijkstra, the runtime  will therefore be:
     $$\tilde O \paren{\abs{S_{0}}\cdot \paren{\abs{E_{S_{1}}}+\abs{S_{0}}\cdot \abs{S_{2}}} + \abs{S_{1}}\cdot \paren{\abs{E_{S_{2}}}+\abs{S_{0}}\cdot \abs{S_{2}}} + \abs{S_{2}}\cdot \paren{\abs{E_{S_{3}}}+\abs{S_{0}}\cdot \abs{S_{2}}}}  = \tilde O \paren{n^{2+\beta} + n^{2+\gamma} + n^{3-\beta-\gamma}}$$ 
     
     By selecting $\beta=\gamma=\frac{1}{3}$, the runtime becomes $\tilde O \paren{n^{\frac{7}{3}}}$.
\end{proof}

\section{Additive $+2\msum{W_{i}}{i=1}{k+1}$-APASP}\label{+2Wi}

In this section we introduce \algref{+2Wi}, a generalization of \algref{+2W1}, which was presented in \secref{+2W1}. Recall that \algref{+2W1} computed a $+2W_{1}$-APASP and had a runtime of $\tilde O \paren{n^{\frac{7}{3}}}$.  \algref{+2Wi} is a determinsitic algorithm that extends this approach to compute a $+2\msum{W_{i}}{i=1}{k+1}$-APASP, requiring $\tilde O \paren{n^{2+\frac{1}{3k+2}}}$ runtime\footnote{The case where $k=0$ does not conform to the general runtime expression for $k>1$, analogous to the distinction observed between the $+2$-APASP and the $+2\cdot \paren{k+1}$-APASP algorithms of Dor, Halperin and Zwick \cite{DorHalZwi2000}.}.

\algref{+2Wi} uses a parameter $k$ that indicates a trade-off between the runtime and  the approximation and parameters $\beta_{1},\ldots,\beta_{3k+2}\in \paren{0,1}$, whose values will be  determined as part of the runtime analysis. Our algorithm works as follows: for each  $i\in\bracke{3k+2}$ we set $\alpha_{i}=\msum{\beta_{j}}{j=1}{i}$ and compute a hitting set $S_{i}$ to the sets $\Gamma\paren{u,n^{\alpha_{i}}}$, over all $u\in V$. For each $S_{i}$ we also compute its corresponding $E_{S_{i}}$. We set $S_{0}=V$ and $S_{3k+3}=\varnothing$, hence $E_{S_{3k+3}}=E$.

\begin{algorithm}[t]
\caption{$+2\sum_{i=1}^{k+1}{W_{i}}$-APASP$\left(G=\left(V,E,w\right)\right)$}\label{alg:+2Wi}

\textbf{Input:} An undirected non-negative weighted graph $G=\left(V,E,w\right)$ and $k\in \mathbb{N}$.

\textbf{Output:} A $+2\sum_{i=1}^{k+1}{W_{i}}$-APASP. 

\medskip

Let $\beta_{1},\ldots, \beta_{3k+2} \in \left( 0 , 1 \right)$ to be fixed later

Initialize $d$ as the edges' weights and $\infty$ otherwise

\For{$i \in \bracke{3k+2}$}
{
    $\alpha_{i} \leftarrow \msum{\beta_{j}}{j=1}{i} $

    \For{$u\in V$}
    {
        Compute $\Gamma\paren{u,n^{\alpha_{i}}}$
    }

    Compute $S_{i}$ to be a hitting set to $\bracce{\Gamma\paren{u,n^{\alpha_{i}}} \mid u\in V}$

    Compute the distances from any $u\in V$ to its pivot $p_{i}\paren{u}$ 

    Construct the set of edges $E_{S_{i}}$
}

\For{$i \in \pair{3k+2,\ldots,1,0}$}
{
    $F_{i} \leftarrow \varnothing$

    \For{$j \in \bracce{0,1,\ldots,3k+2}$}
    {
        \For{$\ell \in \bracce{3k+1-i-j,3k+2-i-j,\ldots,3k+2}$}
        {
            $F_{i} \leftarrow F_{i} \cup \paren{S_{\ell}\times S_{j}}$
        }
    }

    \For{$s\in  S_{i}$}
    {
        Invoke SSSP from $s$ on $ G_{i} \paren{s} = \left( V,E_{ 
 S_{i+1}} \cup \paren{\bracce{s}\times V} \cup H \cup F_{i} , d \right)$ and update $d$ accordingly
    }
}

\Return $d$

\end{algorithm}
\vspace{1mm}

We initialize our distance approximation matrix $d:V\times V \rightarrow \mathbb{R}^{\geq 0}$ with edges' weights or $\infty$ otherwise. For $i\in\bracke{3k+2}$ and $u\in V$ we find the pivot $p_{i}\paren{u}$ and set $d\bracke{u,p_{i}\paren{u}} \leftarrow \delta\paren{u,p_{i}\paren{u}}$ by a single SSSP invocation (See: \obvref{pivotsdistance}). We proceed with SSSP invocations from all $s\in S_{i}$ over the weighted graph $G_{i}\paren{s} = \paren{V,E_{S_{i+1}} \cup \paren{\bracce{s}\times V} \cup H \cup F_i,d}$, where:

\begin{itemize}
    \item $E_{S_{i+1}}$ - The union, over \textbf{all} $u\in V$, of edges $\paren{u,v}$ such that $w\paren{u,v} < \delta\paren{u,p_{i+1}\paren{u}}$,
    \item $\bracce{s} \times V$ - Auxiliary edges from $s$ to  $V$, representing distances previously computed,
    \item $H =  \bracce{\paren{u,p_{j}\paren{u}} \midline u \in V}$ - Auxiliary edges from \textbf{any} $u$ to its pivots $p_{j}\paren{u}$ for $j\in \bracke{3k+2}$,
    \item $F_{i}$ - Auxiliary edges from any $s_{j}\in S_{j}$ to any $s_{\ell}\in  S_{\ell}$, for indices $j,\ell$ s.t. $i+j+\ell \geq 3k+1$.
\end{itemize}

The weights of these edges are given by $d:V\times V \rightarrow \mathbb{R}^{\geq 0}$, which is a weight function over $G_{i}\paren{s}$. These SSSP invocations are performed in descending order, from $i\leftarrow 3k+2$ to $0$, since they rely on auxiliary edges whose weights were previously computed. To properly understand the reason behind this, consider a $u\in V$ and let $i\in\bracke{3k+2}$ be the largest index such that $u\in S_{i}$. Whenever  $d\bracke{p_{i+1}\paren{v},u}$ is updated, we also update $d\bracke{u,p_{i+1}\paren{v}}$. Therefore, when \algref{+2Wi} will invoke SSSP  from $u\in S_{i}$, the auxiliary edge $\paren{u,p_{i+1}\paren{v}}$ will be considered as well, and its weight is the already computed value $d\bracke{u,p_{i+1}\paren{v}}$.

\begin{figure}[t]
\centering

\begin{tikzpicture}[x=0.75pt,y=0.75pt,yscale=-1,xscale=1]

\draw  [color={rgb, 255:red, 0; green, 0; blue, 0 }  ,draw opacity=1 ][fill={rgb, 255:red, 0; green, 0; blue, 0 }  ,fill opacity=1 ] (194.4,154.17) .. controls (194.4,152.99) and (195.36,152.03) .. (196.53,152.03) .. controls (197.71,152.03) and (198.67,152.99) .. (198.67,154.17) .. controls (198.67,155.34) and (197.71,156.3) .. (196.53,156.3) .. controls (195.36,156.3) and (194.4,155.34) .. (194.4,154.17) -- cycle ;
\draw  [color={rgb, 255:red, 245; green, 166; blue, 35 }  ,draw opacity=1 ][dash pattern={on 4.5pt off 4.5pt}] (129.5,139.06) .. controls (129.5,117.67) and (148.27,100.33) .. (171.42,100.33) .. controls (194.57,100.33) and (213.33,117.67) .. (213.33,139.06) .. controls (213.33,160.44) and (194.57,177.78) .. (171.42,177.78) .. controls (148.27,177.78) and (129.5,160.44) .. (129.5,139.06) -- cycle ;
\draw  [color={rgb, 255:red, 189; green, 16; blue, 224 }  ,draw opacity=1 ][dash pattern={on 4.5pt off 4.5pt}] (106.5,132.97) .. controls (106.5,101.97) and (134,76.83) .. (167.92,76.83) .. controls (201.84,76.83) and (229.33,101.97) .. (229.33,132.97) .. controls (229.33,163.98) and (201.84,189.11) .. (167.92,189.11) .. controls (134,189.11) and (106.5,163.98) .. (106.5,132.97) -- cycle ;
\draw  [color={rgb, 255:red, 35; green, 150; blue, 245 }  ,draw opacity=1 ][dash pattern={on 4.5pt off 4.5pt}] (146,147.75) .. controls (146,135.65) and (155.74,125.83) .. (167.75,125.83) .. controls (179.76,125.83) and (189.5,135.65) .. (189.5,147.75) .. controls (189.5,159.85) and (179.76,169.67) .. (167.75,169.67) .. controls (155.74,169.67) and (146,159.85) .. (146,147.75) -- cycle ;

\draw (192.48,157.03) node [anchor=north west][inner sep=0.75pt]  [font=\scriptsize,color={rgb, 255:red, 0; green, 0; blue, 0 }  ,opacity=1 ,xslant=-0.03] [align=left] {$\displaystyle u$};
\draw (152.5,104.5) node [anchor=north west][inner sep=0.75pt]  [font=\scriptsize,color={rgb, 255:red, 245; green, 166; blue, 35 }  ,opacity=1 ] [align=left] {\textcolor[rgb]{0.18,0.5,0.89}{ }\textcolor[rgb]{0.18,0.5,0.89}{$\displaystyle \textcolor[rgb]{0.96,0.65,0.14}{S}\textcolor[rgb]{0.96,0.65,0.14}{_{\kappa ( u)}}$}};
\draw (151,79.5) node [anchor=north west][inner sep=0.75pt]  [font=\scriptsize,color={rgb, 255:red, 189; green, 16; blue, 224 }  ,opacity=1 ] [align=left] {\textcolor[rgb]{0.18,0.5,0.89}{ }\textcolor[rgb]{0.18,0.5,0.89}{$\displaystyle \textcolor[rgb]{0.74,0.06,0.88}{S}\textcolor[rgb]{0.74,0.06,0.88}{_{\kappa ( u) -1}}$}};
\draw (148.5,136) node [anchor=north west][inner sep=0.75pt]  [font=\scriptsize,color={rgb, 255:red, 245; green, 166; blue, 35 }  ,opacity=1 ] [align=left] {\textcolor[rgb]{0.18,0.5,0.89}{ }\textcolor[rgb]{0.18,0.5,0.89}{$\displaystyle \textcolor[rgb]{0.14,0.59,0.96}{S}\textcolor[rgb]{0.14,0.59,0.96}{_{\kappa ( u) +1}}$}};

\end{tikzpicture}

\caption[.]{\label{fig:+2Wi_kappa} The (smallest hitting set) largest index $i\in \bracke{3k+2}$ s.t. $u\in S_{i}$ .}
\vspace{1mm}
\end{figure}

\input{\apaspImagesPath +2Wi_lambda}

\begin{figure}[t]
\centering
\begin{tikzpicture}[x=0.75pt,y=0.75pt,yscale=-1,xscale=1]

\draw    (248,50.17) -- (306.5,50.17) ;
\draw  [color={rgb, 255:red, 0; green, 0; blue, 0 }  ,draw opacity=1 ][fill={rgb, 255:red, 0; green, 0; blue, 0 }  ,fill opacity=1 ] (197.9,50.03) .. controls (197.9,47.82) and (199.69,46.03) .. (201.9,46.03) .. controls (204.11,46.03) and (205.9,47.82) .. (205.9,50.03) .. controls (205.9,52.24) and (204.11,54.03) .. (201.9,54.03) .. controls (199.69,54.03) and (197.9,52.24) .. (197.9,50.03) -- cycle ;
\draw  [color={rgb, 255:red, 0; green, 0; blue, 0 }  ,draw opacity=1 ][fill={rgb, 255:red, 0; green, 0; blue, 0 }  ,fill opacity=1 ] (428.83,51.33) .. controls (428.83,49.12) and (430.62,47.33) .. (432.83,47.33) .. controls (435.04,47.33) and (436.83,49.12) .. (436.83,51.33) .. controls (436.83,53.54) and (435.04,55.33) .. (432.83,55.33) .. controls (430.62,55.33) and (428.83,53.54) .. (428.83,51.33) -- cycle ;
\draw [color={rgb, 255:red, 0; green, 0; blue, 0 }  ,draw opacity=1 ]   (205.15,50.63) .. controls (213.88,42.15) and (204.61,59.05) .. (214.76,50.11) ;
\draw [color={rgb, 255:red, 0; green, 0; blue, 0 }  ,draw opacity=1 ]   (214.76,50.11) .. controls (223.99,42.12) and (214.3,59.26) .. (224.45,50.32) ;
\draw [color={rgb, 255:red, 0; green, 0; blue, 0 }  ,draw opacity=1 ]   (224.45,50.32) .. controls (233.18,41.84) and (223.98,59.46) .. (234.13,50.53) ;
\draw [color={rgb, 255:red, 0; green, 0; blue, 0 }  ,draw opacity=1 ]   (234.13,50.53) .. controls (242.86,42.05) and (233.66,59.67) .. (243.81,50.73) ;

\draw  [color={rgb, 255:red, 0; green, 0; blue, 0 }  ,draw opacity=1 ][fill={rgb, 255:red, 0; green, 0; blue, 0 }  ,fill opacity=1 ] (244,50.17) .. controls (244,47.96) and (245.79,46.17) .. (248,46.17) .. controls (250.21,46.17) and (252,47.96) .. (252,50.17) .. controls (252,52.38) and (250.21,54.17) .. (248,54.17) .. controls (245.79,54.17) and (244,52.38) .. (244,50.17) -- cycle ;
\draw  [color={rgb, 255:red, 0; green, 0; blue, 0 }  ,draw opacity=1 ][fill={rgb, 255:red, 0; green, 0; blue, 0 }  ,fill opacity=1 ] (306.5,50.17) .. controls (306.5,47.96) and (308.29,46.17) .. (310.5,46.17) .. controls (312.71,46.17) and (314.5,47.96) .. (314.5,50.17) .. controls (314.5,52.38) and (312.71,54.17) .. (310.5,54.17) .. controls (308.29,54.17) and (306.5,52.38) .. (306.5,50.17) -- cycle ;
\draw  [color={rgb, 255:red, 216; green, 65; blue, 3 }  ,draw opacity=1 ] (306,62) -- (438.17,62) -- (438.17,55.83) ;
\draw  [color={rgb, 255:red, 216; green, 65; blue, 3 }  ,draw opacity=1 ] (438.17,62) -- (306,62) -- (306,55.83) ;

\draw [color={rgb, 255:red, 0; green, 0; blue, 0 }  ,draw opacity=1 ]   (314.55,51.43) .. controls (323.28,42.95) and (314.01,59.85) .. (324.16,50.91) ;
\draw [color={rgb, 255:red, 0; green, 0; blue, 0 }  ,draw opacity=1 ]   (324.16,50.91) .. controls (333.39,42.92) and (323.7,60.06) .. (333.85,51.12) ;
\draw [color={rgb, 255:red, 0; green, 0; blue, 0 }  ,draw opacity=1 ]   (333.85,51.12) .. controls (342.58,42.64) and (333.38,60.26) .. (343.53,51.33) ;
\draw [color={rgb, 255:red, 0; green, 0; blue, 0 }  ,draw opacity=1 ]   (343.53,51.33) .. controls (352.26,42.85) and (343.06,60.47) .. (353.21,51.53) ;

\draw [color={rgb, 255:red, 0; green, 0; blue, 0 }  ,draw opacity=1 ]   (307.83,53.38) .. controls (296.61,48.65) and (311.38,61.03) .. (298.68,56.38) ;
\draw [color={rgb, 255:red, 0; green, 0; blue, 0 }  ,draw opacity=1 ]   (298.68,56.38) .. controls (287.19,52.28) and (302.43,64.74) .. (289.73,60.09) ;
\draw [color={rgb, 255:red, 0; green, 0; blue, 0 }  ,draw opacity=1 ]   (289.73,60.09) .. controls (278.51,55.35) and (293.48,68.44) .. (280.78,63.79) ;
\draw [color={rgb, 255:red, 0; green, 0; blue, 0 }  ,draw opacity=1 ]   (280.78,63.79) .. controls (269.57,59.06) and (284.54,72.15) .. (271.83,67.5) ;

\draw [color={rgb, 255:red, 0; green, 0; blue, 0 }  ,draw opacity=1 ]   (390.28,51.1) .. controls (399.02,42.62) and (389.75,59.51) .. (399.9,50.58) ;
\draw [color={rgb, 255:red, 0; green, 0; blue, 0 }  ,draw opacity=1 ]   (399.9,50.58) .. controls (409.12,42.59) and (399.43,59.72) .. (409.58,50.79) ;
\draw [color={rgb, 255:red, 0; green, 0; blue, 0 }  ,draw opacity=1 ]   (409.58,50.79) .. controls (418.31,42.3) and (409.11,59.93) .. (419.26,50.99) ;
\draw [color={rgb, 255:red, 0; green, 0; blue, 0 }  ,draw opacity=1 ]   (419.26,50.99) .. controls (428,42.51) and (418.79,60.14) .. (428.94,51.2) ;

\draw [color={rgb, 255:red, 0; green, 0; blue, 0 }  ,draw opacity=1 ]   (352.57,52.04) .. controls (360.89,43.14) and (352.44,60.47) .. (362.15,51.05) ;
\draw [color={rgb, 255:red, 0; green, 0; blue, 0 }  ,draw opacity=1 ]   (362.15,51.05) .. controls (370.97,42.63) and (362.12,60.21) .. (371.83,50.79) ;
\draw [color={rgb, 255:red, 0; green, 0; blue, 0 }  ,draw opacity=1 ]   (371.83,50.79) .. controls (380.14,41.9) and (371.81,59.95) .. (381.51,50.53) ;
\draw [color={rgb, 255:red, 0; green, 0; blue, 0 }  ,draw opacity=1 ]   (381.51,50.53) .. controls (389.83,41.64) and (381.49,59.69) .. (391.19,50.27) ;

\draw  [color={rgb, 255:red, 0; green, 0; blue, 0 }  ,draw opacity=1 ][fill={rgb, 255:red, 0; green, 0; blue, 0 }  ,fill opacity=1 ] (263.83,67.5) .. controls (263.83,65.29) and (265.62,63.5) .. (267.83,63.5) .. controls (270.04,63.5) and (271.83,65.29) .. (271.83,67.5) .. controls (271.83,69.71) and (270.04,71.5) .. (267.83,71.5) .. controls (265.62,71.5) and (263.83,69.71) .. (263.83,67.5) -- cycle ;

\draw (219.75,16.78) node [anchor=north west][inner sep=0.75pt]  [font=\footnotesize,xslant=-0.03] [align=left] {$\displaystyle \varphi ( j,u\sim v)$};
\draw (283.75,16) node [anchor=north west][inner sep=0.75pt]  [font=\footnotesize,xslant=-0.03] [align=left] {$\displaystyle \psi ( j,u\sim v)$};
\draw (195.32,37.37) node [anchor=north west][inner sep=0.75pt]  [font=\footnotesize,color={rgb, 255:red, 0; green, 0; blue, 0 }  ,opacity=1 ,xslant=-0.03] [align=left] {$\displaystyle u$};
\draw (428.75,37.17) node [anchor=north west][inner sep=0.75pt]  [font=\footnotesize,xslant=-0.03] [align=left] {$\displaystyle v$};
\draw (242.92,37.33) node [anchor=north west][inner sep=0.75pt]  [font=\footnotesize,xslant=-0.03] [align=left] {$\displaystyle x$};
\draw (305.25,36.17) node [anchor=north west][inner sep=0.75pt]  [font=\footnotesize,xslant=-0.03] [align=left] {$\displaystyle y$};
\draw (360.33,65.33) node [anchor=north west][inner sep=0.75pt]  [font=\scriptsize] [align=left] {\textcolor[rgb]{0.18,0.5,0.89}{$\displaystyle \textcolor[rgb]{0.85,0.25,0.01}{E_{S}{}_{j+1}}$}};
\draw (313.07,25.91) node [anchor=north west][inner sep=0.75pt]  [font=\footnotesize,rotate=-90.82,xslant=-0.03] [align=left] {$\displaystyle =$};
\draw (250.07,26.58) node [anchor=north west][inner sep=0.75pt]  [font=\footnotesize,rotate=-90.82,xslant=-0.03] [align=left] {$\displaystyle =$};
\draw (243.92,67.5) node [anchor=north west][inner sep=0.75pt]  [font=\footnotesize,xslant=-0.03] [align=left] {$\displaystyle p_{\zeta ( j,u\sim v)}( y)$};

\end{tikzpicture}

\caption[.]{\label{fig:+2Wi_xi_phi_psi_zeta} From $v$'s perspective, the subpath $y\squiggly v\subseteq E_{S_{j+1}}\cup H$ while $\paren{x,y}\notin E_{S_{j+1}}\cup H$. We denote $\varphi\paren{j,u\squiggly v}=x$, $\psi\paren{j,u\squiggly v}=y$ and $\xi\paren{j,u\squiggly v}=\paren{x,y}$. Note that $\delta\paren{y,p_{j+1}\paren{y}}\leq w\paren{x,y}$, and let $\zeta\paren{j,u\squiggly v}$ be the largest index s.t. $\delta\paren{y,p_{\zeta\paren{j,u\squiggly v}}\paren{y}}\leq w\paren{x,y}$.}

\end{figure}

To prove an upper-bound of $ +2\msum{W_{i}}{i=1}{k+1}$, we must first fix a $u\squiggly v$. We then show  $d\bracke{u,v}\leq \delta\paren{u,v} + 2\cdot \paren{w\paren{e_{1}}+\ldots + w\paren{e_{k+1}}}$ for \textbf{distinct} $k+1$ edges $e_{1},\ldots,e_{k+1}\in u\squiggly v$. That is, our main argument would be that $e_{i}\neq e_{j}$, i.e.: no edge is taken into consideration in the upper-bound more than once. Let:

\begin{itemize}
    \item $\kappa\paren{u}$ be the largest index $i\in\bracce{0}\cup\bracke{3k+2}$ for which $u\in S_{i}$. As the graph is undirected, we can assume w.l.o.g that $\kappa\paren{v} \leq \kappa\paren{u}$ (See: \figref{+2Wi_kappa}).
    \item $\lambda\paren{u\squiggly v}$  be smallest $j\in \bracke{3k+3}$ s.t.  $u\squiggly v \subseteq E_{S_{j}}\cup H$,  $u\squiggly v \not\subseteq E_{S_{j-1}} \cup H$ (See: \figref{+2Wi_lambda}).
    \item $\xi\paren{j,u\squiggly v}$ be the edge $\paren{x,y}\in u\squiggly v$ for which $y\squiggly v\subseteq E_{S_{j+1}}$ and $\paren{x,y}\not\in E_{S_{j}}\cup H$. 
    In other words, it is the first edge from $v$'s side that does not belong to $E_{S_{j+1}}\cup H$. Additionally:
    \begin{itemize}
        \item  $\varphi\paren{j,u\squiggly v} = x$ and $\psi\paren{j,u\squiggly v} = y$ are the vertices of this edge (See: \figref{+2Wi_xi_phi_psi_zeta}).
        \item $\zeta\paren{j,u\squiggly v}$ is the maximal index $\ell\in \bracke{3k+2}$ for which: $\delta\paren{y,p_{\ell}\paren{y}} \leq w\paren{x,y}$. The edge $\paren{x,y}\notin E_{S_{j+1}}$, hence $\delta\paren{y,p_{j+1}\paren{y}}\leq w\paren{x,y}$. Hence,  such $j+1\leq \ell \leq 3k+2$ exists. 
    \end{itemize}
    \item $\xi\paren{u\squiggly v}$ be simply $\xi\paren{\lambda\paren{u\squiggly v}-1, u\squiggly v}$. That is, we view the first edge $\paren{x,y}$, from $v$'s side, s.t. $\paren{x,y}\notin E_{S_{\lambda\paren{u\squiggly v}}}\cup H$ while $y\squiggly v\subseteq E_{S_{\lambda\paren{u\squiggly v}}}\cup H$. Note such an edge exists due to the definition of $\lambda\paren{u\squiggly v}$. We denote the endpoints of this edge by $\varphi\paren{u\squiggly v}$ $\psi\paren{j,u\squiggly v}$, and define: $\zeta\paren{u\squiggly v} = \zeta\paren{\lambda\paren{u\squiggly v}-1,u\squiggly v}$ (Replace $j$ by $\lambda\paren{u\squiggly v}-1$ in \figref{+2Wi_xi_phi_psi_zeta}).
    \item $\Delta\paren{u\squiggly v}$ be an upper-bound on the difference between the computed distance estimation $d\bracke{u,v}$ and the real distance $\delta\paren{u,v}$, where $d\bracke{u,v}$ is computed due to the path $u\squiggly v$.
    \item $\abs{u\squiggly v}$ be the number of edges in $u\squiggly v$.
    \item $\mu\paren{x\straight y,u\squiggly v}$ be the\footnote{If the weights of two edges on a certain shortest path are equal, we can break ties.} integer $j\in\bracke{\abs{u\squiggly v}}$ such that $W_{j} \paren{u\squiggly v} = w \paren{x,y}$. In other words, it is an index specifying \qoute{how heavy} is the edge $\paren{x,y}$.
    \item $\mu\paren{j,u\squiggly v}$ be equal to $\mu\paren{\xi\paren{j,u\squiggly v},u\squiggly v}$. That is, we ask \qoute{how heavy} is the edge $\xi\paren{j,u\squiggly v}$, which is the first edge, from $v$'s side, that is not in $E_{S_{j+1}} \cup H$. 
    \item and $\mu\paren{u\squiggly v}$ be $\mu\paren{\xi\paren{u\squiggly v},u\squiggly v}$ -- \qoute{how heavy} is the edge $\xi\paren{u\squiggly v}$, which is the first edge, from $v$'s side, that is not in $E_{S_{\lambda\paren{u\squiggly v}}}\cup H$.
\end{itemize}

To provide a motivation for the recurrence relation of $\Delta\paren{u\squiggly v}$ that will be presented in \lemref{+2Wi_formula} and for the usage of the notions defined above, it is essential to comprehend where the gap between our distance estimate $d$ and the exact distance $\delta$ originates from. Depending on which portions of the path $u\squiggly v$ are included or omitted during an SSSP invocation from $u\in S_{\kappa\paren{u}}$, this gap may be eliminated, bounded by a simple additive term, or defined recursively by the gap along a detour path.

To precisely characterize the gap $\Delta \paren{u \squiggly v}$, we now delve into technical details, using the notions previously defined. Before providing a full proof and stating \lemref{+2Wi_formula}, we would like to analyze three distinct conditions that characterize the manner in which this gap is defined:

\begin{itemize}
    \item The path $u\squiggly v$ lies within the subgraph $G_{\kappa\paren{u}}\paren{u}$. Therefore, invoking SSSP from $u$ takes all edges on $u \squiggly v$ into consideration. Hence, the exact distance $d\bracke{u,v}\leftarrow \delta\paren{u,v}$ will be computed. This possibility is referred to as \cond{A}.
    \item There exists an edge from $\paren{x,y}\in u\squiggly v$ that is not included in $G_{i}\paren{u}$. We consider two possibilities:
        \begin{itemize}
        \item If the auxiliary edge $\paren{p_{\lambda\paren{u\squiggly v}-1}\paren{x},v}$ is available, then an SSSP invocation from $u$ includes the path $u\squiggly x \squiggly p_{\lambda\paren{u\squiggly v}-1}\paren{x} \squiggly x \straight y \squiggly v$, whose total weight is bounded by $\delta\paren{u,v}+2w\paren{x,y}$, by the definition of the edge $\paren{x,y}$. By definition of  $\mu\paren{u\squiggly v}$: $\delta\paren{u,v}+2w\paren{x,y} = \delta\paren{u,v}+2W_{\mu\paren{u\squiggly v}}\paren{x,y}$. We refer to this possibility as \cond{B}.

        We observe that the availability of the auxiliary edge $\paren{p_{\lambda\paren{u\squiggly v}-1}\paren{x},v}$ depends on whether or not $\lambda\paren{u\squiggly v} \geq 3k+1-\kappa\paren{u}-\kappa\paren{v}$, as this is the range of indices $\ell$ that \algref{+2Wi} takes into consideration. 

        \item Otherwise, when the auxiliary edge $\paren{p_{\lambda\paren{u\squiggly v}-1}\paren{x},v}$ is unavailable, we consider the more intricate path $u\squiggly  x \straight y  \squiggly p_{\zeta\paren{\kappa\paren{u}+1,u\squiggly v}}\paren{y} \squiggly y \squiggly v$. By the same reasoning as in \cond{B}, we later show that the gap $\Delta \paren{ u \squiggly v}$ equals $2w\paren{x,y}+\Delta\paren{p_{\zeta\paren{\kappa\paren{u}+1,u\squiggly v}}\paren{y}\squiggly y \straight x \squiggly u}$. See \figref{+2Wi_error}. This possibility is referred to as \cond{C} and requires a more thorough analysis than \cond{A} and \cond{B}. 
        \end{itemize}
\end{itemize}

\begin{figure}[t]
\centering

\begin{tikzpicture}[x=0.75pt,y=0.75pt,yscale=-1,xscale=1]

\draw [color={rgb, 255:red, 45; green, 127; blue, 226 }  ,draw opacity=1 ]   (277.65,153.13) .. controls (286.38,144.65) and (277.11,161.55) .. (287.26,152.61) ;
\draw [color={rgb, 255:red, 45; green, 127; blue, 226 }  ,draw opacity=1 ]   (287.26,152.61) .. controls (296.49,144.62) and (286.8,161.76) .. (296.95,152.82) ;
\draw [color={rgb, 255:red, 45; green, 127; blue, 226 }  ,draw opacity=1 ]   (296.95,152.82) .. controls (305.68,144.34) and (296.48,161.96) .. (306.63,153.03) ;
\draw [color={rgb, 255:red, 45; green, 127; blue, 226 }  ,draw opacity=1 ]   (306.63,153.03) .. controls (315.36,144.55) and (306.16,162.17) .. (316.31,153.23) ;

\draw [color={rgb, 255:red, 93; green, 165; blue, 16 }  ,draw opacity=1 ]   (250.43,213.43) -- (274.67,157) ;
\draw [color={rgb, 255:red, 208; green, 2; blue, 27 }  ,draw opacity=1 ] [dash pattern={on 4.5pt off 4.5pt}]  (246.43,213.43) -- (195.9,152.53) ;
\draw  [color={rgb, 255:red, 0; green, 0; blue, 0 }  ,draw opacity=1 ][fill={rgb, 255:red, 0; green, 0; blue, 0 }  ,fill opacity=1 ] (191.9,152.53) .. controls (191.9,150.32) and (193.69,148.53) .. (195.9,148.53) .. controls (198.11,148.53) and (199.9,150.32) .. (199.9,152.53) .. controls (199.9,154.74) and (198.11,156.53) .. (195.9,156.53) .. controls (193.69,156.53) and (191.9,154.74) .. (191.9,152.53) -- cycle ;
\draw  [color={rgb, 255:red, 0; green, 0; blue, 0 }  ,draw opacity=1 ][fill={rgb, 255:red, 0; green, 0; blue, 0 }  ,fill opacity=1 ] (316,152.67) .. controls (316,150.46) and (317.79,148.67) .. (320,148.67) .. controls (322.21,148.67) and (324,150.46) .. (324,152.67) .. controls (324,154.88) and (322.21,156.67) .. (320,156.67) .. controls (317.79,156.67) and (316,154.88) .. (316,152.67) -- cycle ;
\draw  [color={rgb, 255:red, 45; green, 127; blue, 226 }  ,draw opacity=1 ][fill={rgb, 255:red, 45; green, 127; blue, 226 }  ,fill opacity=1 ] (246.43,213.43) .. controls (246.43,211.22) and (248.22,209.43) .. (250.43,209.43) .. controls (252.64,209.43) and (254.43,211.22) .. (254.43,213.43) .. controls (254.43,215.64) and (252.64,217.43) .. (250.43,217.43) .. controls (248.22,217.43) and (246.43,215.64) .. (246.43,213.43) -- cycle ;
\draw  [color={rgb, 255:red, 0; green, 0; blue, 0 }  ,draw opacity=1 ][fill={rgb, 255:red, 0; green, 0; blue, 0 }  ,fill opacity=1 ] (238,152.67) .. controls (238,150.46) and (239.79,148.67) .. (242,148.67) .. controls (244.21,148.67) and (246,150.46) .. (246,152.67) .. controls (246,154.88) and (244.21,156.67) .. (242,156.67) .. controls (239.79,156.67) and (238,154.88) .. (238,152.67) -- cycle ;
\draw  [color={rgb, 255:red, 0; green, 0; blue, 0 }  ,draw opacity=1 ][fill={rgb, 255:red, 0; green, 0; blue, 0 }  ,fill opacity=1 ] (270.67,153) .. controls (270.67,150.79) and (272.46,149) .. (274.67,149) .. controls (276.88,149) and (278.67,150.79) .. (278.67,153) .. controls (278.67,155.21) and (276.88,157) .. (274.67,157) .. controls (272.46,157) and (270.67,155.21) .. (270.67,153) -- cycle ;
\draw    (246,152.67) -- (274.67,153) ;
\draw [color={rgb, 255:red, 0; green, 0; blue, 0 }  ,draw opacity=1 ]   (199.65,152.63) .. controls (208.38,144.15) and (199.11,161.05) .. (209.26,152.11) ;
\draw [color={rgb, 255:red, 0; green, 0; blue, 0 }  ,draw opacity=1 ]   (209.26,152.11) .. controls (218.49,144.12) and (208.8,161.26) .. (218.95,152.32) ;
\draw [color={rgb, 255:red, 0; green, 0; blue, 0 }  ,draw opacity=1 ]   (218.95,152.32) .. controls (227.68,143.84) and (218.48,161.46) .. (228.63,152.53) ;
\draw [color={rgb, 255:red, 0; green, 0; blue, 0 }  ,draw opacity=1 ]   (228.63,152.53) .. controls (237.36,144.05) and (228.16,161.67) .. (238.31,152.73) ;

\draw (193.98,158.53) node [anchor=north west][inner sep=0.75pt]  [font=\footnotesize,color={rgb, 255:red, 0; green, 0; blue, 0 }  ,opacity=1 ,xslant=-0.03] [align=left] {$\displaystyle u$};
\draw (259.36,224.03) node  [font=\footnotesize,color={rgb, 255:red, 45; green, 127; blue, 226 }  ,opacity=1 ] [align=left] {\begin{minipage}[lt]{17.59pt}\setlength\topsep{0pt}
\textcolor[rgb]{0.29,0.56,0.89}{$\displaystyle \textcolor[rgb]{0.18,0.5,0.89}{p}\textcolor[rgb]{0.18,0.5,0.89}{_{\kappa ( u) +1}}\textcolor[rgb]{0.18,0.5,0.89}{(}\textcolor[rgb]{0.18,0.5,0.89}{y}\textcolor[rgb]{0.18,0.5,0.89}{)}$}
\end{minipage}};
\draw (318.08,158.67) node [anchor=north west][inner sep=0.75pt]  [font=\footnotesize,xslant=-0.03] [align=left] {$\displaystyle v$};
\draw (240.08,158.67) node [anchor=north west][inner sep=0.75pt]  [font=\footnotesize,xslant=-0.03] [align=left] {$\displaystyle x$};
\draw (272.75,159) node [anchor=north west][inner sep=0.75pt]  [font=\footnotesize,xslant=-0.03] [align=left] {$\displaystyle y$};
\draw (23,170) node [anchor=north west][inner sep=0.75pt]  [font=\scriptsize] [align=left] {\textcolor[rgb]{0.18,0.5,0.89}{Edges from }\textcolor[rgb]{0.18,0.5,0.89}{$\displaystyle E_{S}{}_{\kappa ( u) +1}$}};
\draw (23,185) node [anchor=north west][inner sep=0.75pt]  [font=\scriptsize] [align=left] {\textcolor[rgb]{0.36,0.65,0.06}{Edges from }\textcolor[rgb]{0.36,0.65,0.06}{$\displaystyle H$}};
\draw (23,200) node [anchor=north west][inner sep=0.75pt]  [font=\scriptsize] [align=left] {\textcolor[rgb]{0.82,0.01,0.11}{Missing distance to be computed}};

\end{tikzpicture}

\caption[.]{\label{fig:+2Wi_error} The difference $\Delta$ between $d$ and $\delta$. For a simpler, yet inaccurate, understanding, we illustrate $p_{\kappa\paren{u}+1}\paren{y}$. However, the correct pivot to consider is $p_{\zeta\paren{\kappa\paren{u}+1,u\squiggly v}}\paren{y}$.}

\vspace{2mm}
\end{figure}

We are now ready to present  \lemref{+2Wi_formula}, which succinctly  summarizes the three possible outcomes, namely \cond{A}, \cond{B} and \cond{C}:

\begin{lemma}~\label{lem:+2Wi_formula}\Copy{lem:+2Wi_formula}{
   $\Delta\paren{u\squiggly v} \leq 
  \begin{cases}
    0 &  \,\,\,\lambda\paren{u\squiggly v}\leq \kappa\paren{u}+1 \longspace\textnormal{:(A)} \dashedlinesep
    2W_{\mu\paren{u\squiggly v}} &  \begin{array}{l@{}}
        \lambda\paren{u\squiggly v} >\kappa\paren{u}+1 \textnormal{ and }\\  \lambda\paren{u\squiggly v}\geq 3k+1-\kappa\paren{u}-\kappa\paren{v} 
      \end{array} \,\, \textnormal{:(B)}\dashedlinesep
    
    \begin{array}{l@{}}
        2W_{\mu\paren{\kappa\paren{u}+1,u\squiggly v}}+\\  \Delta\paren{p_{\zeta\paren{\kappa\paren{u}+1, u\squiggly v}}\paren{y}\squiggly y \straight x \squiggly u}
      \end{array} & \begin{array}{l@{}}
        \lambda\paren{u\squiggly v} >\kappa\paren{u}+1 \textnormal{ and }\\  \lambda\paren{u\squiggly v}< 3k+1-\kappa\paren{u}-\kappa\paren{v} 
      \end{array} \,\, \textnormal{:(C)}
  \end{cases}$ 
 
  Where $x=\varphi\paren{\kappa\paren{u}+1,u\squiggly v}$ and $y=\psi\paren{\kappa\paren{u}+1,u\squiggly v}$.}
\end{lemma}

\begin{proof}
Consider \cond{A}, where $\lambda\paren{u\squiggly v} \leq \kappa\paren{u}+1$. Hence, $S_{\kappa\paren{u}+1}\subseteq S_{\lambda\paren{{u\squiggly v}}}$. As $u\in S_{\kappa\paren{u}}$, we invoke SSSP on $G_{\kappa\paren{u}}\paren{u}$ which includes the set of edges $E_{S_{\kappa\paren{u}+1}}$. By \obvref{ESiwithinESi+1} we know that $E_{S_{\lambda\paren{u\squiggly v}}} \subseteq E_{S_{\kappa\paren{u}+1}} $. By the definition of $\lambda\paren{u\squiggly v}$ we know that $u\squiggly v \subseteq E_{S_{\kappa\paren{u}+1}}$. We conclude that the exact distance $d\bracke{u,v}=\delta\paren{u,v}$ will be computed, resulting in $\Delta\paren{u\squiggly v}=0$.

    If \cond{A} does not hold we know that $\kappa\paren{u}< \lambda\paren{u\squiggly v}-1$. Recall that the SSSP invocations are performed in a decreasing order. Hence, when considering the SSSP invocation from either $u$ or $v$, the distance estimations from $S_{\lambda\paren{u\squiggly v}-1}$ to all the vertices have already been computed in a previous iteration.

    We now move on to \cond{B}, where additionally: $ \lambda\paren{u\squiggly v}\geq 3k+1-\kappa\paren{u}-\kappa\paren{v}$. Let us consider $i=\kappa\paren{u}$, $j=\kappa\paren{v}$ and $\ell = \lambda\paren{u\squiggly v}$. Let $w=\varphi\paren{u\squiggly v}$ and $z=\psi\paren{u\squiggly v}$. Note that $w\squiggly p_{\lambda\paren{u\squiggly v}-1}\paren{w} \subseteq E_{S_{\lambda\paren{u\squiggly v}}}$ by definition, hence the concatenated path $p_{\lambda\paren{u\squiggly v}-1}\paren{w}\squiggly w \straight z \squiggly u$,$p_{\lambda\paren{u\squiggly v}}\paren{w}\squiggly w \squiggly v \subseteq E_{S_{\lambda\paren{u\squiggly v}}}$. Thus, the difference remains $\Delta\paren{p_{\lambda\paren{u\squiggly v}-1}\paren{w}\squiggly w \straight z \squiggly u}=\Delta\paren{p_{\lambda\paren{u\squiggly v}-1}\paren{w}\squiggly w \squiggly v} =0$. In other words,  $d\bracke{p_{\lambda\paren{u\squiggly v}-1}\paren{w},u}=\delta\paren{p_{\lambda\paren{u\squiggly v}-1}\paren{w},u}$  and $d\bracke{p_{\lambda\paren{u\squiggly v}-1}\paren{w},v}=\delta\paren{p_{\lambda\paren{u\squiggly v}-1}\paren{w},v}$. Recall that $\ell \geq 3k+1-i-j$, which means that when we invoke SSSP from $u$ the edges $\bracce{u}\times S_{\ell}$ and $S_{\ell}\times S_{j}$ will be considered as well. Hence: 

    \begin{alignat*}{3}
        d\bracke{u,v} &\leq \textubrace{d\bracke{u,p_{\lambda\paren{u\squiggly v}-1}\paren{w}}+d\bracke{p_{\lambda\paren{u\squiggly v}-1}\paren{w},v}}{Due to SSSP from $u$}
                && = \textubrace{ \delta\paren{u,p_{\lambda\paren{u\squiggly v}-1}\paren{w}}+\delta\paren{p_{\lambda\paren{u\squiggly v}-1}\paren{w},v}}{Our preceding discussion}\\ 
                & \leq \textubrace{ \delta\paren{u,w} + 2\cdot \delta\paren{w,p_{\lambda\paren{u\squiggly v}-1}\paren{w}}+\delta\paren{w,v}}{Triangle inequality}
               &&= \textubrace{ \delta\paren{u,v}+2\cdot\delta\paren{w,p_{\lambda\paren{u\squiggly v}-1}\paren{w}}}{$u\squiggly w \squiggly v$ is a shortest path}\\
               & \leq \textubrace{ \delta\paren{u,v}+2\cdot w\paren{z,w}}{By definition of $\lambda\paren{u\squiggly v}$}
               && =  \textubrace{ \delta\paren{u,v}+2W_{\mu\paren{u\squiggly v}}}{By definition of $\mu\paren{u\squiggly v}$}
        \end{alignat*}

    It follows that $\Delta\paren{u\squiggly v} = d\bracke{u,v}-\delta\paren{u,v} \leq  2W_{\mu\paren{u\squiggly v}}$.

    Finally, we are left with \cond{C}, where we assume that $ \lambda\paren{u\squiggly v}< 3k+1-\kappa\paren{u}-\kappa\paren{v}$. In other words, during an SSSP invocation from $u$ we cannot use the auxiliary edges of $p_{\lambda\paren{u\squiggly v}-1}$. We will provide a recursive expression for $\Delta\paren{u\squiggly v}$. Let $x=\varphi\paren{\kappa\paren{u}+1,u\squiggly v}$ and $y=\psi\paren{\kappa\paren{u}+1,u\squiggly v}$, as stated in this lemma. We observe that if we had previously computed $d\bracke{p_{\kappa\paren{u}+1}\paren{y},u}$, then an SSSP invocation from $u$ can consider the following path:
    
    \begin{enumerate}
        \item Utilize the auxiliary edge $\paren{u,p_{\kappa\paren{u}+1}\paren{y}}$, whose weight is simply $d\bracke{p_{\kappa\paren{u}+1}\paren{y},u}$.
        \item Use the auxiliary edge $\paren{p_{\kappa\paren{u}+1}\paren{y}, y}$ whose weight is exactly $d\bracke{y,p_{\kappa\paren{u}+1}\paren{y}}=\delta\paren{y,p_{\kappa\paren{u}+1}\paren{y}}$.
        \item Continue on $y\squiggly v$, as $y=\psi\paren{\kappa\paren{u}+1,u\squiggly v}$. 
    \end{enumerate}  

    The weight of such a path is upper-bounded  $d\bracke{u,v} \leq d\bracke{u,p_{\kappa\paren{u}+1}\paren{y}}+\delta\paren{p_{\kappa\paren{u}+1}\paren{y},y} +\delta\paren{y,v}$. Hence, $\Delta\paren{u\squiggly v}  \leq \Delta\paren{p_{\kappa\paren{u}+1}\paren{y}\squiggly y \straight x \squiggly u}+2\cdot w\paren{x,y}$. Recall that $\zeta\paren{\kappa\paren{u}+1,u\squiggly v} \geq \kappa\paren{u}+1$ is defined to be the largest index for which the same edge $\paren{x,y}$ is at least as heavy as the distance to $y$'s pivot. As a consequence, the above holds as well when instead of selecting $p_{\kappa\paren{u}+1}\paren{y}$, we select the pivot $p_{\zeta\paren{\kappa\paren{u}+1,u\squiggly v}}\paren{y}$. Thus: $\Delta\paren{u\squiggly v}  \leq \Delta\paren{p_{\zeta\paren{\kappa\paren{u}+1,u\squiggly v}}\paren{y}\squiggly y \straight x \squiggly u}+2\cdot w\paren{x,y} $. 

    The edge $\paren{x,y}$ is the first edge from $v$'s side on the path $u\squiggly v$ that does not belong to $E_{S_{\kappa\paren{u}+1}}$, which means that its the $\mu\paren{\kappa\paren{u}+1,u\squiggly v}$ heaviest edge on the path $u\squiggly v$. We can therefore conclude that: $\Delta\paren{u\squiggly v}  \leq \Delta\paren{p_{\zeta\paren{\kappa\paren{u}+1,u\squiggly v}}\paren{y}\squiggly y \straight x \squiggly u}+2\cdot W_{\mu\paren{\kappa\paren{u}+1,u\squiggly v}} $.
\end{proof}

Having proven \lemref{+2Wi_formula}, we now use it to derive a non-recursive upper bound for  $\Delta\paren{u\squiggly v}$ in \lemref{+2Wi_iterations}. The core of \algref{+2Wi}'s correctness lies within  \lemref{+2Wi_iterations}, where, by applying \lemref{+2Wi_formula}, we prove two key claims: First, that no edge is accounted for more than once in the upper bound, and second, that the recursion depth is bounded by at most $k+1$ iterations before reaching a halting condition -- either \cond{A} or \lemcond{B}{+2Wi_formula}. Together, these claims imply that at most $k+1$ \textbf{distinct} edges $e_{1}, \ldots, e_{k+1} \in u \squiggly v$ contribute to the upper bound of $\Delta\paren{u\squiggly v}$.

Fix a $u\squiggly v$. As $G$ is undirected we can assume, w.l.o.g., that $\kappa\paren{v}\leq \kappa\paren{u}$. Let  $\paren{x,y}=\xi\paren{\kappa\paren{u}+1,u\squiggly v}$ be the first edge, from $v$'s perspective, not in $E_{S_{\kappa\paren{u}+1}}$. If we had an estimate $d\bracke{u,y}$, invoking SSSP from $u\in S_{\kappa\paren{u}}$ would result in $d\bracke{u,v}\leq d\bracke{u,y}+\delta\paren{y,v}$, as $y\squiggly v \subseteq E_{S_{\kappa\paren{u}+1}}$. 

To estimate $d\bracke{u,y}$, we would need an index $j\in \bracke{3k+2}$ s.t.  $u \squiggly p_{j}\paren{y}\squiggly y$ would be a good estimate on $\delta\paren{u,y}$. To avoid having  $\paren{x,y}$ considered more than once in the upper-bound, we select the pivot $p_{\zeta\paren{\kappa\paren{u}+1, u\squiggly v}}\paren{y}$ as in \lemcond{C}{+2Wi_formula} (See: \figref{+2Wi_error}). 

As for the number of recurrences, note that \cond{A} and \lemcond{B}{+2Wi_formula} are the halting conditions of the recurrence  relation. We prove two upper-bounds on the number of recurrences of \lemcond{C}{+2Wi_formula}: one until reaching \cond{A} and the other until reaching \cond{B}. 

\begin{lemma}~\label{lem:+2Wi_iterations}\Copy{lem:+2Wi_iterations}{ 
   $\Delta\paren{u\squiggly v} \leq 2\msum{W_{i}}{i=1}{k+1} $ }
\end{lemma}

\begin{proof}
If we are either at \cond{A} or \lemcond{B}{+2Wi_formula}, the claim is trivial. In \cond{C}: $\Delta\paren{u\squiggly v} = 2W_{\mu\paren{\kappa\paren{u}+1,u\squiggly v}}+  \Delta\paren{p_{\zeta\paren{\kappa\paren{u}+1, u\squiggly v}}\paren{y}\squiggly y \straight x \squiggly u}$, where $x=\varphi\paren{\kappa\paren{u}+1,u\squiggly v}$ and $y=\psi\paren{\kappa\paren{u}+1,u\squiggly v}$. We want to show:
    \begin{enumerate}
        \item No edge is being taken into consideration in $\Delta\paren{u\squiggly v}$ more than once.
        \item The number of recurrences   until (including) we reached a halting condition is at most $k+1$. 
    \end{enumerate}
We start by showing no edge is considered twice. Let $\hat u = p_{\zeta\paren{\kappa\paren{u}+1, u\squiggly v}}\paren{y}$ and $\hat v = u$. Hence, in \cond{C}: $\Delta\paren{p_{\zeta\paren{\kappa\paren{u}+1, u\squiggly v}}\paren{y}\squiggly y \straight x \squiggly u} = \Delta\paren{\hat u \squiggly y \straight x \squiggly \hat v}$. Observe that $\kappa\paren{\hat u}=\kappa\paren{p_{\zeta\paren{\kappa\paren{u}+1, u\squiggly v}}\paren{y}} = \zeta\paren{\kappa\paren{u}+1, u\squiggly v} \geq \kappa\paren{u}+1$. By the definition of $\zeta\paren{\kappa\paren{u}+1,u\squiggly v}$ we know that $y\straight x \subseteq E_{S_{\zeta\paren{\kappa\paren{u}+1,u\squiggly v}+1}} = E_{S_{\kappa\paren{\hat u}+1}}$. Moreover, the added sub-path $\hat u \squiggly y\subseteq H$, as ${\hat u}$ is pivot of $y$. Consequently, $\hat u \squiggly y \straight x \subseteq E_{S_{\kappa\paren{\hat u}+1}}\cup H$. That is, the edge  $\xi\paren{\kappa\paren{{\hat u}}+1,\hat u \squiggly y \straight x \squiggly \hat v}$ is from the path $x\squiggly \hat v$. Ergo, the recurrence in \cond{C} considers a different edge each time.  

We now prove the second claim: for the number of recurrences of \cond{C}. As we showed in each recurrence a different edge is considered, all that remains is to bound the number of edges, which is at most one more than the number of recurrences (if we halted due to \cond{B}). We consider two options:  $\kappa\paren{u}=\kappa\paren{v}$ and $\kappa\paren{u}>\kappa\paren{v}$. We begin with the first, and by the end briefly consider the second option. We provide two bounds, one when the relation halted due to \cond{B} and the other when it halted due to \cond{A}.

For \cond{B}, observe that $\kappa\paren{\hat u}\geq \kappa\paren{u}+1$ and $\kappa\paren{\hat v} = \kappa\paren{u} = \kappa\paren{v}$. Consider the vertices $\hat{ \hat u}=p_{\zeta\paren{\kappa\paren{\hat u}+1,\hat u \squiggly y \straight x \squiggly \hat v}} \paren{\psi\paren{\kappa\paren{\hat u}+1,\hat u \squiggly y \straight x \squiggly \hat v}}$ and $\hat{\hat v}=\hat u$. Therefore, $\kappa(\hat{\hat u}) \geq \kappa\paren{\hat u}+1$ and $\kappa(\hat{\hat v})=\kappa\paren{\hat u} \geq \kappa\paren{u}+1 > \kappa\paren{u}=\kappa\paren{\hat v}$. In other words, in the first recurrence of \cond{C}, $\kappa\paren{\hat u}$ increased by at least $1$ compared to $\kappa\paren{u}$ and $\kappa\paren{\hat v}$ is equal to $\kappa\paren{v}$. By the second invocation, both $\kappa(\hat{\hat u})$ and $\kappa(\hat{\hat v})$ increase by at least $1$ compared to $\kappa\paren{\hat u}$ and $\kappa\paren{\hat v}$, respectively. The value $\lambda\paren{u\squiggly v}=\lambda\paren{\hat u \squiggly y \straight x \squiggly \hat v}$ because the only edges we add to the path are from $H$ and we neglect edges that belong to $E_{S_{\kappa\paren{u}+1}}\subseteq E_{S_{\lambda\paren{u\squiggly v}-1}}$, assuming $\lambda\paren{u\squiggly v}>\kappa\paren{u}+1$. Therefore, there would be at most $\ceil{\frac{3k+1-\lambda\paren{u\squiggly v}}{2}}-\kappa\paren{u}$ recurrences, which means at most $\ceil{\frac{3k+1-\lambda\paren{u\squiggly v}}{2}}-\kappa\paren{u}+1 < \frac{3k+1-\lambda\paren{u\squiggly v}}{2}-\kappa\paren{u}+2$ distinct edges. 

For \cond{A}, we bound the number of recurrences by $\lambda\paren{u\squiggly v}-\kappa\paren{u}-2$, as $\kappa\paren{u}$ increases by at least $1$ during each recurrence. This is also the number of distinct edges. 

As both upper-bounds on the number of distinct edges hold, we compare them to find the actual bound: $ \lambda\paren{u\squiggly v}-\kappa\paren{u} -2 = \frac{3k+1-\lambda\paren{u\squiggly v}}{2}-\kappa\paren{u} +2$. This holds for $\lambda\paren{u\squiggly v}=k+3$. The number of edges  is  at most $\lambda\paren{u\squiggly v}-\kappa\paren{u}-2=k+1-\kappa\paren{u} \leq k+1$, as $\kappa\paren{u}\geq 0$.

As for the second option, $\kappa\paren{u} \geq \kappa\paren{v}$, the only change  
is that $\kappa\paren{\hat u}$ and $\kappa\paren{\hat v}$ increase by at least $1$ from the first iteration rather than the second, which serves in our favor. 
\end{proof}

We are now finally ready to state and prove the main theorem, \thmref{+2Wi}, completing the correctness and runtime analysis of \algref{+2Wi}.

\begin{theorem}~\label{thm:+2Wi}\Copy{thm:+2Wi}{
   \algref{+2Wi} computes a $+2\msum{W_{i}}{i=1}{k+1}$-APASP in $\tilde O\paren{n^{2+\frac{1}{3k+2}}}$ time. }
\end{theorem}

\begin{proof}
Recall that $\Delta\paren{u\squiggly v} = d\bracke{u,v}-\delta\paren{u,v}$. By \lemref{+2Wi_iterations} we know that $\Delta\paren{u\squiggly v} \leq 2\msum{W_{i}}{i=1}{k+1} $. It follows that $d\bracke{u,v} \leq \delta\paren{u,v} + 2\msum{W_{i}}{i=1}{k+1}$. Observe that $d\bracke{u,v}$ always represents the weight of a path between $u$ and $v$, hence $\delta\paren{u,v}\leq d\bracke{u,v}$. 

We now turn to consider the runtime. Initializing $d$ takes $O\paren{n^{2}}$ time. Let $i\in\bracke{3k+2}$. Computing, for all $u\in V$, the sets $\Gamma\paren{u,n^{\alpha_{i}}}$ requires $ O \paren{n^{1+\alpha_{i}}}$ time by \obvref{ball}. By definition: $\alpha_{i}<\alpha_{i+1}$. It follows that the total cost so far, for all $i$, is $ O \paren{k\cdot \paren{n^{1+\alpha_{3k+2}}}}$. Computing the hitting set $S_{i}$ can be done deterministically in $O\paren{n^{1+\alpha_{i}}}$ time due to \obvref{hs}, which means that the total cost remains $ O \paren{k\cdot \paren{n^{1+\alpha_{3k+2}}}}$. Finding the $i^{\textnormal{th}}$ pivot and computing its distance, for all vertices, can be done via a single Dijkstra invocation in $\tilde O \paren{m}$ time by \obvref{pivotsdistance}. This adds $\tilde O \paren{k\cdot m}$ for all $i\in \bracke{3k+2}$. Computing the edge set $E_{S_{i}}$ can be done in  $\tilde O \paren{n^{1+\alpha_{i}}}$ time due to \obvref{edgesruntime}, which means that the runtime remains $\tilde O \paren{k\cdot\paren{n^{1+\alpha_{3k+2}}}}$. As $k$ is constant, all the above remains within $\tilde O\paren{n^{2}}$ runtime. 

By \obvref{hs} the size of $S_{i}$ is $\tilde O \paren{n^{1-\alpha_{i}}}$ and by \obvref{edgessize} the size of $E_{S_{i+1}}$ is $\tilde O \paren{n^{1+\alpha_{i+1}}}$. Note that $\alpha_{i+1}-\alpha_{i}=\beta_{i+1}$. Hence, without taking into account the edges of $F_{i}$, all of the SSSP invocations from $S_{i}$ require $\tilde O \paren{n^{2+\beta_{i+1}}}$ runtime. 

We proceed to bound $\abs{F_i}$. Recall that $F_i$ consists of $S_{\ell}\times S_{j}$ for all indices $j,\ell$ such that $j+\ell \in \bracce{3k+1-i,3k+2-i,\ldots, 3k+2}$. The corresponding cost of invoking Dijkstra from $S_{i}$ over $F_{i}$ is simply $\abs{S_{i}\times S_{\ell}\times S_{j}} \in \tilde O\paren{n^{3-\alpha_{i}-\alpha_{j}-\alpha_{\ell}}}$.

Fix $\beta_{i}=\beta\in\paren{0,1}$. Hence, $\alpha_{i}=i\cdot \beta$.  Therefore, this runtime becomes $\tilde O \paren{n^{3-\paren{i+j+\ell}\cdot \beta}}$. Since $j+\ell\geq 3k+1-i$, we have $i+j+\ell \geq 3k+1$, implying that this runtime bounded by $\tilde O \paren{n^{3-\paren{3k+1}\cdot \beta}}$. We choose $\beta$ such that:

$$ 2+\beta= 3-\paren{3k+1}\cdot \beta $$

Which holds for $\beta=\frac{1}{3k+2}$. Therefore, the overall runtime of \algref{+2Wi} becomes $\tilde O \paren{n^{2+\frac{1}{3k+2}}}$. 
\end{proof}

\section{Multiplicative $\paren{\frac{7}{3}+\varepsilon}$-APASP}\label{7over3+eps}

In this section we present \algref{7over3+eps} that computes, for an arbitrary $\varepsilon>0$, a $\paren{\frac{7}{3}+\varepsilon}$-APASP and has a runtime of $\tilde O \paren{mn^{\beta}+n^{2+\gamma}+\paren{\frac{1}{\varepsilon}}^{O\paren{1}}\cdot n^{\omega\paren{1-\beta-\gamma}}\cdot \log W}$, where $\beta,\gamma\in\paren{0,1}$ are parameters whose value will be determined when we will analyze the runtime. 
Since the algorithm processes the bunch $\ball{u}{S_1}{S_2}$, we sample each vertex of $S_1$ independently into $S_2$ with probability $n^{-\gamma}$. This guarantees that $\mathbb{E}\bracke{\abs{\ball{u}{S_1}{S_2}}}\in\tilde O\paren{n^{\gamma}}$, keeping the cost of processing these bunches within the desired running time.

Our algorithm works as follows: we randomly and independently sample each member of $V$ into $S_{1}$ with probability $n^{-\beta}$. We then repeat the process for vertices of $S_{1}$ to form $S_{2}$ with probability $n^{-\gamma}$.  We compute $E_{S_{1}}$ (resp. $E_{S_{2}}$). We also set $S_{0}=V$ and $S_{3}=\varnothing$. 

We initialize our distance approximation matrix $d:V\times V \rightarrow \mathbb{R}^{\geq 0}$ to hold either the weight of the edge or $\infty$. We find the pivots $p_{1}\paren{u}$ (resp. $p_{2}\paren{u}$ ) for all $u\in V$ and the exact distance $d\bracke{u,p_{1}\paren{u}} \leftarrow \delta\paren{u,p_{1}\paren{u}}$ (resp. $d\bracke{u,p_{2}\paren{u}}$) by a single SSSP invocation (See: \obvref{pivotsdistance}). We then update distances for $\paren{x,y}\in E$ to the pivots $p_{1}\paren{v}$ and $p_{2}\paren{v}$ of all $v\in \ball{y}{S_{1}}$. To know whether $v\in \ball{y}{S_{1}}$, we simply check whether $d\bracke{y,v}<d\bracke{y,p_{1}\paren{y}}=\delta\paren{y,p_{1}\paren{y}}$. 

Similarly, for each vertex $u\in V$ and $s\in S_{1}$ we add $s$ to $\ball{u}{S_{1}}{S_{2}}$ if the condition $d\bracke{u,s} < d\bracke{u,p_{2}\paren{u}} =\delta\paren{u,p_{2}\paren{u}}$ holds. This can be efficiently computed in less than $\tilde O \paren{n^{2}}$ time.

Let $i\in \bracce{0,1}$. We begin with an SSSP invocation from all members $s\in S_{i}$ over the weighted graph $G_{i}\paren{s} = \paren{V,E_{S_{i+1}} \cup \paren{\bracce{s}\times V} \cup H ,d}$, where:

\begin{itemize}
    \item $E_{S_{i+1}}$ - The union, over \textbf{all} $u\in V$, of edges $\paren{u,v}$ such that $w\paren{u,v} < \delta\paren{u,p_{i+1}\paren{u}}$,
    \item $\bracce{s} \times V$ - Auxiliary edges from $s$ to  $V$, representing distances previously computed,
    \item $H =  \bracce{\paren{u,p_{j}\paren{u}} \midline u \in V}$ - Auxiliary edges from \textbf{any} $u$ to its pivots $p_{j}\paren{u}$ for $j\in\bracce{1,2}$.
\end{itemize}

\begin{algorithm}[H]
\caption{$\paren{\frac{7}{3}+\varepsilon}$-APASP$\left(G=\left(V,E,w\right)\right)$}\label{alg:7over3+eps}

\textbf{Input:} An undirected non-negative weighted graph $G=\left(V,E,w\right)$ and an  $\varepsilon>0$.

\textbf{Output:} A $\paren{\frac{7}{3}+\varepsilon}$-APASP. 

\medskip

Let $\beta,\gamma \in \left( 0 , 1 \right)$ to be fixed later

Initialize $d$ as the edges' weights and $\infty$ otherwise



Set $S_{0}\leftarrow V$ and $S_{3}\leftarrow \varnothing$

Sample each vertex of $V$ with probability $n^{-\beta}$ to $S_{1}$

Sample each vertex of $S_{1}$ with probability $n^{-\gamma}$ to $S_{2}$

Compute the distances from any $u\in V$ to its pivot $p_{1}\paren{u}$ (resp. $p_{2}\paren{u}$)

Construct the set of edges $E_{S_{1}}$ (resp. $E_{S_{2}}$)

Compute the sets $\ball{u}{S_{1}}{S_{2}}$ and $\ball{u}{S_{1}}$ for every $u\in V$

\For{$\paren{x,y}\in E$}
{
    \For{$v\in \ball{y}{S_{1}}$}
    {      
        \For{$i \in \bracce{0,1,2}$}
        {
            $d\bracke{p_{i}\paren{v},x}\leftarrow \mmin{d\bracke{p_{i}\paren{v},v}+d\bracke{v,y}+d\bracke{x,y}, d\bracke{p_{i}\paren{v},x}}$
        }
    }
}

\For{$i \in \bracce{0,1}$}
{
    \For{$s\in  S_{i}$}
    {
        Invoke SSSP from $s$ on $G_{i} \paren{s} = \left( V,E_{ 
 S_{i+1}} \cup \paren{\bracce{s}\times V} \cup H  , d \right)$ and update $d$ accordingly
    }
}

Compute a $\paren{1+\varepsilon}$-MSASP from $S_{2}$ on $G=\paren{V,E,w}$ and update $d$ accordingly

\For{$u,v\in V$}
{
    \For{$s\in \ball{u}{S_{1}}{S_{2}}$}
    {
        $d\bracke{u,v}\leftarrow \mmin{d\bracke{u,s}+d\bracke{s,v},d\bracke{u,v}}$
    }
}

\For{$u,v\in V$} 
{
    \For{$i \in \bracke{1,2}$}
    {
        $d\bracke{u,v}\leftarrow \mmin{d\bracke{u,p_{i}\paren{u}}+d\bracke{p_{i}\paren{u},v},d\bracke{u,v}}$
    }
}

\Return $d$

\end{algorithm}

\vspace{2mm}

The weight of each such edge is given by $d:V\times V \rightarrow \mathbb{R}^{\geq 0}$, which is a weight function over $G_{i}\paren{s}$. We then invoke a $\paren{1+\varepsilon}$-MSASP from $S_{2}$. In contrast to \algref{+2Wi}, the order between the invocations of SSSP and MSASP here holds no importance. 

Finally, we perform two distance updates. The first is  $d\bracke{u,v}\leftarrow \mmin{d\bracke{u,s}+d\bracke{s,v},d\bracke{u,v}}$ for any vertex $s\in\ball{u}{S_{1}}{S_{2}}$; the second is by passing through any pivot of either $u$ or $v$, that is: $d\bracke{u,v}\leftarrow \mmin{d\bracke{u,p_{i}\paren{u}}+d\bracke{p_{i}\paren{u},v},d\bracke{u,v}}$.

To prove a multiplicative stretch of $\paren{\frac{7}{3}+\varepsilon}$, we follow the steps of the proof of the $\frac{7}{3}$-APASP algorithm of Baswana and Kavitha \cite{BasKav2010}. As we performed changes to their algorithm, we provide the reader with a full proof and a runtime analysis of \algref{7over3+eps}. 

Let $u,v\in V$ and consider a $u\squiggly v$. We begin by laying out the case-tree, which organizes the possible cases that can occur for $u\squiggly v$. We prove each case separately and outline the changes that arise due to the $\paren{1+\varepsilon}$-MSASP from $S_{2}$, when compared to the $\frac{7}{3}$-APASP algorithm of Baswana and Kavitha \cite{BasKav2010}. 

We start the analysis by considering the three main cases, determined by the relation between $\delta\paren{u,v}$ and the sum of distances $\delta\paren{u,p_{i}\paren{u}}+\delta\paren{v,p_{i}\paren{v}}$, where $i\in\bracce{1,2}$.

\begin{figure}[t]
\centering
\begin{tikzpicture}[x=0.75pt,y=0.75pt,yscale=-1,xscale=1]

\draw  [color={rgb, 255:red, 126; green, 211; blue, 33 }  ,draw opacity=0.73 ][fill={rgb, 255:red, 126; green, 211; blue, 33 }  ,fill opacity=0.35 ] (269.26,209.86) .. controls (238.32,182.67) and (235.15,136.32) .. (262.18,106.33) .. controls (289.21,76.35) and (336.2,74.08) .. (367.14,101.27) .. controls (398.08,128.46) and (401.24,174.81) .. (374.22,204.79) .. controls (347.19,234.78) and (300.2,237.05) .. (269.26,209.86) -- cycle ;
\draw  [color={rgb, 255:red, 74; green, 144; blue, 226 }  ,draw opacity=0.55 ][fill={rgb, 255:red, 74; green, 144; blue, 226 }  ,fill opacity=0.29 ] (163.58,210.86) .. controls (132.64,183.67) and (129.47,137.32) .. (156.5,107.33) .. controls (183.53,77.35) and (230.52,75.08) .. (261.46,102.27) .. controls (292.39,129.45) and (295.56,175.8) .. (268.53,205.79) .. controls (241.51,235.78) and (194.52,238.05) .. (163.58,210.86) -- cycle ;
\draw  [color={rgb, 255:red, 0; green, 0; blue, 0 }  ,draw opacity=1 ][fill={rgb, 255:red, 0; green, 0; blue, 0 }  ,fill opacity=1 ] (200.9,165.53) .. controls (200.9,163.32) and (202.69,161.53) .. (204.9,161.53) .. controls (207.11,161.53) and (208.9,163.32) .. (208.9,165.53) .. controls (208.9,167.74) and (207.11,169.53) .. (204.9,169.53) .. controls (202.69,169.53) and (200.9,167.74) .. (200.9,165.53) -- cycle ;
\draw  [color={rgb, 255:red, 0; green, 0; blue, 0 }  ,draw opacity=1 ][fill={rgb, 255:red, 0; green, 0; blue, 0 }  ,fill opacity=1 ] (322,165.67) .. controls (322,163.46) and (323.79,161.67) .. (326,161.67) .. controls (328.21,161.67) and (330,163.46) .. (330,165.67) .. controls (330,167.88) and (328.21,169.67) .. (326,169.67) .. controls (323.79,169.67) and (322,167.88) .. (322,165.67) -- cycle ;
\draw  [color={rgb, 255:red, 45; green, 127; blue, 226 }  ,draw opacity=1 ][fill={rgb, 255:red, 45; green, 127; blue, 226 }  ,fill opacity=1 ] (131.57,122.2) .. controls (131.57,119.99) and (133.36,118.2) .. (135.57,118.2) .. controls (137.78,118.2) and (139.57,119.99) .. (139.57,122.2) .. controls (139.57,124.41) and (137.78,126.2) .. (135.57,126.2) .. controls (133.36,126.2) and (131.57,124.41) .. (131.57,122.2) -- cycle ;
\draw  [color={rgb, 255:red, 93; green, 165; blue, 16 }  ,draw opacity=1 ][fill={rgb, 255:red, 93; green, 165; blue, 16 }  ,fill opacity=1 ] (378.1,219.63) .. controls (378.1,217.42) and (379.89,215.63) .. (382.1,215.63) .. controls (384.31,215.63) and (386.1,217.42) .. (386.1,219.63) .. controls (386.1,221.84) and (384.31,223.63) .. (382.1,223.63) .. controls (379.89,223.63) and (378.1,221.84) .. (378.1,219.63) -- cycle ;
\draw [color={rgb, 255:red, 45; green, 127; blue, 226 }  ,draw opacity=1 ] [dash pattern={on 4.5pt off 4.5pt}]  (139.5,124.33) -- (200.5,163) ;
\draw [color={rgb, 255:red, 93; green, 165; blue, 16 }  ,draw opacity=1 ] [dash pattern={on 4.5pt off 4.5pt}]  (378.93,216.53) -- (329.25,169) ;
\draw [color={rgb, 255:red, 0; green, 0; blue, 0 }  ,draw opacity=1 ]   (208.16,166.14) .. controls (216.67,157.43) and (207.84,174.56) .. (217.75,165.36) ;
\draw [color={rgb, 255:red, 0; green, 0; blue, 0 }  ,draw opacity=1 ]   (217.75,165.36) .. controls (226.76,157.14) and (217.52,174.52) .. (227.44,165.32) ;
\draw [color={rgb, 255:red, 0; green, 0; blue, 0 }  ,draw opacity=1 ]   (227.44,165.32) .. controls (235.95,156.62) and (227.21,174.47) .. (237.12,165.28) ;
\draw [color={rgb, 255:red, 0; green, 0; blue, 0 }  ,draw opacity=1 ]   (237.12,165.28) .. controls (245.63,156.57) and (236.89,174.43) .. (246.81,165.23) ;

\draw [color={rgb, 255:red, 0; green, 0; blue, 0 }  ,draw opacity=1 ]   (245.96,166.13) .. controls (254.47,157.43) and (245.64,174.56) .. (255.55,165.36) ;
\draw [color={rgb, 255:red, 0; green, 0; blue, 0 }  ,draw opacity=1 ]   (255.55,165.36) .. controls (264.57,157.14) and (255.32,174.52) .. (265.24,165.32) ;
\draw [color={rgb, 255:red, 0; green, 0; blue, 0 }  ,draw opacity=1 ]   (265.24,165.32) .. controls (273.75,156.62) and (265.01,174.47) .. (274.92,165.28) ;
\draw [color={rgb, 255:red, 0; green, 0; blue, 0 }  ,draw opacity=1 ]   (274.92,165.28) .. controls (283.44,156.58) and (274.69,174.43) .. (284.61,165.24) ;

\draw [color={rgb, 255:red, 0; green, 0; blue, 0 }  ,draw opacity=1 ]   (283.77,166.38) .. controls (292.16,157.57) and (283.55,174.81) .. (293.35,165.49) ;
\draw [color={rgb, 255:red, 0; green, 0; blue, 0 }  ,draw opacity=1 ]   (293.35,165.49) .. controls (302.26,157.15) and (293.24,174.64) .. (303.03,165.32) ;
\draw [color={rgb, 255:red, 0; green, 0; blue, 0 }  ,draw opacity=1 ]   (303.03,165.32) .. controls (311.43,156.51) and (302.92,174.48) .. (312.72,165.15) ;
\draw [color={rgb, 255:red, 0; green, 0; blue, 0 }  ,draw opacity=1 ]   (312.72,165.15) .. controls (321.12,156.34) and (312.6,174.31) .. (322.4,164.99) ;

\draw (200.25,173.83) node [anchor=north west][inner sep=0.75pt]  [font=\footnotesize,color={rgb, 255:red, 0; green, 0; blue, 0 }  ,opacity=1 ,xslant=-0.03] [align=left] {$\displaystyle u$};
\draw (130.23,107.27) node  [font=\footnotesize,color={rgb, 255:red, 45; green, 127; blue, 226 }  ,opacity=1 ] [align=left] {\begin{minipage}[lt]{17.59pt}\setlength\topsep{0pt}
\textcolor[rgb]{0.29,0.56,0.89}{$\displaystyle \textcolor[rgb]{0.18,0.5,0.89}{p}\textcolor[rgb]{0.18,0.5,0.89}{_{1}}\textcolor[rgb]{0.18,0.5,0.89}{(}\textcolor[rgb]{0.18,0.5,0.89}{u}\textcolor[rgb]{0.18,0.5,0.89}{)}$}
\end{minipage}};
\draw (321.42,174.5) node [anchor=north west][inner sep=0.75pt]  [font=\footnotesize,xslant=-0.03] [align=left] {$\displaystyle v$};
\draw (186.47,92.01) node [anchor=north west][inner sep=0.75pt]  [font=\footnotesize,color={rgb, 255:red, 45; green, 127; blue, 226 }  ,opacity=1 ] [align=left] {$\displaystyle \ball{u}{S_{1}}$};
\draw (294.75,91.67) node [anchor=north west][inner sep=0.75pt]  [font=\footnotesize,color={rgb, 255:red, 93; green, 165; blue, 16 }  ,opacity=1 ] [align=left] {$\displaystyle \ball{v}{S_{1}}$};
\draw (395.03,230.23) node  [font=\footnotesize,color={rgb, 255:red, 93; green, 165; blue, 16 }  ,opacity=1 ] [align=left] {\begin{minipage}[lt]{17.59pt}\setlength\topsep{0pt}
\textcolor[rgb]{0.29,0.56,0.89}{$\displaystyle \textcolor[rgb]{0.36,0.65,0.06}{p}\textcolor[rgb]{0.36,0.65,0.06}{_{1}}\textcolor[rgb]{0.36,0.65,0.06}{( v}\textcolor[rgb]{0.36,0.65,0.06}{)}$}
\end{minipage}};

\end{tikzpicture}

\caption[.]{\label{fig:7over3+eps_case_a1} \case{a}{1} of \algref{7over3+eps}.}

\end{figure}

\input{\apaspImagesPath 7over3+eps_case_a2}

\begin{figure}[t]
\centering
\begin{tikzpicture}[x=0.75pt,y=0.75pt,yscale=-1,xscale=1]

\draw  [color={rgb, 255:red, 188; green, 164; blue, 16 }  ,draw opacity=0.73 ][fill={rgb, 255:red, 188; green, 164; blue, 16 }  ,fill opacity=0.35 ] (347.26,209.86) .. controls (316.32,182.67) and (313.15,136.32) .. (340.18,106.33) .. controls (367.21,76.35) and (414.2,74.08) .. (445.14,101.27) .. controls (476.08,128.46) and (479.24,174.81) .. (452.22,204.79) .. controls (425.19,234.78) and (378.2,237.05) .. (347.26,209.86) -- cycle ;
\draw  [color={rgb, 255:red, 130; green, 28; blue, 188 }  ,draw opacity=0.55 ][fill={rgb, 255:red, 130; green, 28; blue, 188 }  ,fill opacity=0.29 ] (161.08,210.86) .. controls (130.14,183.67) and (126.97,137.32) .. (154,107.33) .. controls (181.03,77.35) and (228.02,75.08) .. (258.96,102.27) .. controls (289.89,129.45) and (293.06,175.8) .. (266.03,205.79) .. controls (239.01,235.78) and (192.02,238.05) .. (161.08,210.86) -- cycle ;
\draw  [color={rgb, 255:red, 0; green, 0; blue, 0 }  ,draw opacity=1 ][fill={rgb, 255:red, 0; green, 0; blue, 0 }  ,fill opacity=1 ] (398,165.67) .. controls (398,163.46) and (399.79,161.67) .. (402,161.67) .. controls (404.21,161.67) and (406,163.46) .. (406,165.67) .. controls (406,167.88) and (404.21,169.67) .. (402,169.67) .. controls (399.79,169.67) and (398,167.88) .. (398,165.67) -- cycle ;
\draw  [color={rgb, 255:red, 130; green, 28; blue, 188 }  ,draw opacity=1 ][fill={rgb, 255:red, 130; green, 28; blue, 188 }  ,fill opacity=1 ] (132.57,203.2) .. controls (132.57,200.99) and (134.36,199.2) .. (136.57,199.2) .. controls (138.78,199.2) and (140.57,200.99) .. (140.57,203.2) .. controls (140.57,205.41) and (138.78,207.2) .. (136.57,207.2) .. controls (134.36,207.2) and (132.57,205.41) .. (132.57,203.2) -- cycle ;
\draw  [color={rgb, 255:red, 188; green, 164; blue, 16 }  ,draw opacity=1 ][fill={rgb, 255:red, 188; green, 164; blue, 16 }  ,fill opacity=1 ] (475.6,137.13) .. controls (475.6,134.92) and (477.39,133.13) .. (479.6,133.13) .. controls (481.81,133.13) and (483.6,134.92) .. (483.6,137.13) .. controls (483.6,139.34) and (481.81,141.13) .. (479.6,141.13) .. controls (477.39,141.13) and (475.6,139.34) .. (475.6,137.13) -- cycle ;
\draw [color={rgb, 255:red, 130; green, 28; blue, 188 }  ,draw opacity=1 ][fill={rgb, 255:red, 130; green, 28; blue, 188 }  ,fill opacity=1 ] [dash pattern={on 4.5pt off 4.5pt}]  (136.57,203.2) -- (204.9,165.53) ;
\draw [color={rgb, 255:red, 0; green, 0; blue, 0 }  ,draw opacity=1 ]   (208.16,166.14) .. controls (216.67,157.43) and (207.84,174.56) .. (217.75,165.36) ;
\draw [color={rgb, 255:red, 0; green, 0; blue, 0 }  ,draw opacity=1 ]   (217.75,165.36) .. controls (226.76,157.14) and (217.52,174.52) .. (227.44,165.32) ;
\draw [color={rgb, 255:red, 0; green, 0; blue, 0 }  ,draw opacity=1 ]   (227.44,165.32) .. controls (235.95,156.62) and (227.21,174.47) .. (237.12,165.28) ;
\draw [color={rgb, 255:red, 0; green, 0; blue, 0 }  ,draw opacity=1 ]   (237.12,165.28) .. controls (245.63,156.57) and (236.89,174.43) .. (246.81,165.23) ;

\draw [color={rgb, 255:red, 0; green, 0; blue, 0 }  ,draw opacity=1 ]   (245.96,166.13) .. controls (254.47,157.43) and (245.64,174.56) .. (255.55,165.36) ;
\draw [color={rgb, 255:red, 0; green, 0; blue, 0 }  ,draw opacity=1 ]   (255.55,165.36) .. controls (264.57,157.14) and (255.32,174.52) .. (265.24,165.32) ;
\draw [color={rgb, 255:red, 0; green, 0; blue, 0 }  ,draw opacity=1 ]   (265.24,165.32) .. controls (273.75,156.62) and (265.01,174.47) .. (274.92,165.28) ;
\draw [color={rgb, 255:red, 0; green, 0; blue, 0 }  ,draw opacity=1 ]   (274.92,165.28) .. controls (283.44,156.58) and (274.69,174.43) .. (284.61,165.24) ;

\draw [color={rgb, 255:red, 0; green, 0; blue, 0 }  ,draw opacity=1 ]   (359.75,165.88) .. controls (368.38,157.29) and (359.32,174.3) .. (369.36,165.23) ;
\draw [color={rgb, 255:red, 0; green, 0; blue, 0 }  ,draw opacity=1 ]   (369.36,165.23) .. controls (378.48,157.13) and (369.01,174.39) .. (379.04,165.32) ;
\draw [color={rgb, 255:red, 0; green, 0; blue, 0 }  ,draw opacity=1 ]   (379.04,165.32) .. controls (387.67,156.73) and (378.69,174.47) .. (388.73,165.4) ;
\draw [color={rgb, 255:red, 0; green, 0; blue, 0 }  ,draw opacity=1 ]   (388.73,165.4) .. controls (397.35,156.81) and (388.38,174.55) .. (398.41,165.49) ;

\draw [color={rgb, 255:red, 0; green, 0; blue, 0 }  ,draw opacity=1 ]   (283.82,166.01) .. controls (292.46,157.44) and (283.37,174.43) .. (293.43,165.38) ;
\draw [color={rgb, 255:red, 0; green, 0; blue, 0 }  ,draw opacity=1 ]   (293.43,165.38) .. controls (302.56,157.3) and (293.06,174.53) .. (303.11,165.49) ;
\draw [color={rgb, 255:red, 0; green, 0; blue, 0 }  ,draw opacity=1 ]   (303.11,165.49) .. controls (311.75,156.91) and (302.74,174.64) .. (312.79,165.59) ;
\draw [color={rgb, 255:red, 0; green, 0; blue, 0 }  ,draw opacity=1 ]   (312.79,165.59) .. controls (321.43,157.01) and (312.43,174.74) .. (322.48,165.69) ;

\draw [color={rgb, 255:red, 0; green, 0; blue, 0 }  ,draw opacity=1 ]   (321.63,166.42) .. controls (330.09,157.66) and (321.36,174.85) .. (331.22,165.59) ;
\draw [color={rgb, 255:red, 0; green, 0; blue, 0 }  ,draw opacity=1 ]   (331.22,165.59) .. controls (340.18,157.31) and (331.04,174.74) .. (340.9,165.49) ;
\draw [color={rgb, 255:red, 0; green, 0; blue, 0 }  ,draw opacity=1 ]   (340.9,165.49) .. controls (349.36,156.73) and (340.73,174.64) .. (350.59,165.39) ;
\draw [color={rgb, 255:red, 0; green, 0; blue, 0 }  ,draw opacity=1 ]   (350.59,165.39) .. controls (359.05,156.63) and (350.41,174.54) .. (360.27,165.28) ;

\draw  [color={rgb, 255:red, 0; green, 0; blue, 0 }  ,draw opacity=1 ][fill={rgb, 255:red, 0; green, 0; blue, 0 }  ,fill opacity=1 ] (200.9,165.53) .. controls (200.9,163.32) and (202.69,161.53) .. (204.9,161.53) .. controls (207.11,161.53) and (208.9,163.32) .. (208.9,165.53) .. controls (208.9,167.74) and (207.11,169.53) .. (204.9,169.53) .. controls (202.69,169.53) and (200.9,167.74) .. (200.9,165.53) -- cycle ;
\draw [color={rgb, 255:red, 188; green, 164; blue, 16 }  ,draw opacity=1 ] [dash pattern={on 4.5pt off 4.5pt}]  (479.6,137.13) -- (402,165.67) ;

\draw (200.25,173.83) node [anchor=north west][inner sep=0.75pt]  [font=\footnotesize,color={rgb, 255:red, 0; green, 0; blue, 0 }  ,opacity=1 ,xslant=-0.03] [align=left] {$\displaystyle u$};
\draw (131.23,218.27) node  [font=\footnotesize,color={rgb, 255:red, 45; green, 127; blue, 226 }  ,opacity=1 ] [align=left] {\begin{minipage}[lt]{17.59pt}\setlength\topsep{0pt}
\textcolor[rgb]{0.29,0.56,0.89}{$\displaystyle \textcolor[rgb]{0.51,0.11,0.74}{p}\textcolor[rgb]{0.51,0.11,0.74}{_{2}}\textcolor[rgb]{0.51,0.11,0.74}{(}\textcolor[rgb]{0.51,0.11,0.74}{u}\textcolor[rgb]{0.51,0.11,0.74}{)}$}
\end{minipage}};
\draw (397.42,174.5) node [anchor=north west][inner sep=0.75pt]  [font=\footnotesize,xslant=-0.03] [align=left] {$\displaystyle v$};
\draw (179.47,92.51) node [anchor=north west][inner sep=0.75pt]  [font=\footnotesize,color={rgb, 255:red, 130; green, 28; blue, 188 }  ,opacity=1 ] [align=left] {$\displaystyle \ball{u}{S_{2}}$};
\draw (365.75,91.67) node [anchor=north west][inner sep=0.75pt]  [font=\footnotesize,color={rgb, 255:red, 188; green, 164; blue, 16 }  ,opacity=1 ] [align=left] {$\displaystyle \ball{v}{S_{2}}$};
\draw (490.53,148.73) node  [font=\footnotesize,color={rgb, 255:red, 188; green, 164; blue, 16 }  ,opacity=1 ] [align=left] {\begin{minipage}[lt]{17.59pt}\setlength\topsep{0pt}
\textcolor[rgb]{0.29,0.56,0.89}{$\displaystyle \textcolor[rgb]{0.74,0.64,0.06}{p_{2}( v)}$}
\end{minipage}};

\end{tikzpicture}

\caption[.]{\label{fig:7over3+eps_case_a3} \case{a}{3} of \algref{7over3+eps}.}

\end{figure}

\begin{itemize}
    \item \case{a}{1}: $\delta\paren{u,v}<\delta\paren{u,p_{1}\paren{u}}+\delta\paren{v,p_{1}\paren{v}}$ (See: \figref{7over3+eps_case_a1}),
    \item \case{a}{2}: $\delta\paren{u,p_{1}\paren{u}}+\delta\paren{v,p_{1}\paren{v}}\leq \delta\paren{u,v}<\delta\paren{u,p_{2}\paren{u}}+\delta\paren{v,p_{2}\paren{v}}$ (See: \figref{7over3+eps_case_a2}),
    \item \case{a}{3}: $\delta\paren{u,p_{2}\paren{u}}+\delta\paren{v,p_{2}\paren{v}}\leq \delta\paren{u,v}$ (See: \figref{7over3+eps_case_a3}).
\end{itemize}
 
\case{a}{1} and \case{a}{3} follow relatively quickly, whereas \case{a}{2} is more meticulous and requires a thorough analysis and a partition to several sub-cases. To tackle \case{a}{2}, we present a deeper case-tree, which will take into consideration the existence of an edge $\paren{x,y}\in u\squiggly v$ with the following properties:
\begin{lemma}~\label{lem:7over3+eps_case_a2_properties}\Copy{lem:7over3+eps_case_a2_properties}{ 
   If \case{a}{2} holds then either $d\bracke{u,v}\leq 2\cdot \delta\paren{u,v}$ or there exists a single edge $\paren{x,y}\in u\squiggly v$ such that  $\paren{x,y}\notin E_{S_{2}}$. Additionally, the following conditions hold:
   
   \begin{minipage}{0.8\textwidth}
        \begin{enumerate}
            \setlength{\itemsep}{0pt}
            \setlength{\parskip}{0pt}
            \begin{multicols}{2}
                \item $x\in\ball{u}{S_{2}}$,
                \item $y\in\ball{v}{S_{2}}$,
                \item $x\notin\ball{y}{S_{2}}$,
                \item $y\notin\ball{x}{S_{2}}$.
            \end{multicols}
        \end{enumerate}
    \end{minipage}}
\end{lemma}

\begin{proof} 
    If $u\squiggly v \subseteq E_{S_{2}}$ then, when we invoke SSSP from $p_{1}\paren{u}$, we will compute $d\bracke{p_{1}\paren{u}} = \delta\paren{p_{1}\paren{u}}$. Hence:
    \begin{alignat*}{3}
        d\bracke{u,v} &\leq \textubrace{d\bracke{u,p_{1}\paren{u}}+  d\bracke{p_{1}\paren{u},v}}{Updating through the pivot of $u$}
        && \leq \textubrace{\delta\paren{u,p_{1}\paren{u}} + \delta\paren{p_{1}\paren{u},v}}{Our preceding discussion} 
        && \leq \textubrace{ \delta\paren{u,v } + 2 \cdot \delta\paren{u,p_{2}\paren{u}} }{Triangle inequality}
    \end{alignat*}
    Symmetrically, $d\bracke{u,v}\leq \delta\paren{u,v } + 2 \cdot \delta\paren{v,p_{2}\paren{v}}$. Hence: $d\bracke{u,v} \leq \delta\paren{u,v} +\delta\paren{u,p_{1}\paren{u}}+\delta\paren{v,p_{1}\paren{v}} \leq 2\cdot\delta\paren{u,v}$.  
    
    If the path $u\squiggly v \not\subseteq E_{S_{2}}$, it follows that $u\notin \ball{v}{S_{2}}$ and $v\notin \ball{u}{S_{2}}$. We argue that any vertex $z\in u \squiggly v$ belongs either to $\ball{u}{S_{2}}$ or $\ball{v}{S_{2}}$. Assume by contradiction there exists a $z\in u \squiggly v$ such that $z\notin \ball{u}{S_{2}}\cup \ball{v}{S_{2}}$. Hence: $\delta\paren{u,v}=\delta\paren{u,z}+\delta\paren{z,v} \geq \delta\paren{u,p_{2}\paren{u}}+\delta\paren{v,p_{2}\paren{v}} > \delta\paren{u,v}$, which is a contradiction. Moreover, if there is a vertex $z$ that belongs to  $\ball{u}{S_{2}}\cap \ball{v}{S_{2}}$, then both $u\squiggly z \subseteq E_{S_{2}}$ and $z\squiggly v \subseteq E_{S_{2}}$, thereby $u\squiggly v \subseteq E_{S_{2}}$,  contradicting our assumption. Ergo, any $z\in u\squiggly v$ belongs to \textbf{exactly} one of  $\ball{u}{S_{2}}$ and $\ball{v}{S_{2}}$.

    We now argue that there is exactly one single edge $\paren{x,y}\in u\squiggly v$ such that $\paren{x,y}\notin E_{S_{2}}$. As $u\squiggly v \not\subseteq E_{S_{2}}$, it follows there exists such an edge. We assume, for the sake of contradiction, the existence of two edges $u\squiggly a \straight b \squiggly c \straight d \squiggly v$ such that $\paren{a,b},\paren{c,d}\notin E_{S_{2}}$.
    
    We observe that $b\notin \ball{u}{S_{2}}$ and $c\notin \ball{v}{S_{2}}$. By our previous argument, each vertex on $u\squiggly v$ must belong to either $\ball{u}{S_{2}}$ or $\ball{v}{S_{2}}$. Therefore, $b\in \ball{v}{S_{2}}$ and $c\in \ball{u}{S_{2}}$. By \obvref{pathinball} both  $c\straight d \squiggly v\subseteq E_{S_{2}}$ and $u\squiggly a \straight b\subseteq E_{S_{2}}$, contradicting $\paren{c,d}\notin E_{S_{2}}$ and $\paren{a,b}\notin E_{S_{2}}$, respectively\footnote{Even of those would already imply a contradiction.}.

    Ergo, there is only a single edge $u\squiggly x \straight y \squiggly v$ such that $\paren{x,y}\notin E_{S_{2}}$. We observe that both $u\squiggly x\subseteq E_{S_{2}}$ and $y\squiggly v\subseteq E_{S_{2}}$. As mentioned before, $x$  belongs to \textbf{exactly} one of $\ball{u}{S_{2}}$ or $\ball{v}{S_{2}}$. If $x\in \ball{v}{S_{2}}$, then $u\squiggly x \squiggly v \subseteq E_{S_{2}}$, which is a contradiction. Therefore, $x\in \ball{u}{S_{2}}$. By the same argument, $y\in \ball{v}{S_{2}}$. 

    Additionally, if either $x\in\ball{y}{S_{2}}$ or $y\in\ball{x}{S_{2}}$, then $x\straight y \subseteq E_{S_{2}}$, which is a contradiction as well. This concludes the proof.
\end{proof}

We proved the existence of the edge $\paren{x,y}$, with its properties, as in \lemref{7over3+eps_case_a2_properties}. We can now extend our case analysis by presenting a full version of our case-tree, considering the following cases:

\begin{itemize}
    \item \case{a}{1}: $\delta\paren{u,v}<\delta\paren{u,p_{1}\paren{u}}+\delta\paren{v,p_{1}\paren{v}}$,
    \item \case{a}{2}: $\delta\paren{u,p_{1}\paren{u}}+\delta\paren{v,p_{1}\paren{v}}\leq \delta\paren{u,v}<\delta\paren{u,p_{2}\paren{u}}+\delta\paren{v,p_{2}\paren{v}}$
    \begin{itemize}
        \item \notcase{b}: $u\notin \ball{x}{S_{1}}$
        \begin{itemize}
            \item \notcase{b}{1}: $p_{1}\paren{x}\in \ball{u}{S_{1}}{S_{2}}$
            \item \notcase{b}{2}: $p_{1}\paren{x}\notin \ball{u}{S_{1}}{S_{2}}$
        \end{itemize}
        \item \case{b}: $u\in \ball{x}{S_{1}}$
        \begin{itemize}
            \item \notcase{c}: $v\in\ball{y}{S_{1}}$
            \item \case{c}: $v\notin\ball{y}{S_{1}}$
            \begin{itemize}
                \item \notcase{d}: $p_{1}\paren{y}\in\ball{v}{S_{1}}{S_{2}}$
                \item \case{d}: $p_{1}\paren{y}\notin\ball{v}{S_{1}}{S_{2}}$
            \end{itemize}
        \end{itemize}
    \end{itemize}
    \item \case{a}{3}: $\delta\paren{u,p_{2}\paren{u}}+\delta\paren{v,p_{2}\paren{v}}\leq \delta\paren{u,v}$
\end{itemize}

Utilizing the case-tree provided above, we now prove an upper-bound of $\frac{7}{3}+\varepsilon$. \case{a}{1} and \case{a}{3} are addressed in \lemref{7over3+eps_case_a1} and \lemref{7over3+eps_case_a3}, respectively. The more intricate \case{a}{2} is handled by a detailed examination of its sub-cases, as elaborated in \clmref{7over3+eps_case_notb1} through \clmref{7over3+eps_case_d}. The proof of \thmref{7over3+eps} thus consists of combining the upper bounds obtained by all individual cases, ensuring that every possibility is taken within consideration.

\begin{lemma}~\label{lem:7over3+eps_case_a1} 
   If \case{a}{1} holds then $d\bracke{u,v}  = \delta\paren{u,v}$.
\end{lemma}

\begin{proof} 
    If there exists a vertex $a\in u\squiggly v$ such that  $a\notin\ball{u}{S_{1}}\cup\ball{v}{S_{1}}$ then $\delta\paren{u,v}=\delta\paren{u,a}+\delta\paren{a,v}\geq \delta\paren{u,p_{1}\paren{u}}+\delta\paren{v,p_{1}\paren{v}}$, which is a contradiction to the condition in this case. Hence, all vertices $a\in u\squiggly v$ are $a\in\ball{u}{S_{1}}\cup\ball{v}{S_{1}}$. 

    If all edges are  $u\squiggly v \subseteq E_{S_{1}}$ then the exact distance will be computed when invoking, in the first time, SSSP from $u\in V=S_{0}$. If there are at least two edges $u\squiggly a\straight b \squiggly c \straight d \squiggly v$ such that $\paren{a,b},\paren{c,d}\notin E_{S_{1}}$ then $b\notin \ball{u}{S_{1}}$ and $c\notin \ball{v}{S_{1}}$. This means that $b\notin \ball{v}{S_{1}}$ and  $c\notin \ball{u}{S_{1}}$. Both are a contradiction to our previous statement that all vertices on $u\squiggly v$ must belong to $\ball{u}{S_{1}}\cup\ball{v}{S_{1}}$. 

    Therefore, there can only be at most one edge $\paren{a,b}\notin E_{S_{1}}$. This means that when invoking SSSP from $a\in V$ in the first time we will compute $d\bracke{a,v}=\delta\paren{a,v}$ due to the edge $\paren{a,b}$. When invoking SSSP for the second time, the vertex $v$ will make use of the auxiliary edge $\paren{v,a}$ whose weight is $\delta\paren{a,v}$. The sub-path $u\squiggly a\subseteq E_{S_{1}}$ and hence the exact distance $d\bracke{u,v}=\delta\paren{u,v}$ will be computed.
\end{proof}

\begin{lemma}~\label{lem:7over3+eps_case_a3} 
   If \case{a}{3} holds then $d\bracke{u,v}  \leq \paren{2+\varepsilon}\cdot \delta\paren{u,v}$.
\end{lemma}

\begin{proof} 
    W.l.o.g. $\delta\paren{u,p_{2}\paren{u}}\leq\delta\paren{v,p_{2}\paren{v}}$. Hence: $\delta\paren{u,p_{2}\paren{u}} \leq \frac{\delta\paren{u,v}}{2}$. As we compute MSASP from  $S_{2}$: $d\bracke{u,p_{2}\paren{u}}=\paren{1+\varepsilon}\cdot \delta\paren{u,p_{2}\paren{u}}$ and $d\bracke{p_{2}\paren{u},v}=\paren{1+\varepsilon}\cdot \delta\paren{p_{2}\paren{u},v}$. As we consider $p_{2}\paren{u}$ as a pivot for $d\bracke{u,v}$:

    \begin{alignat*}{3}
        d\bracke{u,v} &\leq \textubrace{d\bracke{u,p_{2}\paren{u}}+  d\bracke{p_{2}\paren{u},v}}{Updating through the pivot of $u$}
        && \leq \textubrace{ \paren{1+\varepsilon}\cdot \paren{\delta\paren{u,p_{2}\paren{u}} + \delta\paren{p_{2}\paren{u},v}}}{Our preceding discussion} \\
        & \leq \textubrace{ \paren{1+\varepsilon}\cdot\paren{\delta\paren{u,v } + 2 \cdot \delta\paren{u,p_{2}\paren{u}} }}{Triangle inequality}
        && \leq \textubrace{ \paren{1+\varepsilon}\cdot\paren{\delta\paren{u,v } + \delta\paren{u,v}}}{W.l.o.g. assumption}\\
        &= \textubrace{ \paren{2+2\varepsilon}\cdot \delta\paren{u,v }}{Simply summing}
        \end{alignat*}
By selecting  $\varepsilon' = \frac{\varepsilon}{2}$, we complete the proof.
\end{proof}

We can now delve within \case{a}{2}. Let us recall the vertices $x,y$ from \lemref{7over3+eps_case_a2_properties} and the detailed case-tree previously presented. 

We begin by considering the \qoute{easier} sub-cases, namely \notcase{b}{1} and \notcase{b}{2}, and then conclude \notcase{b}. This is done in \clmref{7over3+eps_case_notb1}, \clmref{7over3+eps_case_notb2} and \clmref{7over3+eps_case_notb}, respectively. 

From there, we turn to \case{b}. Starting here, each time, we examine a certain condition. If it holds, we provide a brief proof. If it fails, we consider that sub-case and split it into additional sub-cases, repeating the approach, until we reach \case{d} and \notcase{d}. 

We begin now by addressing \notcase{b}{1}. Throughout all the following claims, we assume, without loss of generality, that $\delta\paren{u,x} \leq \delta\paren{y,v}$.

\begin{claim2}~\label{clm:7over3+eps_case_notb1} 
   If \notcase{b}{1} holds then $d\bracke{u,v}  \leq 2\cdot \delta\paren{u,v}$.
\end{claim2}

\begin{proof} 
Recall that after all the initializations, we update $d\bracke{p_{1}\paren{x},y}\leq \delta\paren{x,p_{1}\paren{x}}+w\paren{x,y}$. Therefore, when we invoke SSSP from $p_{1}\paren{x}\in S_{1}$ we will consider the auxiliary edge $\paren{p_{1}\paren{x},y}$. The rest of the path $y\subseteq v\subseteq E_{S_{2}}$. Hence, we would reach from $p_{1}\paren{x}$ to both $u$ and $v$. 

\begin{alignat*}{3}
        d\bracke{u,v} &\leq \textubrace{d\bracke{u,p_{1}\paren{x}}+  d\bracke{p_{1}\paren{x},v}}{Updating by $p_{1}\paren{x} \in \ball{u}{S_{1}}{S_{2}}$}
        && \leq \textubrace{\delta\paren{u,p_{1}\paren{x}} + \delta\paren{p_{1}\paren{x},v}}{Our preceding discussion} 
        && \leq \textubrace{ \delta\paren{u,x}+2\delta\paren{x,p_{1}\paren{x}}+\delta\paren{x,v}}{Triangle inequality}\\
        & = \textubrace{ \delta\paren{u,v}+2\delta\paren{x,p_{1}\paren{x}}}{$u\squiggly x \squiggly v$ is a shortest path}
        &&\leq  \textubrace{ \delta\paren{u,v}+2\delta\paren{u,x}}{Due to \notcase{b}}
        &&\leq  \textubrace{ 2\delta\paren{u,v}}{By our w.l.o.g. assumption}\\
\end{alignat*}
\end{proof}

\begin{claim2}~\label{clm:7over3+eps_case_notb2} 
   If \notcase{b}{2} holds then $d\bracke{u,v}  \leq \paren{\frac{7}{3}+\varepsilon}\cdot \delta\paren{u,v}$.
\end{claim2}

\begin{proof} 
Recall that we invoke $\paren{1+\varepsilon}$-MSASP from $S_{2}$. In particular, $d\bracke{p_{2}\paren{u},v}\leq \paren{1+\varepsilon}\cdot \delta\paren{p_{2}\paren{u},v}$. We also update the distance through $p_{2}\paren{u}$. Therefore: 

\begin{alignat*}{3}
        d\bracke{u,v} &\leq \textubrace{d\bracke{u,p_{2}\paren{u}}+  d\bracke{p_{2}\paren{u},v}}{Updating due to $p_{2}\paren{u}$}
        && \leq \textubrace{\paren{1+\varepsilon}\cdot\paren{\delta\paren{u,p_{2}\paren{u}} + \delta\paren{p_{2}\paren{u},v}}}{$\paren{1+\varepsilon}$-MSASP from $S_{2}$} \\
        & = \textubrace{ \paren{1+\varepsilon}\cdot\paren{\delta\paren{u,v}+2\delta\paren{u,p_{2}\paren{u}}}}{Triangle inequality}
        && \leq \textubrace{ \paren{1+\varepsilon}\cdot\paren{\delta\paren{u,v}+2\delta\paren{u,p_{1}\paren{x}}}}{Due to \notcase{b}{2}}\\
        &\leq  \textubrace{ \paren{1+\varepsilon}\cdot\paren{\delta\paren{u,v}+2\delta\paren{u,x}+2\delta\paren{x,p_{1}\paren{x}}}}{Triangle inequality} \\
\end{alignat*}
 To finalize our analysis, we need to establish an upper-bound for $\delta\left(x,p_{1}\left(x\right)\right)$. Let us recall what we know: 
        \begin{itemize}
            \item The set $S_{2}\subseteq S_{1}$ and $y\notin \ball{x}{S_{2}}$, hence: $\delta\left(x,p_{1}\left(x\right)\right)\leq \delta\left(x,p_{2}\left(x\right)\right) \leq w\left(x,y\right)$,
            \item $u\notin \ball{x}{S_{1}}$ and therefore $\delta\left(x,p_{1}\left(x\right)\right) \leq \delta\left(u,x\right)$,
            \item Due to our w.l.o.g. assumption: $\delta\left(x,p_{1}\left(x\right)\right) \leq\delta\left(y,v\right)$.
        \end{itemize}

         Consider the path $u\squiggly x \straight y \squiggly v$. The distance $\delta\left(x,p_{1}\left(x\right)\right)$ is smaller than each of the three components of this path: $u\squiggly x$, $x\straight y$ and $y\squiggly v$. Therefore, $\delta\left(x,p_{1}\left(x\right)\right)\leq \frac{1}{3}\delta\left(u,v\right)$.
Hence: 
\begin{alignat*}{3}
        d\bracke{u,v} &\leq  \textubrace{ \paren{1+\varepsilon}\cdot\paren{\delta\paren{u,v}+2\delta\paren{u,x}+\delta\paren{x,p_{1}\paren{x}}+\delta\paren{x,p_{2}\paren{x}}}}{$\delta\paren{x,p_{1}\paren{x}}\leq\delta\paren{x,p_{2}\paren{x}}$}\\
        &\leq  \textubrace{ \paren{1+\varepsilon}\cdot\paren{\delta\paren{u,v}+\delta\paren{u,x}+\delta\paren{y,v}+w\paren{x,y}+\delta\paren{x,p_{1}\paren{x}}}}{W.l.o.g. assumption and $y\notin\ball{x}{S_{2}}$}\\
        &\leq  \textubrace{ \paren{1+\varepsilon}\cdot\paren{\delta\paren{u,v}+\delta\paren{u,v}+\frac{\delta\paren{u,v}}{3}}}{Our preceding discussion}\\
        &\leq  \textubrace{ \paren{1+\varepsilon}\cdot\frac{7\delta\paren{u,v}}{3}}{Simply summing} \leq \textubrace{ \paren{\frac{7}{3}+\frac{7\varepsilon}{3}}\cdot\delta\paren{u,v}}{Simply summing}\\
\end{alignat*}

We conclude the proof by selecting $\varepsilon^{'} = \frac{7}{3}\cdot \varepsilon$. 
\end{proof}

Having proven both \notcase{b}{1} and \notcase{b}{2}, we now know deduce \clmref{7over3+eps_case_notb}:

\begin{claim2}~\label{clm:7over3+eps_case_notb} 
   If \notcase{b} holds then $d\bracke{u,v}  \leq \paren{\frac{7}{3}+\varepsilon}\cdot \delta\paren{u,v}$.
\end{claim2}

\begin{proof} 
    We have shown \clmref{7over3+eps_case_notb1}  and \clmref{7over3+eps_case_notb2}. As either \notcase{b}{1} or \notcase{b}{2} hold, we conclude that $d\bracke{u,v}  \leq \paren{\frac{7}{3}+\varepsilon}\cdot \delta\paren{u,v}$.
\end{proof}

We now delve within \case{b}. We begin by proving \notcase{c}, in \clmref{7over3+eps_case_notc}:

\begin{claim2}~\label{clm:7over3+eps_case_notc} 
   If \notcase{c} holds then $d\bracke{u,v}  \leq 2\cdot \delta\paren{u,v}$.
\end{claim2}

\begin{proof} 
Recall that after all the initializations, we update $d\bracke{p_{1}\paren{v},x}\leq \delta\paren{v,p_{1}\paren{v}}+\delta\paren{x,v}$. Therefore, when we invoke SSSP from $p_{1}\paren{v}\in S_{1}$ we will consider the auxiliary edge $\paren{p_{1}\paren{v},x}$. The rest of the path $u\squiggly x\subseteq E_{S_{2}}$. Hence, we would reach from $p_{1}\paren{v}$ to $u$. 

\begin{alignat*}{3}
        d\bracke{u,v} &\leq \textubrace{d\bracke{u,p_{1}\paren{v}}+  d\bracke{p_{1}\paren{v},v}}{Update via $p_{1}\paren{v}$}
        && \leq \textubrace{\delta\paren{u,p_{1}\paren{v}} + \delta\paren{p_{1}\paren{v},v}}{Our preceding discussion} 
        && \leq \textubrace{ \delta\paren{u,v}+2\delta\paren{v,p_{1}\paren{v}}}{Triangle inequality}\\
\end{alignat*}
As \notcase{b} holds as well, a symmetrical argument yields that $d\bracke{u,v}\leq \delta\paren{u,v}+2\delta\paren{u,p_{1}\paren{u}}$. Combining both:
\begin{alignat*}{3}
        d\bracke{u,v} &\leq \textubrace{\delta\paren{u,v}+\delta\paren{u,p_{1}\paren{u}}+\delta\paren{v,p_{1}\paren{v}}}{Averaging both inequalities}
        && \leq \textubrace{\delta\paren{u,v}+\delta\paren{u,v}}{By \case{a}{2}} 
        && \leq \textubrace{ 2\cdot \delta\paren{u,v}}{Simply summing}\\
\end{alignat*}
\end{proof}

We continue with \clmref{7over3+eps_case_notd}:

\begin{claim2}~\label{clm:7over3+eps_case_notd} 
   If \notcase{d} holds then $d\bracke{u,v}  \leq 2\cdot \delta\paren{u,v}$.
\end{claim2}

\begin{proof} 
Recall that after all the initializations, we update $d\bracke{p_{1}\paren{y},x}\leq \delta\paren{y,p_{1}\paren{y}}+w\paren{x,y}$. Therefore, when we invoke SSSP from $p_{1}\paren{y}\in S_{1}$ we will consider the auxiliary edge $\paren{p_{1}\paren{y},x}$. The rest of the path $y\subseteq v\subseteq E_{S_{2}}$. Hence, we would reach from $p_{1}\paren{y}$ to both $u$ and $v$. 

\begin{alignat*}{3}
        d\bracke{u,v} &\leq \textubrace{d\bracke{u,p_{1}\paren{y}}+  d\bracke{p_{1}\paren{y},v}}{Updating by $p_{1}\paren{y} \in \ball{v}{S_{1}}{S_{2}}$}
        && \leq \textubrace{\delta\paren{u,p_{1}\paren{y}} + \delta\paren{p_{1}\paren{y},v}}{Our preceding discussion} 
        && \leq \textubrace{ \delta\paren{u,y}+2\delta\paren{y,p_{1}\paren{y}}+\delta\paren{y,v}}{Triangle inequality}\\
        & = \textubrace{ \delta\paren{u,v}+2\delta\paren{y,p_{1}\paren{y}}}{$u\squiggly y \squiggly v$ is a shortest path}
        &&\leq  \textubrace{ \delta\paren{u,v}+\delta\paren{y,v} + \delta\paren{y,p_{2}\paren{y}}}{Due to \notcase{c}}
        &&\leq  \textubrace{ \delta\paren{u,v}+\delta\paren{y,v}+w\paren{x,y}}{By \lemref{7over3+eps_case_a2_properties}}\\
        &\leq  \textubrace{ \delta\paren{u,v}+\delta\paren{x,v}}{$u\squiggly y\squiggly v$ is a shortest path}
        &&\leq  \textubrace{ 2\cdot \delta\paren{u,v}}{Simply summing}
\end{alignat*}
\end{proof}

This leaves us with \case{d} from \clmref{7over3+eps_case_d}:

\begin{claim2}~\label{clm:7over3+eps_case_d} 
   If \case{d} holds then $d\bracke{u,v}  \leq \paren{\frac{7}{3}+\varepsilon}\cdot \delta\paren{u,v}$.
\end{claim2}

\begin{proof} 
Recall that we invoke $\paren{1+\varepsilon}$-MSASP from $S_{2}$. In particular, $d\bracke{p_{2}\paren{v},u}\leq \paren{1+\varepsilon}\cdot \delta\paren{p_{2}\paren{v},u}$. We also update the distance through $p_{2}\paren{v}$. Therefore: 

\begin{alignat*}{3}
        d\bracke{u,v} &\leq \textubrace{d\bracke{u,p_{2}\paren{v}}+  d\bracke{p_{2}\paren{v},v}}{Updating due to $p_{2}\paren{v}$}
        && \leq \textubrace{\paren{1+\varepsilon}\cdot\paren{\delta\paren{u,p_{2}\paren{v}} + \delta\paren{p_{2}\paren{v},v}}}{$\paren{1+\varepsilon}$-MSASP from $S_{2}$} \\
        & = \textubrace{ \paren{1+\varepsilon}\cdot\paren{\delta\paren{u,v}+2\delta\paren{v,p_{2}\paren{v}}}}{Triangle inequality}
        && \leq \textubrace{ \paren{1+\varepsilon}\cdot\paren{\delta\paren{u,v}+2\delta\paren{v,p_{1}\paren{y}}}}{Due to \case{d}}\\
        &\leq  \textubrace{ \paren{1+\varepsilon}\cdot\paren{\delta\paren{u,v}+2\delta\paren{y,v}+2\delta\paren{y,p_{1}\paren{v}}}}{Triangle inequality} \\
        &\leq  \textubrace{ \paren{1+\varepsilon}\cdot\paren{\delta\paren{u,v}+4\delta\paren{y,v}}}{Due to \case{c}} \\
\end{alignat*} 

We pause and prove that $d\bracke{u,v}\leq \delta\paren{u,v}+2\delta\paren{u,p_{1}\paren{u}}$. Recall that after all the initializations, we update $d\bracke{p_{1}\paren{u},x}\leq \delta\paren{u,p_{1}\paren{u}}+\delta\paren{u,x}$. When we invoke SSSP from $p_{1}\paren{u}\in S_{1}$ we will consider the auxiliary edge $\paren{p_{1}\paren{u},y}$. The rest of the path $y\squiggly v\subseteq E_{S_{2}}$. Hence, we would reach from $p_{1}\paren{u}$ to $v$. 

\begin{alignat*}{3}
        d\bracke{u,v} &\leq \textubrace{d\bracke{u,p_{1}\paren{u}}+  d\bracke{p_{1}\paren{u},v}}{Update via $p_{1}\paren{u}$}
        && \leq \textubrace{\delta\paren{u,p_{1}\paren{u}} + \delta\paren{p_{1}\paren{u},v}}{Our preceding discussion} 
        && \leq \textubrace{ \delta\paren{u,v}+2\delta\paren{u,p_{1}\paren{u}}}{Triangle inequality}\\
\end{alignat*}

We can now return to our analysis. Let us sum both inequalities:

\begin{alignat*}{3}
        3 \cdot d\bracke{u,v} &\leq \textubrace{ \paren{1+\varepsilon}\cdot\paren{\delta\paren{u,v}+4\delta\paren{y,v}}+2\delta\paren{u,v}+4\delta\paren{u,p_{1}\paren{u}}}{Simply summing} \\
        & \leq \textubrace{\paren{1+\varepsilon}\cdot\paren{3\cdot\delta\paren{u,v}+4\cdot \paren{\delta\paren{y,v} + \delta\paren{u,p_{1}\paren{u}}} }}{Simply summing} \\
        & \leq \textubrace{ \paren{1+\varepsilon}\cdot\paren{3\cdot\delta\paren{u,v}+4\cdot \paren{\delta\paren{y,v} + \delta\paren{u,p_{2}\paren{u}}}}}{$\delta\paren{u,p_{1}\paren{u}}\leq\delta\paren{u,p_{2}\paren{u}}$} \\ 
        & \leq \textubrace{ \paren{1+\varepsilon}\cdot\paren{3\cdot\delta\paren{u,v}+4\cdot \paren{\delta\paren{y,v} + w\paren{x,y}}}}{By \lemref{7over3+eps_case_a2_properties}} \\
        &=  \textubrace{ \paren{1+\varepsilon}\cdot\paren{3\cdot\delta\paren{u,v}+4\cdot \delta\paren{x,v}}}{$ x\straight y \squiggly v$ is a shortest path} \\
        &\leq  \textubrace{ \paren{1+\varepsilon}\cdot 7\cdot\delta\paren{u,v}}{$ u\squiggly x \squiggly v$ is a shortest path} \leq  \textubrace{ \paren{7+7\varepsilon}\cdot\delta\paren{u,v}}{Simply summing} \\
\end{alignat*} 

By selecting $\varepsilon ' = \frac{3\cdot \varepsilon}{7}$, we get that: $3\cdot d\bracke{u,v} \leq \paren{7+3\cdot\varepsilon} \cdot \delta\paren{u,v}$.
Hence, it emanates that $d\bracke{u,v}\leq \paren{\frac{7}{3}+\varepsilon}\cdot \delta\paren{u,v}$.
\end{proof}

This indicates that for \case{c} in \clmref{7over3+eps_case_c}:

\begin{claim2}~\label{clm:7over3+eps_case_c} 
   If \case{c} holds then $d\bracke{u,v}  \leq \paren{\frac{7}{3}+\varepsilon}\cdot \delta\paren{u,v}$.
\end{claim2}

\begin{proof} 
    Note that either \case{d} or \notcase{d} hold. We have shown \clmref{7over3+eps_case_notd}  and \clmref{7over3+eps_case_d}. Hence, we conclude that either way $d\bracke{u,v}  \leq \paren{\frac{7}{3}+\varepsilon}\cdot \delta\paren{u,v}$.
\end{proof}

Going back to \case{b} from \clmref{7over3+eps_case_b}:
\begin{claim2}~\label{clm:7over3+eps_case_b} 
   If \case{b} holds then $d\bracke{u,v}  \leq \paren{\frac{7}{3}+\varepsilon}\cdot \delta\paren{u,v}$.
\end{claim2}
\begin{proof} 
    This follows by \clmref{7over3+eps_case_notc}  and \clmref{7over3+eps_case_c}.
\end{proof}

We have covered all sub-cases of \case{a}{2}. Hence we can prove \lemref{7over3+eps_case_a2}: 
\begin{lemma}~\label{lem:7over3+eps_case_a2} 
   If \case{a}{2} holds then $d\bracke{u,v}  \leq \paren{\frac{7}{3}+\varepsilon}\cdot \delta\paren{u,v}$.
\end{lemma}
\begin{proof} 
    By \clmref{7over3+eps_case_notb}  and \clmref{7over3+eps_case_b} we conclude that $d\bracke{u,v}  \leq \paren{\frac{7}{3}+\varepsilon}\cdot \delta\paren{u,v}$.
\end{proof} 

This indicates the end of our case-analysis. We have covered all three possible cases. We now move to our main theorem, stating both the correctness and runtime of \algref{7over3+eps}:

\begin{theorem}~\label{thm:7over3+eps}\Copy{thm:7over3+eps}{
   \algref{7over3+eps} computes a $\paren{\frac{7}{3}+\varepsilon}$-APASP in expected $\tilde O\paren{\paren{\frac{1}{\varepsilon}}^{O\paren{1}}\cdot n^{2.15135313}\cdot \log W}$ time, for an arbitrary $\varepsilon>0$ and weights of the set $\bracke{W}\cup\bracce{0,\infty}$.}
\end{theorem}

\begin{proof}
The correctness follows by the case analysis and \lemref{7over3+eps_case_a1}, \lemref{7over3+eps_case_a2} and \lemref{7over3+eps_case_a3}. We consider the runtime. Initializing $d$ takes $O\paren{n^{2}}$ time. Computing $\ball{u}{S_{1}}$ and 
$\ball{u}{S_{1}}{S_{2}}$ requires less than $\tilde O\paren{n^{2}}$ and can be performed by a preprocess that computes $d\bracke{u,v}$ over $E_{S_1}$ and $d\bracke{u,s}=d\bracke{s,u}$ for $s\in S_{1}$ over $E_{S_2}$. These values hold correct value
by \obvref{pathinball}. 
Finding both pivots of each vertex and computing the distances to both pivots requires $\tilde O \paren{m}$ by \obvref{pivotsdistance}. Construcitng the edge set $E_{S_{1}}$ (respectively, $E_{S_{2}}$) requires $\tilde O \paren{n^{1+\beta}}$ (respectively, $\tilde O \paren{n^{1+\beta+\gamma}}$) by \obvref{edgesruntime}.

As members are selected by probability independently, the expected size of $S_{1}$ is $\tilde O \paren{n^{1-\beta}}$ and of $S_{2}$ is $\tilde O \paren{n^{1-\beta-\gamma}}$. By \obvref{edgessize} the size of $E_{S_{1}}$ is $\tilde O \paren{n^{1+\beta}}$  and the size of $E_{S_{2}}$ is $\tilde O \paren{n^{1+\beta+\gamma}}$. As there are $n$ vertices, the size of $\paren{\bracce{s}\times V }\cup H$ is $\tilde O \paren{n}$, which is smaller than any $E_{S_{i+1}}$. Implementing the SSSP invocation by Dijkstra, the runtime  will therefore be $\tilde O \paren{\abs{S_{0}}\cdot \abs{E_{S_{1}}}+\abs{S_{1}}\cdot \abs{E_{S_{2}}}}  = \tilde O \paren{n^{2+\beta} + n^{2+\gamma}}$. Recall that $\abs{S_{2}} \in \tilde O \paren{n^{1-\beta-\gamma}}$. Invoking the $\paren{1+\varepsilon}$-MSASP  from $S_{2}$ requires $\tilde O \paren{m^{1+o\paren{1}} + \paren{ \frac{1}{\varepsilon} }^{O\paren{1}} \cdot n^{\omega\paren{1-\beta-\gamma}} \cdot  \log W}$.

Additional operations require: $\abs{E}\cdot n^{\beta} \in \tilde O\paren{mn^{\beta}}$ time for the updates in the for-loop straight after the initialization; $\tilde O\paren{n^{2+\gamma}}$ time for the updates through $s\in \ball{u}{S_{1}}{S_{2}}$ by \obvref{bunchsize}; $\tilde O\paren{n^{2}}$ for the final updates through all the pivots of either $u$ or $v$.  This concludes the runtime cost of each operation. The total runtime is therefore $\tilde O\paren{n^{2+\beta}+n^{2+\gamma}+\paren{ \frac{1}{\varepsilon} }^{O\paren{1}} \cdot n^{\omega\paren{1-\beta-\gamma}} \cdot  \log W}$. We can  \cite{complexityBalancer} select $\beta=\gamma \approx 0.15135313$. Hence the runtime will be: $\tilde O\paren{\paren{\frac{1}{\varepsilon}}^{O\paren{1}}\cdot n^{2.15135313}\cdot \log W}$. 

To address the sparse case, we can slightly adjust the algorithm to depend on the graph's density. Instead of simply considering $S_{1}$, we let $S_{1,1}\supseteq S_{1,2} \supseteq \ldots \supseteq S_{1,t}$ for $t\in O\paren{\log n}$. These sub-sets are sampled from one another hierarchically; a vertex from $S_{1,i}$ joins $S_{1,i+1}$ with probability $n^{-\frac{ \beta}{t}}$. Hence, a vertex $v\in V$ belongs to $S_{1,i}$ with probability $n^{-\frac{ i\cdot \beta}{t}}$. This  results in reducing the runtime for the initial SSSP invocations from all $s\in S_{0} =V$. 

Instead of invoking SSSP over the edges of $E_{S_{1}}$, we would invoke SSSP from all $s\in S_{1,i}$ over $E_{S_{1,i+1}}$. The runtime for each of the $\log n$ invocations would be $\tilde O \paren{n^{2}}$ instead of the previous single invocation which required $\tilde O\paren{n^{2+\beta}}$. 

We observe that the first for-loop iteration, right after the initialization, would still require $\tilde O \paren{mn^{\beta}}$. This reduces the component of $n^{2+\beta}$ to $mn^{\beta}$. As we still update through $s\in \ball{u}{S_{1}}{S_{2}}$ our runtime still consists of the component of $n^{2+\gamma}$. Hence, repeating this idea for $S_{2}$ would not result in any runtime improvement. The runtime now becomes $\tilde O \paren{mn^{\beta} + n^{2+\gamma} +\left(\frac{1}{\varepsilon} \right)^{O\left(1\right)} \cdot n^{\omega\left(1-\beta-\gamma\right)} \cdot  \log W}$ instead of $\tilde O\paren{n^{2+\beta}+n^{2+\gamma}+\paren{ \frac{1}{\varepsilon} }^{O\paren{1}} \cdot n^{\omega\paren{1-\beta-\gamma}} \cdot  \log W}$.
\end{proof}

\section{Multiplicative $\paren{\frac{3\ell+4}{\ell+2}+\varepsilon}$-APASP}\label{3ellplus4overellplus2+eps}

In this section we present \algref{3ellplus4overellplus2+eps} which is a framework of random algorithms for $\paren{\frac{3\ell+4}{\ell+2}+\varepsilon}$-APASP problems, where $\ell \in \mathbb{N}$. We observe that this is both a generalization of \algref{7over3+eps}, which is a $\paren{\frac{7}{3}+\varepsilon}$-APASP algorithm, and also a runtime improvement, for dense graphs, of the framework of the random $\frac{3\ell+4}{\ell+2}$-APASP algorithms presented by Akav and Roditty \cite{AkaRod2021}.

The runtime of \algref{3ellplus4overellplus2+eps} is $\tilde O\paren{mn^{\beta_{1}}+\msum{n^{2+\beta_{i}}}{i=2}{\ell+1}+\paren{\frac{1}{\varepsilon}}^{O\paren{1}}\cdot n^{\omega\paren{1-\alpha_{\ell+1}}}\cdot \log W}$.
The previously mentioned parameter $\ell$  indicates the trade-off between the runtime and the multiplicative stretch, while the values of the parameters $\beta_{1},\ldots,\beta_{\ell+1} \in \paren{0,1}$ will depend on graph's density, and will be set during the runtime analysis of \algref{3ellplus4overellplus2+eps}.

Our algorithm works as follows: for $i\in\bracke{\ell+1}$ we
sample each vertex of $S_{i-1}$ independently into $S_{i}$
with probability $n^{-\beta_{i}}$, where $S_{0}=V$. We compute the corresponding $E_{S_{i}}$ and set $S_{\ell+2}=\varnothing$, hence $E_{S_{\ell+2}}=E$. We also denote $\alpha_{i}=\msum{\beta_{j}}{j=1}{i}$. 
We now initialize the distance approximation matrix $d:V\times V \rightarrow \mathbb{R}^{\geq 0}$ to hold all edges' weights or $\infty$ else. We find the pivot  $p_{i}\paren{u}$  for all $u\in V$ and update the distance $d\bracke{u,p_{i}\paren{u}} \leftarrow \delta\paren{u,p_{i}\paren{u}}$ by a single SSSP invocation (See: \obvref{pivotsdistance}). 

Following the initialization, we update the distances, for $i\in\bracke{\ell+1}$, from any $\paren{x,y}\in E$ to $p_{i}\paren{v}$ of $v\in \ball{y}{S_{i-1}}{S_{i}}$. That is, $d\bracke{p_{i}\paren{v},x}\leftarrow \mmin{d\bracke{p_{i}\paren{v},v}+d\bracke{v,y}+d\bracke{x,y}, d\bracke{p_{i}\paren{v},x}}$. We now consider an $i\in \bracce{0}\cup \bracke{\ell}$ and proceed with an SSSP invocation from all $s\in S_{i}$ over the weighted graph $G_{i}\paren{s} = \paren{V,E_{S_{i+1}} \cup \paren{\bracce{s}\times V} \cup H ,d}$, where:

\begin{itemize}
    \item $E_{S_{i+1}}$ - The union, over \textbf{all} $u\in V$, of edges $\paren{u,v}$ such that $w\paren{u,v} < \delta\paren{u,p_{i+1}\paren{u}}$,
    \item $\bracce{s} \times V$ - Auxiliary edges from $s$ to  $V$, representing distances previously computed,
    \item $H =  \bracce{\paren{u,p_{j}\paren{u}} \midline u \in V}$ - Auxiliary edges from \textbf{any} $u$ to its pivots $p_{j}\paren{u}$ for $j\in\bracke{\ell+1}$.
\end{itemize}

The weight of each is given by $d:V\times V \rightarrow \mathbb{R}^{\geq 0}$. Observe that $d$ is a weight function over $G_{i}\paren{s}$. We  proceed with a $\paren{1+\varepsilon}$-MSASP  from the set of sources $S_{\ell+1}$ over the entire graph. Contrary to \algref{+2Wi} and similarly to \algref{7over3+eps}, the order of invocations of SSSP and MSASP here holds no importance.

\begin{algorithm}[H]
\caption{$\paren{\frac{3\ell+4}{\ell+2}+\varepsilon}$-APASP$\left(G=\left(V,E,w\right)\right)$}\label{alg:3ellplus4overellplus2+eps}

\textbf{Input:} An undirected graph $G=\left(V,E\right)$, a positive weight function $w:E\rightarrow\mathbb{R}^+$, an integer $\ell\in\mathbb{N}$ and an arbitrary $\varepsilon>0$.

\textbf{Output:} A matrix $d$ which is a $\paren{\frac{3\ell+4}{\ell+2}+\varepsilon}$-APASP. 

\medskip

Let $\beta_{1},\ldots,\beta_{\ell+1} \in \paren{0,1}$ to be fixed later

Initialize $d$ as the edges' weights and $\infty$ otherwise

Initialize $S_{0}\leftarrow V$ and set $S_{\ell+2} \leftarrow \varnothing$

\For{$i \in \bracke{\ell+1}$}
{


    \For{$s\in S_{i-1}$}
    {
        Sample $s$ independently into $S_{i}$ with probability $n^{-\beta_{i}}$
    }



    Compute the distances from any $u\in V$ to its pivot $p_{i}\paren{u}$ 

    Construct the set of edges $E_{S_{i}}$

    Compute the set $\ball{u}{S_{i-1}}{S_{i}}$ for all $u\in V$
}

\For{$\paren{x,y}\in E$}
{
    \For{$i \in \bracke{\ell+1}$}
    {
         \For{$v\in \ball{y}{S_{i-1}}{S_{i}}$}
        { 
            $d\bracke{p_{i}\paren{v},x}\leftarrow \mmin{d\bracke{p_{i}\paren{v},v}+d\bracke{v,y}+d\bracke{x,y}, d\bracke{p_{i}\paren{v},x}}$        
        }
    }
}
\For{$i \in \bracce{0}\cup \bracke{\ell}$}
{
    \For{$s\in  S_{i}$}
    {
        Invoke SSSP from $s$ on $G_{i} \paren{s} = \paren{ V,E_{ 
 S_{i+1}} \cup \paren{\bracce{s}\times V} \cup H, d }$ and update $d$ accordingly
    }
}

Compute a $\paren{1+\varepsilon}$-MSASP from $S_{\ell+1}$ on $G=\paren{V,E,w}$ and update $d$ accordingly

\For{$u,v\in V$}
{
    \For{$i \in \bracke{\ell+1}$}
    {
        \For{$s\in \ball{u}{S_{i-1}}{S_{i}}$}
        {
            $d\bracke{u,v}\leftarrow \mmin{d\bracke{u,s}+d\bracke{s,v},d\bracke{u,v}}$
        }
    }
}

\For{$u,v\in V$} 
{           
  $d\bracke{u,v}\leftarrow\mmin{d\bracke{u,p_{i}\paren{u}}+d\bracke{p_{i}\paren{u},v},d\bracke{u,v}}{i=0}{\ell+1}$
}

\Return $d$
\end{algorithm}

Finally, we apply two types of distance updates. Let $u,v\in V$ and $i \in \bracce{0}\cup \bracke{\ell}$. First,  $d\bracke{u,v}\leftarrow \mmin{d\bracke{u,s}+d\bracke{s,v},d\bracke{u,v}}$ through any $s\in\ball{u}{S_{i}}{S_{i+1}}$. The second is from all the $2\cdot\paren{\ell+1}$ pivots\footnote{Recall that $d\bracke{u,v}$ is updated whenever $d\bracke{v,u}$ is updated.} of both $u$ and $v$, namely: $d\bracke{u,v}\leftarrow \mmin{d\bracke{u,p_{i}\paren{u}}+d\bracke{p_{i}\paren{u},v},d\bracke{u,v}}$.

To prove the correctness of our algorithm, we follow an analogous proof to the $\frac{3\ell+4}{\ell+2}$-APASP of Akav and Roditty \cite{AkaRod2021}. As we performed changes to their algorithm, we provide a full proof of \algref{3ellplus4overellplus2+eps}, indicating where the additional $\varepsilon$ affects.
We begin with \clmref{3ellplus4overellplus2+eps_properties}, where we state certain properties, which reminds us of the properties of the edge $\paren{x,y}$ stated in \lemref{7over3+eps_case_a2_properties} in \secref{7over3+eps}. 

\begin{claim2}~\label{clm:3ellplus4overellplus2+eps_properties}
There exists an index $i\in \bracce{0}\cup \bracke{\ell}$ such that it is the largest index for which $\delta\paren{u,p_{i}\paren{u}} +\delta\paren{v,p_{i}\paren{v}} \leq \delta\paren{u,v} < \delta\paren{u,p_{i+1}\paren{u}} +\delta\paren{v,p_{i+1}\paren{v}}$. Additionally, either $d\bracke{u,v}\leq\paren{2+\varepsilon}\cdot\delta\paren{u,v}$ or there exists exactly one edge $\paren{a,b}\in u\squiggly v$ such that $\paren{a,b}\notin E_{S_{i+1}}$, $a\in \ball{u}{S_{i+1}}$,
$b\in \ball{v}{S_{i+1}}$,
$a\notin \ball{b}{S_{i+1}}$
and
$b\notin \ball{a}{S_{i+1}}$.
\end{claim2}

\begin{proof}
$S_{0}=V$ and $S_{\ell+2}=\varnothing$ imply that $\delta\paren{u,p_{0}\paren{u}} = \delta\paren{v,p_{0}\paren{v}} = 0$ and $\delta\paren{u,p_{\ell+2}\paren{u}} = \delta\paren{v,p_{\ell+2}\paren{v}} = \infty$. Since $\delta\paren{u,v}$ is neither zero nor infinite, such an index $i\in \bracke{0}\cup\bracce{\ell}$ exists. We may therefore consider the largest such index $i$.

If $u\squiggly v \subseteq E_{S_{i+1}}$ then, once SSSP is invoked from $p_{i}\paren{u}\in S_{i}$, the path $u\squiggly v$ is taken into consideration. Hence, $d\bracke{p_{i}\paren{u},v} \leq \delta\paren{p_{i}\paren{u},v} \leq \delta\paren{u,p_{i}\paren{u}}+\delta\paren{u,v}$. As $d\bracke{u,p_{i}\paren{u}}=\delta\paren{u,p_{i}\paren{u}}$, it follows that: $d\bracke{u,v}\leq d\bracke{u,p_{i}\paren{u}}+d\bracke{p_{i}\paren{u},v} \leq 2\cdot \delta\paren{u,p_{i}\paren{u}}+\delta\paren{u,v}$. By a symmetric argument for $p_{i}\paren{v}$, we can show that: $d\bracke{u,v}\leq d\bracke{u,p_{i}\paren{v}}+d\bracke{p_{i}\paren{v},v} \leq 2\cdot \delta\paren{v,p_{i}\paren{v}}+\delta\paren{u,v}$. By the selection of the index $i$ it follows that: 

$$2d\bracke{u,v} \leq 2\cdot\delta\paren{u,v} + 2\cdot \delta\paren{u,p_{i}\paren{u}} + 2\cdot \delta\paren{v,p_{i}\paren{v}} \leq 4\cdot \delta\paren{u,v}$$

It follows that $d\bracke{u,v}\leq 2\cdot \delta\paren{u,v} < \paren{2+\varepsilon }\cdot \delta\paren{u,v}$. We now consider the case where $u\squiggly v \not\subseteq E_{S_{i+1}}$. Therefore, there exists at least one edge $\paren{a,b}\in u\squiggly v$ such that $\paren{a,b}\notin E_{S_{i+1}}$. 

We now show that each vertex on $u\squiggly v $ must belong to either $\ball{u}{S_{i+1}}$ or $\ball{v}{S_{i+1}}$. Having even one vertex $z\in u\squiggly v$ such that $z\notin\ball{u}{S_{i+1}}\cup\ball{v}{S_{i+1}}$, would imply that $\delta\paren{u,v} = \delta\paren{u,z} + \delta\paren{z,v} \geq \delta\paren{u,p_{i+1}\paren{u}}+\delta\paren{v,p_{i+1}\paren{v}}$, which is a contradiction to the selection of the index $i$. We observe that having a vertex $z\in \ball{u}{S_{i+1}}\cap\ball{v}{S_{i+1}} $ would imply that $u\squiggly z \subseteq E_{S_{i+1}}$ and also $z\squiggly v \subseteq E_{S_{i+1}}$, which contradicts our current assumption that states  $u\squiggly v \not\subseteq E_{S_{i+1}}$.

We now assume, for the sake of contradiction, that there are at least two edges $u\squiggly a\straight b \squiggly c\straight d \squiggly v$ such that $\paren{a,b},\paren{c,d}\notin E_{S_{i+1}}$. We remark that if $b\in \ball{u}{S_{i+1}}$, then by \obvref{pathinball} it would follow that $u\squiggly a \squiggly b \subseteq E_{S_{i+1}}$, contradicting our selection of $\paren{a,b}$. Therefore, $b\notin \ball{u}{S_{i+1}}$. By a symmetric argument $c\notin \ball{v}{S_{i+1}}$. 

We previously argued that each vertex on $u\squiggly v$ must belong to either $\ball{u}{S_{i+1}}$
or $\ball{v}{S_{i+1}}$. It follows that $b\in \ball{v}{S_{i+1}}$
and
$c\in \ball{u}{S_{i+1}}$. By \obvref{pathinball} it follows that $b\squiggly c \straight d \squiggly v \subseteq E_{S_{i+1}}$ and $u\squiggly a \straight b \squiggly c \subseteq E_{S_{i+1}}$. These observations contradict that $\paren{c,d}\notin E_{S_{i+1}}$ and $\paren{a,b}\notin E_{S_{i+1}}$, respectively\footnote{This part is nearly identical to the proof of \lemref{7over3+eps_case_a2_properties}.}.

It follows that there is only a single edge $\paren{a,b}\notin E_{S_{i+1}}$. If either $a\in \ball{b}{S_{i+1}}$ or  $b\in \ball{a}{S_{i+1}}$ it would contradict $\paren{a,b}\notin E_{S_{i+1}}$. The same contradiction would occur if  $a\in \ball{v}{S_{i+1}}$ or  $b\in \ball{u}{S_{i+1}}$. As either vertex must be within  $\ball{u}{S_{i+1}}$ or  $ \ball{v}{S_{i+1}}$, it follows that  $a\in \ball{u}{S_{i+1}}$ and  $b\in \ball{v}{S_{i+1}}$. 
We briefly observe that $i<\ell+1$. Indeed, since $\paren{a,b}\notin E_{S_{i+1}}$, it must be that $E_{S_{i+1}}\neq E$. As $E= E_{S_{\ell+2}}$, it follows that $i+1<\ell+2$ or $i\leq \ell$. 
\end{proof}

Having proven the properties of the index $i$ and the edge $\paren{a,b}$, we can now 
establish two upper-bounds for $d\bracke{u,v}$. Combining them would result in the overall approximation. We begin with the first upper-bound:

\begin{lemma}~\label{lem:3ellplus4overellplus2+eps_minus}\Copy{lem:3ellplus4overellplus2+eps_minus}{
$d\bracke{u,v}\leq \paren{1+\varepsilon}\cdot \mmin{3\cdot \delta\paren{u,v}-2\cdot\delta\paren{b,v},3\cdot \delta\paren{u,v}-2\cdot\delta\paren{u,a}}$}
\end{lemma}

\begin{proof}
By \clmref{3ellplus4overellplus2+eps_properties} we know that $i\leq \ell$. Hence, as $E_{S_{\ell+2}}=E$, there exists an index $i+2\leq j \leq \ell+2$ such that $\paren{a,b} \in E_{S_{j}}$. As this is a shortest path $a\in \ball{b}{S_{j-1}}$. Hence, as we may increase $j$ or we have reached $j=\ell+2$, we may assume that $j$ is the smallest index for which both $b\in \ball{a}{S_{j-1}}$ and $a\in \ball{b}{S_{j-1}}$. 

Consider $p_{j-1}\paren{a}$ and  $p_{j-1}\paren{b}$. Since $i<\ell+1 \leq j-1$ and $j$ is the smallest index for which $b\in \ball{a}{S_{j}}$ then $b\notin \ball{a}{S_{j-1}}$ which means that $\delta\paren{a,p_{j-1}\paren{a}} \leq w\paren{a,b}$. Symmetrically, $\delta\paren{b,p_{j-1}\paren{b}} \leq w\paren{a,b}$. Therefore:

\begin{alignat*}{3}
        \delta\paren{u,p_{j-1}\paren{u}} & \leq \textubrace{\delta\paren{u,p_{j-1}\paren{a}}}{Pivot definition}
        && \leq \textubrace{ \delta\paren{u,a}+\delta\paren{a,p_{j-1}\paren{a}}}{Triangle inequality} \\
        & \leq \textubrace{ \delta\paren{u,a}+w\paren{a,b}}{Our preceding discussion}
        && = \textubrace{\delta\paren{u,b }}{$u\squiggly a \straight b$ is a shortest path}
\end{alignat*}

Symmetrically: $\delta\paren{v,p_{j-1}\paren{v}} \leq \delta\paren{a,v}$. Note that $\paren{a,b}\in \Gamma\paren{a,n^{\alpha_{j}}}$ and $i<j$, it indicates that $u\squiggly v \subseteq E_{S_{j}}$. Consider two possibilities: $j<\ell+2$ or $j=\ell+2$. 

If $j< \ell+2$ then $j-1\leq \ell$, hence $d\bracke{p_{j-1}\paren{u},v} \leq \delta\paren{u,v}+ \delta\paren{u,p_{j-1}\paren{u}}$ and $d\bracke{p_{j-1}\paren{v},u} \leq \delta\paren{u,v}+ \delta\paren{v,p_{j-1}\paren{v}}$. If $j=\ell+2$ then we invoked a $\paren{1+\varepsilon}$-MSASP from $S_{j-1}=S_{\ell+1}$, hence: $d\bracke{p_{j-1}\paren{u},v} \leq \paren{1+\varepsilon}\cdot \paren{\delta\paren{u,v}+ \delta\paren{u,p_{j-1}\paren{u}}}$ and $d\bracke{p_{j-1}\paren{v},u} \leq \paren{1+\varepsilon}\cdot \paren{\delta\paren{u,v}+ \delta\paren{v,p_{j-1}\paren{v}}}$. In any case, as we consider both $p_{j-1}\paren{u}$ and $p_{j-1}\paren{v}$, it follows that:

\begin{alignat*}{3}
        d\bracke{u,v} & \leq \textubrace{\mmin{d\bracke{u,p_{j-1}\paren{u}}+d\bracke{p_{j-1}\paren{u},v},d\bracke{u,p_{j-1}\paren{v}}+d\bracke{p_{j-1}\paren{v},v}}}{Updating by both pivots}\\
        & \leq \textubrace{ \paren{1+\varepsilon}\cdot \paren{\mmin{\delta\paren{u,v}+ 2\cdot \delta\paren{u,p_{j-1}\paren{u}},\delta\paren{u,v}+ 2\cdot \delta\paren{v,p_{j-1}\paren{v}}} } }{Our preceding discussion} \\
        & \leq \textubrace{ \paren{1+\varepsilon}\cdot \paren{\mmin{\delta\paren{u,v}+ 2\cdot \delta\paren{u,b},\delta\paren{u,v}+ 2\cdot \delta\paren{a,v}} } }{Our pre-preceding discussion} \\
        & = \textubrace{\paren{1+\varepsilon}\cdot \paren{ \mmin{3\cdot \delta\paren{u,v} - 2\cdot \delta\paren{b,v},3\cdot \delta\paren{u,v} - 2\cdot \delta\paren{u,a}} } }{$u\squiggly a \straight b\squiggly v$ is a shortest path}
\end{alignat*}

\end{proof}

Before proceeding to the next bound, we direct the reader to \appref{3ellplus4overellplus2}, where we present the algorithm of Thorup and Zwick \cite{ThoZwi2005}, denoted as \algref{tz}. The correctness of the following bound relies on certain properties of \algref{tz}.

\begin{lemma}~\label{lem:3ellplus4overellplus2+eps_plus}\Copy{lem:3ellplus4overellplus2+eps_plus}{
$d\bracke{u,v}\leq \delta\paren{u,v}+2\cdot\paren{\ell+1}\cdot\delta\paren{u,a}$ or $d\bracke{u,v}\leq  \delta\paren{u,v}+2\cdot\paren{\ell+1}\cdot\delta\paren{b,v}$}
\end{lemma}

\begin{proof}
Let $r\in \bracce{0}\cup\bracke{i}$ be the largest index such that $a\notin \ball{u}{S_r}$
and
$u\notin \ball{a}{S_r}$. Let $q\in \bracce{0}\cup\bracke{i}$ be the largest index such that $b\notin \ball{v}{S_q}$
and
$v\notin \ball{b}{S_q}$. By \clmref{3ellplus4overellplus2+eps_properties} it follows that $a\in \ball{u}{S_{i+1}}$
and
$b\in \ball{v}{S_{i+1}}$. Hence, $r,q<i+1$. 

We distinguish between two cases. At first, we consider when $q\leq r$ and show that $d\bracke{u,v}\leq \paren{1+\varepsilon}\cdot \paren{\delta\paren{u,v}+2\cdot\paren{\ell+1}\cdot\delta\paren{u,a}}$. Else, when $r<q $, we show that $d\bracke{u,v}\leq \paren{1+\varepsilon}\cdot \paren{\delta\paren{u,v}+2\cdot\paren{\ell+1}\cdot\delta\paren{b,v}}$. As the proof for the second case is symmetrical, we omit the details. We therefore focus solely on the case where $q\leq r$. 

Before we continue, let us consider several edge cases. If $u\squiggly v = u\straight v$ then $d\bracke{u,v}=\delta\paren{u,v}$. If $u\squiggly v = u\straight a \straight b=v$ we ignore $q$ and only consider $r$. If $u\squiggly v = u=a\straight b \straight v$, we ignore $r$ and consider only $q$. Either way, the proof below (or its symmetrical version) remains valid. Therefore, let us assume $q\leq r$.

By the definition of $r$ it follows that $a\in \ball{u}{S_{r+1}}$
or
$u\in \ball{a}{S_{r+1}}$. In either case, by \obvref{pathinball}, it follows that $u\squiggly a \subseteq E_{S_{r+1}}$. Symmetrically, $b\squiggly v \subseteq E_{S_{q+1}}$. By \obvref{ESiwithinESi+1} it follows that $E_{S_{q+1}}\subseteq E_{S_{r+1}}$. Hence, $u\squiggly a, b\squiggly v\subseteq E_{S_{r+1}}$. 

Recall \algref{tz}. Let $f=f\paren{u,a,r}$ indicate the index for which \algref{tz} halts for the input $u,a,r$. We now consider several cases regarding the parity of $\ell$, $r$ and $f$. We begin by assuming that $\ell$ is odd, $r$ is even and $f$ is even. 

In this case, $f-r$ is even as well.  By \lemref{tz} it follows that $\delta\paren{u,p_{f}\paren{u}} \leq \paren{f-r}\cdot\delta\paren{u,a}+\delta\paren{u,p_{r}\paren{u}}$. By the definition of $f$ it follows that $p_{f}\paren{u}
\in \ball{a}{S_{f}}{S_{f+1}}$. Hence, in our algorithm, we update the distance for the edge $\paren{a,b}$  to be: $d\bracke{p_{f}\paren{u},b} \leq \delta\paren{p_{f}\paren{u},a}+w\paren{a,b}$. As $r\leq f$ it follows by \obvref{ESiwithinESi+1} that $E_{S_{r}} \subseteq E_{S_{f}}$. When invoking SSSP from $p_{f}\paren{u}\in S_{f}$ the edges of $b\squiggly v \subseteq E_{S_{r+1}} \subseteq E_{S_{f+1}}$. Hence: 

 \begin{alignat*}{3}
        d\bracke{p_{f}\paren{u},v} & \leq \textubrace{d\bracke{p_{f}\paren{u},b} +\delta\paren{b,v}}{SSSP from $p_{f}\paren{u}\in S_{f}$}
        && \leq \textubrace{ \delta\paren{p_{f}\paren{u},a}+w\paren{a,b} + \delta\paren{b,v}}{Our preceding discussion} \\
        & \leq \textubrace{\delta\paren{u,p_{f}\paren{u}} +\delta\paren{u,a}+\delta\paren{a,v}}{Triangle inequality}
        && \leq \textubrace{\delta\paren{u,p_{f}\paren{u}} +\delta\paren{u,v} }{$u\squiggly a \squiggly v$ is a shortest path}
\end{alignat*}

Recall that we update $d\bracke{u,v}$ to pass through $p_{f}\paren{u}$. Ergo:

\begin{alignat*}{3}
        d\bracke{u,v} & \leq \textubrace{d\bracke{u,p_{f}\paren{u}} +d\bracke{p_{f}\paren{u},v}}{Update through $p_{f}\paren{u}$}
        && \leq \textubrace{ \delta\paren{u,v}+2\cdot\delta\paren{u,p_{f}\paren{u}}}{Our preceding inequality} \\
        & \leq \textubrace{\delta\paren{u,v} +2\cdot \paren{\paren{f-r}\cdot\delta\paren{u,a}}+\delta\paren{u,p_{r}\paren{u} }}{Our pre-preceding discussion}
\end{alignat*}

If $r=0$ then $\delta\paren{u,p_{r}\paren{u}}=0$. As $f-r\leq \ell+1$ it follows that: $d\bracke{u,v} \leq \delta\paren{u,v} +2\cdot \paren{\ell+1}\cdot\delta\paren{u,a}$. If $r\geq 1$ then $\delta\paren{u,p_{r}\paren{u}}\leq \delta\paren{u,a}$, as $r$ was selected such at $a\notin \ball{u}{S_r}$. Therefore: $d\bracke{u,v}\leq \delta\paren{u,v} +2\cdot \paren{f-r+1}\cdot\delta\paren{u,a}$. Yet $r\geq 1$, hence: $d\bracke{u,v}\leq \delta\paren{u,v} +2\cdot f\cdot\delta\paren{u,a} \leq \delta\paren{u,v} +2\cdot \paren{\ell+1}\cdot\delta\paren{u,a}$. 

The rest of the proof, for the other parity cases, remains precisely as is in the proof of \lemref{3ellplus4overellplus2_plus} \cite{AkaRod2021}. As we make no changes, we omit the technical details for these cases.
\end{proof}

Having proven both upper-bounds, we can now follow with our main theorem:

\begin{theorem}~\label{thm:3ellplus4overellplus2+eps}\Copy{thm:3ellplus4overellplus2+eps}{ 
   Let $\alpha_{j}=\msum{\beta_{i}}{i=1}{j}$, $\varepsilon>0$ with weights from $\bracke{W}\cup\bracce{0,\infty}$. \algref{3ellplus4overellplus2+eps} computes a $\paren{\frac{3\ell+4}{\ell+2}+\varepsilon}$-APASP in expected $\tilde O\paren{\paren{\frac{1}{\varepsilon}}^{O\paren{1}}\cdot n^{\omega\paren{1-\alpha_{\ell+1}}}\cdot \log W+mn^{\beta_{1}}+\msum{n^{2+\beta_{i}}}{i=2}{\ell+1}}$ time.}
\end{theorem}

\begin{proof}
By \lemref{3ellplus4overellplus2+eps_plus} it follows that either $d\bracke{u,v}\leq \delta\paren{u,v}+2\cdot \paren{\ell+1}\cdot\delta\paren{u,a}$ or $d\bracke{u,v}\leq \delta\paren{u,v}+2\cdot \paren{\ell+1}\cdot\delta\paren{b,v}$. Assume, w.l.o.g., that: $d\bracke{u,v}\leq \delta\paren{u,v}+2\cdot \paren{\ell+1}\cdot\delta\paren{u,a}$.  

Due to \lemref{3ellplus4overellplus2+eps_minus} we know that: $d\bracke{u,v}\leq \paren{1+\varepsilon}\cdot \mmin{3\cdot\delta\paren{u,v}-2\cdot\delta\paren{b,v},3\cdot\delta\paren{u,v}-2\cdot\delta\paren{u,a}}$. Substituting the above yields that:  $d\bracke{u,v}\leq  \mmin{\delta\paren{u,v}+2\cdot \paren{\ell+1}\cdot\delta\paren{u,a},\paren{1+\varepsilon}\cdot\paren{3\cdot\delta\paren{u,v}-2\cdot\delta\paren{u,a}}}$. We consider the following two possibilities:

\begin{enumerate}
    \item When $\delta\paren{u,a} \leq \frac{\delta\paren{u,v}}{\ell+2}$. Then: $d\bracke{u,v} \leq \delta\paren{u,v}+2\cdot \paren{\ell+1}\cdot\delta\paren{u,a} \leq \delta\paren{u,v} + 2\cdot \frac{\ell+1}{\ell+2}\cdot \delta\paren{u,v} = \frac{3\ell+4}{\ell+2}\cdot \delta\paren{u,v}$, 
    \item Otherwise: $\delta\paren{u,a} > \frac{\delta\paren{u,v}}{\ell+2}$, or $-\delta\paren{u,a} < -\frac{\delta\paren{u,v}}{\ell+2}$. Therefore:  $d\bracke{u,v} \leq\paren{1+\varepsilon}\cdot \paren{ 3\cdot \delta\paren{u,v}-2\cdot \delta\paren{u,a}} < \paren{1+\varepsilon}\cdot \paren{3\cdot \delta\paren{u,v} -\frac{2\cdot \delta\paren{u,v}}{\ell+2}} = \paren{1+\varepsilon}\cdot \frac{3\ell+4}{\ell+2}\cdot\delta\paren{u,v} = \paren{\frac{3\ell+4}{\ell+2}+\varepsilon\cdot \frac{3\ell+4}{\ell+2} }\cdot \delta\paren{u,v}$.
\end{enumerate}

By selecting $\varepsilon ' = \frac{\paren{\ell+2}\cdot \varepsilon}{3\ell+4}$, we conclude the correctness of \algref{3ellplus4overellplus2+eps}. 

We now turn to consider its runtime. Initializing $d$ takes $O\paren{n^{2}}$ time. Let $i\in\bracce{0}\cup\bracke{\ell}$. Finding the $i^{\textnormal{th}}$ pivot of each vertex and computing the distances to it requires $\tilde O \paren{m}$ by \obvref{pivotsdistance}. Finding the edge set $E_{S_{i}}$ requires $\tilde O \paren{n^{1+\alpha_{i}}}$ by \obvref{edgesruntime}. 
Using the standard bunch preprocessing procedure, compute
$\ball{u}{S_i}{S_{i+1}}$ and store
$d\bracke{u,s}=\delta\paren{u,s}$ for every
$s\in\ball{u}{S_i}{S_{i+1}}$. By \obvref{pathinball} this will indeed verify whether $s\in \ball{u}{S_{i}}{S_{i+1}}$ or not.
Hence, the total runtime for the initialization is $\tilde O \paren{n^{2}}$, as $\ell \in \tilde O\paren{1}$. 

By definition of $\alpha_{i}$, the expected size of $S_{i}$ is $\tilde O \paren{n^{1-\alpha_{i}}}$. By \obvref{edgessize} the expected size of $E_{S_{i+1}}$ is $\tilde O \paren{n^{1+\alpha_{i+1}}}=\tilde O \paren{n^{1+\alpha_{i}+\beta_{i+1}}}$. As there are $n$ vertices, the size of $\paren{\bracce{s}\times V }\cup H$ is $\tilde O \paren{n}$, which is smaller than any $E_{S_{i+1}}$. Implementing the SSSP invocation by Dijkstra, the runtime  would therefore be $ \tilde O \paren{n^{2+\beta_{i+1}}}$. Recall that $\abs{S_{\ell+1}} \in \tilde O \paren{n^{1-\alpha_{\ell+1}}}$. The $\paren{1+\varepsilon}$-MSASP invocation requires $\tilde O \paren{m^{1+o\paren{1}} + \paren{ \frac{1}{\varepsilon} }^{O\paren{1}} \cdot n^{\omega\paren{1-\alpha_{\ell+1}}} \cdot  \log W}$.

Additional operations require: $\abs{E}\cdot n^{\beta_{i}} \in \tilde O\paren{mn^{\beta_{i}}}$ time for the updates in the for-loop straight after the initialization; $\tilde O\paren{n^{2+\beta_{i+1}}}$ time for the updates through $s\in \ball{u}{S_{i}}{S_{i+1}}$ as the expected size of $\ball{u}{S_{i}}{S_{i+1}}$ is $n^{\beta_{i+1}}$ by \obvref{bunchsize}; $\tilde O\paren{n^{2}}$ for the final updates through all the pivots of either $u$ or $v$. 

The total runtime is therefore $\tilde O\paren{\paren{\frac{1}{\varepsilon}}^{O\paren{1}}\cdot n^{\omega\paren{1-\alpha_{\ell+1}}}\cdot \log W+\msum{n^{2+\beta_{i}}}{i=1}{\ell+1}}$. We can reduce it even more,  to depend on the graph's density. 

Set $S_{1,0}=V$ and construct $S_{1,0}\supseteq S_{1,1}\supseteq\cdots\supseteq S_{1,t}$ by sampling each vertex of $S_{1,i-1}$ independently into
$S_{1,i}$ with probability
$n^{-\frac{\beta_1}{t}}$ for $t\in O\paren{\log n}$. 
Instead of an SSSP invocation from any $s\in S_{0}$ over the edge set $E_{S_{1}}$, we invoke SSSP from any $s\in S_{1,i}$ over $E_{S_{1,i+1}}$. This would require  $\tilde O\paren{n^{2}}$ runtime. 
The additional updates for all $\paren{x,y}\in E$ and $v\in\ball{y}{S_{1}}$ would still require $\tilde O\paren{mn^{\beta_{1}}}$ time. The updates through $s\in \ball{u}{S_{i}}{S_{i+1}}$ would still require $\tilde O\paren{n^{2+\beta_{i+1}}}$ time\footnote{Observe that this is also the case when the runtime optimization that depends on the graph's density occurs in \thmref{7over3+eps}.}. Therefore, there is no use in partitioning $S_{i+1}$ for any $i\in\bracke{\ell}$ into $O\paren{\log n}$ levels, as was done for $S_{1}$. 

For simplicity, we denote: $\beta_{1}=\beta$ and $\gamma_{i}=\beta_{i+1}$ for all $i\in\bracke{\ell}$. The rest of the runtime analysis remains unchanged. We conclude by stating the proper runtime is:

$$\tilde O \paren{mn^{\beta}+\msum{n^{2+\gamma_{i}}}{i=1}{\ell}+\left(\frac{1}{\varepsilon} \right)^{O\left(1\right)} \cdot n^{\omega\left(1-\beta-\gamma_{1}-\ldots-\gamma_{\ell}\right)} \cdot  \log W}$$

The expression above can be simplified by selecting $\beta_{1}=\beta$ and $\beta_{2}=\ldots = \beta_{\ell+1}=\gamma$. The runtime then becomes $\tilde O \paren{mn^{\beta}+n^{2+\gamma}+\paren{\frac{1}{\varepsilon}}^{O\paren{1}}\cdot n^{\omega\paren{1-\beta-\ell\cdot \gamma}} \cdot  \log W}$. When the graph in question is dense, i.e. $m\in \tilde \Theta\paren{n^2}$, we simplify even more by setting, additionally, $\beta=\gamma$.  
\end{proof}

This completes the correctness and runtime analysis of \algref{3ellplus4overellplus2+eps}. To state explicit runtimes, we focus on the dense setting $m\in \tilde \Theta \paren{n^2}$. Given the  value of $\ell$, one can optimize \cite{complexityBalancer} the exact values for $\beta$.

\begin{table}[t]
  \begin{center}
\small\addtolength{\tabcolsep}{-3pt}
    \renewcommand\arraystretch{1.5}
    \begin{tabular}{|c|c|c|} 
      \hline
      \textbf{\small $\ell$} & \textbf{\small Multiplicative Stretch}  & \textbf{\small Runtime} \\
      \hhline{|=|=|=|}
      $0$& $2+\varepsilon$  & $\paren{\frac{1}{\varepsilon}}^{O\paren{1}}\cdot n^{2.21235201}\cdot \log W$ \\
      \hline
      $1$& $\frac{7}{3}+\varepsilon$  & $\paren{\frac{1}{\varepsilon}}^{O\paren{1}}\cdot n^{2.15135313}\cdot \log W$ \\
      \hline
      $2$& $\frac{5}{2}+\varepsilon$  & $\paren{\frac{1}{\varepsilon}}^{O\paren{1}}\cdot n^{2.1185119}\cdot \log W$ \\
      \hline
      $3$& $\frac{13}{5}+\varepsilon$  & $\paren{\frac{1}{\varepsilon}}^{O\paren{1}}\cdot n^{2.09778592}\cdot \log W$ \\
      \hline
      $4$& $\frac{8}{3}+\varepsilon$  & $\paren{\frac{1}{\varepsilon}}^{O\paren{1}}\cdot n^{2.08346083}\cdot \log W$\\
      \hline
    \end{tabular}
    \renewcommand\arraystretch{1}
    \caption{Examples for several $\paren{\frac{3\ell+4}{\ell+2}+\varepsilon}$-APASP for various values of $\ell\in\mathbb{N}$. }~\label{tab:3ellplus4overellplus2+eps_examples}
  \end{center}

    \vspace{-6mm}
\end{table} 

Our framework yields a wide range of results. To perceive them, we set, as examples, several values of $\ell$ (See: \tabref{3ellplus4overellplus2+eps_examples}). When $\ell=0$ we compute a $\paren{2+\varepsilon}$-APASP in $\tilde O \paren{\paren{\frac{1}{\varepsilon}}^{O\paren{1}}\cdot n^{2.21235201}\cdot \log W}$ time, which is the same as Dory, Forster, Kirkpatrick, Nazari, Vassilevska Williams, and de Vos  \cite{DorForKirNazVasVos2023}; when $\ell=1$ our algorithm computes a $\paren{\frac{7}{3}+\varepsilon}$-APASP in $\tilde O \paren{\paren{\frac{1}{\varepsilon}}^{O\paren{1}}\cdot n^{2.15135313}\cdot \log W}$ time -- as in \algref{7over3+eps}; when $\ell=2$ our algorithm computes a $\paren{\frac{5}{2}+\varepsilon}$-APASP in $\tilde O \paren{\paren{\frac{1}{\varepsilon}}^{O\paren{1}}\cdot n^{2.1185119}\cdot \log W}$ time, etc. 

\section{Near additive $\paren{1+\varepsilon,\mmin{2W_{1},4W_{2}}}$-APASP}\label{1+eps,4W2}

In this section we introduce the deterministic \algref{1+eps,4W2}, which considers on two parameters $\beta,\gamma\in\paren{0,1}$, whose values will be determined when we will analyze the runtime. 
\algref{1+eps,4W2} proceeds as follows: we compute the hitting set $S_{1}$ (resp. $S_{2}$) for the sets $\Gamma\paren{u,n^{\beta}}$ (resp. $\Gamma\paren{u,n^{\beta+\gamma}}$), over all $u\in V$. We compute $E_{S_{1}}$ (resp. $E_{S_{2}}$). We also set $S_{0}=V$ and $S_{3}=\varnothing$.

The distance approximation matrix $d:V\times V \rightarrow \mathbb{R}^{\geq 0}$ is initialized to the weight of the edges or $\infty$ else-wise. We find the pivots $p_{1}\paren{u}$ (resp. $p_{2}\paren{u}$) for all $u\in V$ and the exact distance $d\bracke{u,p_{1}\paren{u}} \leftarrow \delta\paren{u,p_{1}\paren{u}}$ (resp. $d\bracke{u,p_{2}\paren{u}}$) by a single SSSP invocation (See: \obvref{pivotsdistance}). 

We proceed by invoking a $\paren{1+\varepsilon}$-MSASP from $S_{2}$ over the entire graph. We then compute a $\paren{1+\varepsilon}$-AMPMM of $d_{0}^{2} \star d_{2}^{0}$, where $d_{i}^{j}$ is a rectangular matrix containing all values of $d$, its rows narrowed down to include only the vertices of $S_{i}$ and its columns to those of $S_{j}$. We note that similarly to \algref{+2Wi}, the order here is of importance. 

The following occurs twice, in decreasing order $i\leftarrow 1$ to $0$ (only two hitting sets): we perform an SSSP invocation from all $s\in S_{i}$ on the graph $G_{i}\paren{s}=\paren{V,E_{S_{i+1}} \cup \paren{\bracce{s}\times V} \cup H,d}$, where:

\begin{itemize}
    \item $E_{S_{i+1}}$ - The union, over \textbf{all} $u\in V$, of edges $\paren{u,v}$ such that $w\paren{u,v} < \delta\paren{u,p_{i+1}\paren{u}}$,
    \item $\bracce{s} \times V$ - Auxiliary edges from $s$ to $V$, representing distances previously computed,
    \item $H =  \bracce{\paren{u,p_{j}\paren{u}} \midline u \in V}$ - Auxiliary edges from \textbf{any} $u$ to its pivots $p_{j}\paren{u}$ for $j\in\bracce{1,2}$.
\end{itemize}

In other words, after we performed an SSSP invocation from all $s\in S_{1}$, we perform one from all $v\in S_{0}=V$, then again from all $s\in S_{1}$ and then finally one more time from all $v\in S_{0}=V$. We observe that $d:V\times V \rightarrow \mathbb{R}^{\geq 0}$ is a weight function over $G_{i}\paren{s}$.

\begin{algorithm}[t]
\caption{$\paren{1+\varepsilon,\mmin{2W_{1},4W_{2}}}$-APASP$\left(G=\left(V,E,w\right)\right)$}\label{alg:1+eps,4W2}

\textbf{Input:} An undirected non-negative weighted graph $G=\left(V,E,w\right)$ and an  $\varepsilon>0$.

\textbf{Output:} A  $\paren{1+\varepsilon,\mmin{2W_{1},4W_{2}}}$-APASP. 

\medskip

Let $\beta,\gamma \in \paren{0,1}$ to be fixed later

Initialize $d$ as the edges' weights and $\infty$ otherwise

Compute $\Gamma\paren{u,n^{\beta}}$ (respectively, $\Gamma\paren{u,n^{\beta+\gamma}}$) for all $u\in V$

Compute $S_{1}$ (respectively, $S_{2}$) to be a hitting set to $\bracce{\Gamma\paren{u,n^{\beta}} \mid u\in V}$ (respectively, $\Gamma\paren{u,n^{\beta+\gamma}}$)

Compute the distances from any $u\in V$ to its pivot $p_{1}\paren{u}$ (respectively, $p_{2}\paren{u}$)

Construct the set of edges $E_{S_{1}}$ (respectively, $E_{S_{2}}$)

Compute $\paren{1+\varepsilon}$-MSASP from $S_{2}$ on $G=\paren{V,E,w}$ and update $d$ accordingly

Compute a $\paren{1+\varepsilon}$-AMPMM $d_{0}^{2} \star d_{2}^{0}$ and update $d$ accordingly

\For{$j \in \bracke{2}$}
{
    \For{$i \in \pair{1,0}$}
    {
        \For{$s\in  S_{i}$}
        {
            Invoke SSSP from $s$ on $G_{i} \paren{s} = \left( V,E_{ 
     S_{i+1}} \cup \paren{\bracce{s}\times V} \cup H  , d \right)$ and update $d$ accordingly
        }
    }
}

Compute a $\paren{1+\varepsilon}$-AMPMM $d_{0}^{1} \star d_{1}^{1}$ and update $d$ accordingly

\For{$u\in  S_{0}=V$}
        {
            Invoke SSSP from $u$ on $G_{0} \paren{u} = \left( V,E_{ 
     S_{1}} \cup \paren{\bracce{u}\times V} \cup H  , d \right)$ and update $d$ accordingly
        }

\Return $d$

\end{algorithm}\vspace{0mm}

We then compute another $\paren{1+\varepsilon}$-AMPMM, this time for the rectangular matrices $d_{0}^{1}\star d_{1}^{1}$. Finally, we perform one more SSSP invocation from all $u\in V = S_{0}$  over the graph $G_{0}\paren{u}=\paren{V,E_{S_{1}} \cup \paren{\bracce{u}\times V} \cup H,d}$.

To prove the \emph{near additive} approximation of $\paren{1+\varepsilon,\mmin{2W_{1},4W_{2}}}$, we begin by presenting a case analysis in a case-tree, as we previously did for \algref{7over3+eps}. 

Let $u,v\in V$ and $u\squiggly v$. Consider $\paren{a,b},\paren{c,d}\in u\squiggly v$ such that $\paren{a,b}$ is a heaviest edge and $\paren{c,d}$ is a second heaviest edge. We consider the following cases and sub-cases:
\begin{itemize}
    \item \case{a}{1}: $u\squiggly v\subseteq E_{S_{1}}$,
    \item \case{a}{2}: There is exactly one edge $\paren{z,t}\in u\squiggly v$ such that $\paren{z,t}\notin E_{S_{1}}$,
    \item \case{a}{3}: There are at least two edges on $u\squiggly v$ that are not members of $E_{S_{1}}$.
    \begin{itemize}
        \item \case{b}{1}: $u\squiggly v \subseteq E_{S_{2}}$,
        \item \case{b}{2}: There is exactly one $\paren{x,y}\in u\squiggly v$ and at least one $\paren{z,t}\in u\squiggly v$ such that $\paren{x,y}\notin E_{S_{2}}$ and $\paren{z,t}\notin E_{S_{1}}$,
        \item \case{b}{3}: There are at least two $\paren{x_1,y_1},\paren{x_2,y_2} \in u\squiggly v$ such that  $\paren{x_1,y_1},\paren{x_2,y_2} \notin E_{S_{2}}$.
    \end{itemize}
\end{itemize}

We observe that the above covers all possible cases. We proceed by tackling \case{a}{1}, \case{a}{2} and finally the non-trivial \case{a}{3}. Let us begin with \case{a}{1}:

\begin{lemma}~\label{lem:1+eps,4W2_case_a1} 
   If \case{a}{1} holds then $d\bracke{u,v} = \delta\paren{u,v}$.
\end{lemma}

\begin{proof} The path $u\squiggly v \subseteq E_{S_{1}}$, hence, when invoking SSSP from $u\in V=S_{0}$ for the first time, the entire path would be considered. Hence, $d\bracke{u,v}=\delta\paren{u,v}$.
\end{proof}

We now move to \case{a}{2}, whose proof is slightly less straightforward:

\begin{lemma}~\label{lem:1+eps,4W2_case_a2} 
   If \case{a}{2} holds then $d\bracke{u,v} = \delta\paren{u,v}$.
\end{lemma}

\begin{proof} There is a single edge $\paren{z,t}\notin E_{S_{1}}$. Hence, $u\squiggly z, t\squiggly v\subseteq E_{S_{1}}$. When invoking SSSP from $t\in V = S_{0}$ for the first time, the neighbouring edge $\paren{z,t}$ will be taken into consideration as well. Hence, $d\bracke{t,u}=\delta\paren{t,u}$. During the second SSSP invocations from $S_{0}=V$, this time from $u\in V$, the distance $d\bracke{u,t}=\delta\paren{u,t}$ would already be computed due to the previous iteration. The rest of $t\squiggly v\in E_{S_{1}}$. Therefore, $d\bracke{u,v}=\delta\paren{u,v}$. 
\end{proof}

We can now delve within \case{a}{3}. To do so, we consider its sub-cases. We begin with the \case{b}{1}: 

\begin{lemma}~\label{lem:1+eps,4W2_case_b1} 
   If \case{b}{1} holds then $d\bracke{u,v} \leq \delta\paren{u,v}+2\cdot w\paren{c,d}$.
\end{lemma}

\begin{proof} Let $\paren{z_{1},t_{1}},\paren{z_{2},t_{2}}\in u\squiggly v$ such that $\paren{z_{1},t_{1}}, \paren{z_{2},t_{2}} \notin E_{S_{1}}$ and $u\squiggly z_{1} \subseteq E_{S_{1}}$ and $t_{2}\squiggly v \subseteq E_{S_{1}}$ -- Namely, these are the first and the last edges on $u\squiggly v$ that do not belong to $E_{S_{1}}$. Due to \case{a}{3}, they are distinct. Assume, w.l.o.g., that $w\paren{z_{2},t_{2}}\leq w\paren{z_{1},t_{1}}$. We know that $u\squiggly v\subseteq E_{S_{2}}$. Hence, when invoking SSSP from $p_{1}\paren{t_{2}}\in S_{1}$, the path $p_{1}\paren{t_{2}} \straight t_{2} \squiggly u$ will be considered. Hence, $d\bracke{p_{1}\paren{t_{2}},u} \leq \delta\paren{u,t_{2}}+\delta\paren{t_{2},p_{1}\paren{t_{2}}} $. When invoking SSSP from $u\in S_0=V$ for the second time, the auxiliary edge $\paren{u,p_{1}\paren{t_{2}}}$ will be considered as well. Its weight is at most $\delta\paren{u,t_{2}}+\delta\paren{t_{2},p_{1}\paren{t_{2}}}$. Recall that $t_{2}\squiggly v \subseteq E_{S_{1}}$. Therefore:

    \begin{alignat*}{3}
        d\bracke{u,v} & \leq \textubrace{d\bracke{u,p_{1}\paren{t_{2}}}+  \delta\paren{p_{1}\paren{t_{2}},t_{2}} + \delta\paren{t_{2},v}}{second SSSP from $u\in V=S_{0}$}
        && \leq \textubrace{ \delta\paren{u,t_{2}}+\delta\paren{t_{2},p_{1}\paren{t_{2}}}+\delta\paren{p_{1}\paren{t_{2}},t_{2}} + \delta\paren{t_{2},v}}{Our preceding discussion} \\
        & \leq \textubrace{ \delta\paren{u,t_{2} } + 2 \cdot w\paren{z_{2},t_{2}} + \delta\paren{t_{2},v}}{$\paren{z_{2},t_{2}}\notin E_{S_{1}}$}
        && = \textubrace{\delta\paren{u,v } + 2 \cdot w\paren{z_{2},t_{2}}}{$u\squiggly t_{2} \squiggly v$ is a shortest path}\\
        & \leq \textubrace{ \delta\paren{u,v } + 2 \cdot w\paren{c,d}}{Our w.l.o.g. assumption}
        \end{alignat*}
\end{proof}

The most meticulous analysis is within \case{b}{2}, when proving \lemref{1+eps,4W2_case_b2}:

\begin{lemma}~\label{lem:1+eps,4W2_case_b2} 
   If \case{b}{2} holds then $d\bracke{u,v} \leq \paren{1+\varepsilon}\cdot \delta\paren{u,v}+\mmin{2\cdot w\paren{a,b},4\cdot w\paren{c,d}}$.
\end{lemma}

\begin{proof} Let $\paren{z,t}\in u\squiggly v$ be the \textbf{single} edge $\paren{z,t}\notin E_{S_{2}}$. Let $\paren{x_{1},y_{1}}\in u\squiggly v$ such that $\paren{x_{1},y_{1}}\notin E_{S_{1}}$. W.l.o.g. $u\squiggly v = u\squiggly z \straight t \squiggly x_{1} \straight y_{1} \squiggly v$. Furthermore, let $\paren{x_{2},y_{2}}\in x_{1}\squiggly v $ such that $\paren{x_{2},y_{2}}\notin E_{S_{1}}$ and $t\squiggly x_{1}, y_{2} \squiggly v \subseteq E_{S_{1}}$. It might be that $\paren{x_{1},y_{1}}=\paren{x_{2},y_{2}}$. However, $\paren{z,t} \neq \paren{x_{1},y_{1}},\paren{x_{2},y_{2}}$ by \case{b}{2}. 

If $w\paren{z,t}\leq w\paren{x_{1},y_{1}}$ or $w\paren{z,t}\leq w\paren{x_{2},y_{2}}$ then $w\paren{z,t}\leq w\paren{c,d}$. Hence, we follow a similar argument as in \lemref{1+eps,4W2_case_b3}.
After invoking MSASP from $p_{2}\paren{t}\in S_{2}$: $d\bracke{p_{2}\paren{t},u} \leq \paren{1+\varepsilon}\cdot\delta\paren{p_{2}\paren{t},u}$ and $d\bracke{p_{2}\paren{t},v}\leq \paren{1+\varepsilon}\cdot\delta\paren{p_{2}\paren{t},v}$. Therefore:

    \begin{alignat*}{3}
        d\bracke{u,v} & \leq \textubrace{\paren{1+\varepsilon}\cdot\paren{d\bracke{u,p_{2}\paren{t}}+  d\bracke{p_{2}\paren{t},v}} }{$\paren{1+\varepsilon}$-AMPMM for $d_{0}^{2} \star d^{2}_{0}$}
        && \leq \textubrace{ \paren{1+\varepsilon}^{2}\cdot\paren{\delta\paren{p_{2}\paren{t},u} + \delta\paren{p_{2}\paren{t},v}}}{Our preceding discussion} \\
        & \leq \textubrace{ \paren{1+\varepsilon}^{2}\cdot\paren{\delta\paren{u,t} + 2\cdot \delta\paren{t,p_{2}\paren{t}}+\delta\paren{t,v}}}{Triangle inequality}
        && \leq \textubrace{\paren{1+\varepsilon}^{2}\cdot\paren{\delta\paren{u,v} + 2\cdot \delta\paren{t,p_{2}\paren{t}}}}{$u\squiggly t\squiggly v$ is a shortest path}\\
        & \leq \textubrace{ \paren{1+2\cdot \varepsilon+\varepsilon^{2}}\cdot\paren{\delta\paren{u,v} + 2\cdot w\paren{z,t}}}{Due to \case{b}{2}}
        && \leq \textubrace{ \paren{1+2\cdot \varepsilon+\varepsilon^{2}}\cdot\paren{\delta\paren{u,v}  +2\cdot w\paren{c,d}}}{Our assumption within this sub-case}\\
        & \leq \textubrace{ \paren{1+4\cdot \varepsilon+2\cdot\varepsilon^{2}}\cdot\delta\paren{u,v}+2\cdot w\paren{c,d}}{$2w\paren{c,d}\leq \delta\paren{u,v}$}
    \end{alignat*}
Select $\varepsilon'= \sqrt{1+\frac{\varepsilon}{2}}-1$ to complete this case. However, it is possible that $w\paren{x_{1},y_{1}},w\paren{x_{2},y_{2}}< w\paren{z,t}$. Hence, the above would only yield an upper-bound of $+2W_{1}$ instead of $+2W_{2}$. To overcome this, we proceed in the following manner: during the first SSSP invocation from $p_{1}\paren{x_{1}}$ the edges $p_1\paren{x_{1}}\straight x_{1} \squiggly y_{2}\straight p_{1}\paren{y_{2}}$ will be considered, as $x_{1} \squiggly v\subseteq E_{S_{2}}$, therefore: $d\bracke{p_{1}\paren{x_{1}},p_{1}\paren{y_{2}}} \leq \delta\paren{x_{1},p_{1}\paren{x_{1}}}+\delta\paren{x_{1},y_{2}}+\delta\paren{y_{2},p_{1}\paren{y_{2}}}$. 

Additionally, during the first SSSP from $z\in S_{0} = V$ the edge $\paren{z,t}$ will be taken into consideration and $t\squiggly x_{1}\subseteq E_{S_{1}}$, hence: $d\bracke{z,p_{1}\paren{x_{1}}}\leq \delta\paren{z,x_{1}}+\delta\paren{x_{1},p_{1}\paren{x_{1}}}$, as we consider edges of $H$ as well. During the second SSSP from $p_{1}\paren{x_{1}}\in S_{1}$, we would use the auxiliary edge $\paren{p_{1}\paren{x_{1}},z}$, whose weight is upper-bounded by $\delta\paren{z,x_{1}}+\delta\paren{x_{1},p_{1}\paren{x_{1}}}$. The edge $\paren{z,t}$ is the \textbf{sole} edge that does not belong to $E_{S_{2}}$, hence $u\squiggly z\subseteq E_{S_{2}}$. Therefore, $d\bracke{p_{1}\paren{x_{1}},u}\leq \delta\paren{u,x_{1}}+\delta\paren{x_{1},p_{1}\paren{x_{1}}}$. 

After the second SSSP we compute $\paren{1+\varepsilon}$-AMPMM for $d_{0}^{1} \star d_{1}^{1}$. Consider $d\bracke{u,p_{1}\paren{x_{1}}}$ and $d\bracke{p_{1}\paren{x_{1}},p_{1}\paren{y_{2}}}$. Therefore: 
 \begin{alignat*}{3}
        d\bracke{u,p_{1}\paren{y_{2}}} & \leq \textubrace{\paren{1+\varepsilon} \cdot\paren{d\bracke{u,p_{1}\paren{x_{1}}}+d\bracke{p_{1}\paren{x_{1}},p_{1}\paren{y_{2}}}}}{$\paren{1+\varepsilon}$-AMPMM for $d_{0}^{1} \star d_{1}^{1}$}\\
        & \leq \textubrace{\paren{1+\varepsilon} \cdot\paren{\delta\paren{u,x_{1}}+\delta\paren{x_{1},p_{1}\paren{x_{1}}}+\delta\paren{x_{1},p_{1}\paren{x_{1}}}+\delta\paren{x_{1},y_{2}}+\delta\paren{y_{2},p_{1}\paren{y_{2}}}}} {By the preceding discussion}\\
        & \leq \textubrace{\paren{1+\varepsilon} \cdot\paren{\delta\paren{u,y_{2}}+2\cdot \delta\paren{x_{1},p_{1}\paren{x_{1}}}+\delta\paren{y_{2},p_{1}\paren{y_{2}}}}} {$u\squiggly x_{1}\squiggly y_{2}$ is a shortest path} \\
\end{alignat*}

We now turn to the third SSSP invocation, this time from $u\in V$. By using the auxiliary edge $\paren{u,p_{1}\paren{y_{2}}}$ whose weight we have just upper-bounded, and by recalling that $y_{2}\squiggly v \subseteq E_{S_{1}}$ we deduce:

\begin{alignat*}{3}
        d\bracke{u,v} & \leq \textubrace{d\bracke{u,p_{1}\paren{y_{2}}}+\delta\paren{p_{1}\paren{y_{2}},y_{2}}+\delta\paren{y_{2},v}}{SSSP from $u$}\\
        & \leq \textubrace{\paren{1+\varepsilon} \cdot\paren{\delta\paren{u,y_{2}}+2\cdot \delta\paren{x_{1},p_{1}\paren{x_{1}}}+\delta\paren{y_{2},p_{1}\paren{y_{2}}}}+\delta\paren{y_{2},p_{1}\paren{y_{2}}}+\delta\paren{y_{2},v}}{By the preceding discussion} \\
        & \leq \textubrace{\paren{1+\varepsilon} \cdot\paren{\delta\paren{u,y_{2}}+2\cdot \delta\paren{x_{1},p_{1}\paren{x_{1}}}+2\cdot\delta\paren{y_{2},p_{1}\paren{y_{2}}}+\delta\paren{y_{2},v}}}{Simply summing}\\
        & \leq \textubrace{\paren{1+\varepsilon} \cdot\paren{\delta\paren{u,v}+2\cdot \delta\paren{x_{1},p_{1}\paren{x_{1}}}+2\cdot\delta\paren{y_{2},p_{1}\paren{y_{2}}}}}{$u\squiggly y_{2} \squiggly v$ is a shortest path} \\
        & \leq \textubrace{\paren{1+\varepsilon} \cdot\paren{\delta\paren{u,v}+2\cdot  w\paren{x_{1},y_{1}}+2\cdot w\paren{x_{2},y_{2}}}}{Definition of $\paren{x_{1},y_{1}}$ and $\paren{x_{2},y_{2}}$} \\
        & \leq \textubrace{\paren{1+\varepsilon} \cdot\paren{\delta\paren{u,v}+4\cdot w\paren{c,d}}}{Because $\paren{z,t}$ is heavier than both} \\
        & \leq \textubrace{\paren{1+4\cdot\varepsilon} \cdot\delta\paren{u,v}+4\cdot w\paren{c,d}}{$w\paren{c,d}\leq \delta\paren{u,v}$} 
\end{alignat*}
We conclude by selecting $\varepsilon'=\frac{\varepsilon}{4}$. This means that if $w\paren{z,t}\leq w\paren{x_{1},y_{1}}$ or $w\paren{z,t}\leq w\paren{x_{2},y_{2}}$ we can upper-bound our approximation by $\paren{1+\varepsilon}\cdot\delta\paren{u,v}+2\cdot w \paren{c,d}$. Otherwise, we can provide two upper-bounds: $\paren{1+\varepsilon}\cdot\delta\paren{u,v}+2\cdot w \paren{a,b}$  and $\paren{1+\varepsilon}\cdot\delta\paren{u,v}+4\cdot w \paren{c,d}$. Hence, the actual upper-bound is $\paren{1+\varepsilon}\cdot\delta\paren{u,v}+\mmin{2\cdot w\paren{a,b},4\cdot w \paren{c,d}}$.

\end{proof}

We are left with:

\begin{lemma}~\label{lem:1+eps,4W2_case_b3} 
   If \case{b}{3} holds then $d\bracke{u,v} \leq \paren{1+\varepsilon}\cdot \delta\paren{u,v}+2\cdot w\paren{c,d}$.
\end{lemma}

\begin{proof} Let $\paren{x_{1},y_{1}},\paren{x_{2},y_{2}}\in u\squiggly v$ such that $\paren{x_{1},y_{1}}, \paren{x_{2},y_{2}} \notin E_{S_{2}}$ and $u\squiggly x_{1} \subseteq E_{S_{2}}$ and $y_{2}\squiggly v \subseteq E_{S_{2}}$. W.l.o.g.: $w\paren{x_{2},y_{2}}\leq w\paren{x_{1},y_{1}}$.  After invoking MSASP from $p_{2}\paren{y_{2}}\in S_{2}$: $d\bracke{p_{2}\paren{y_{2}},u} \leq \paren{1+\varepsilon}\cdot\delta\paren{p_{2}\paren{y_{2}},u}$ and $d\bracke{p_{2}\paren{y_{2}},v}\leq \paren{1+\varepsilon}\cdot\delta\paren{p_{2}\paren{y_{2}},v}$. Ergo:

    \begin{alignat*}{3}
        d\bracke{u,v} & \leq \textubrace{\paren{1+\varepsilon}\cdot\paren{d\bracke{u,p_{2}\paren{y_{2}}}+  d\bracke{p_{2}\paren{y_{2}},v}} }{$\paren{1+\varepsilon}$-AMPMM for $d_{0}^{2} \star d^{2}_{0}$}
        && \leq \textubrace{ \paren{1+\varepsilon}^{2}\cdot\paren{\delta\paren{p_{2}\paren{y_{2}},u} + \delta\paren{p_{2}\paren{y_{2}},v}}}{Our preceding discussion} \\
        & \leq \textubrace{ \paren{1+\varepsilon}^{2}\cdot\paren{\delta\paren{u,y_{2}} + 2\cdot \delta\paren{y_{2},p_{2}\paren{y_{2}}}+\delta\paren{y_{2},v}}}{Triangle inequality}
        && \leq \textubrace{\paren{1+\varepsilon}^{2}\cdot\paren{\delta\paren{u,v} + 2\cdot \delta\paren{y_{2},p_{2}\paren{y_{2}}}}}{$u\squiggly y_{2}\squiggly v$ is a shortest path}\\
        & \leq \textubrace{ \paren{1+2\cdot \varepsilon+\varepsilon^{2}}\cdot\paren{\delta\paren{u,v} + 2\cdot w\paren{x_{2},y_{2}}}}{Due to \case{b}{3}}
        && \leq \textubrace{ \paren{1+2\cdot \varepsilon+\varepsilon^{2}}\cdot\paren{\delta\paren{u,v}  +2\cdot w\paren{c,d}}}{Our w.l.o.g. assumption}\\
        & \leq \textubrace{ \paren{1+4\cdot \varepsilon+2\cdot\varepsilon^{2}}\cdot\delta\paren{u,v}+2\cdot w\paren{c,d}}{$2w\paren{c,d}\leq \delta\paren{u,v}$}
    \end{alignat*}
By selecting $\varepsilon'= \sqrt{1+\frac{\varepsilon}{2}}-1$ we complete the proof.
\end{proof}

This indicates the end of our case-analysis. Having covered all possible cases of our case-tree, we can now prove:

\begin{theorem}~\label{thm:1+eps,4W2}\Copy{thm:1+eps,4W2}{
\algref{1+eps,4W2} runs in $\tilde O\paren{\paren{\frac{1}{\varepsilon}}^{O\paren{1}}\cdot n^{2.15135313}\cdot \log W}$ time and computes a $\paren{1+\varepsilon,\mmin{2W_{1},4W_{2}}}$-APASP, for weights of the range $\bracke{W}\cup\bracce{0,\infty}$.}
\end{theorem}

\begin{proof}
The correctness follows by the case analysis and \lemref{1+eps,4W2_case_a1} to \lemref{1+eps,4W2_case_b3}. Let us analyze the runtime. Initializing $d$ takes $O\paren{n^{2}}$ time. Computing $\Gamma\paren{u,n^{\beta}}$ (respectively, $\Gamma\paren{u,n^{\beta+\gamma}}$) for all vertices requires $\tilde O \paren{m+n^{1+\beta}}$ (respectively, $\tilde O \paren{m+n^{1+\beta+\gamma}}$) by \obvref{ball}. Computing the hitting set $S_{1}$ (respectively, $S_{2}$) can be done deterministically in $O\paren{n^{1+\beta}}$ (respectively, $O\paren{n^{1+\beta+\gamma}}$) by \obvref{hs}. Finding both pivots of each vertex and computing the distances to both pivots requires $\tilde O \paren{m}$ by \obvref{pivotsdistance}. Finding the edge set $E_{S_{1}}$ (respectively, $E_{S_{2}}$) requires $\tilde O \paren{n^{1+\beta}}$ (respectively, $\tilde O \paren{n^{1+\beta+\gamma}}$) by \obvref{edgesruntime}.

By \obvref{hs} the size of $S_{1}$ is $\tilde O \paren{n^{1-\beta}}$ and the size of $S_{2}$ is $\tilde O \paren{n^{1-\beta-\gamma}}$. By \obvref{edgessize} the size of $E_{S_{1}}$ is $\tilde O \paren{n^{1+\beta}}$  and the size of $E_{S_{2}}$ is $\tilde O \paren{n^{1+\beta+\gamma}}$. As there are $n$ vertices, the size of $\paren{\bracce{s}\times V }\cup H$ is $\tilde O \paren{n}$, which is smaller than any $E_{S_{i+1}}$. Implementing the SSSP invocation by Dijkstra, the runtime  will therefore be $\tilde O \paren{\abs{S_{0}}\cdot \abs{E_{S_{1}}}+\abs{S_{1}}\cdot \abs{E_{S_{2}}}}  = \tilde O \paren{n^{2+\beta} + n^{2+\gamma}}$. Recall that $\abs{S_{2}} \in \tilde O \paren{n^{1-\beta-\gamma}}$. The $\paren{1+\varepsilon}$-MSASP invocation requires $\tilde O \paren{m^{1+o\paren{1}} + \paren{ \frac{1}{\varepsilon} }^{O\paren{1}} \cdot n^{\omega\paren{1-\beta-\gamma}} \cdot  \log W}$.

The first $\paren{1+\varepsilon}$-AMPMM of $d_{0}^{2} \star d_{2}^{0}$ requires $\tilde O \paren{\frac{1}{\varepsilon} \cdot n^{\omega\paren{1-\beta-\gamma}} \cdot \log W}  $ by the $\paren{1+\varepsilon}$-AMPMM algorithm of Zwick \cite{Zwick2002}. Similarly, the second $\paren{1+\varepsilon}$-AMPMM invocation, this time of $d_{0}^{1}\star d_{1}^{1}$, requires $\tilde O \paren{\frac{1}{\varepsilon} \cdot n^{\omega\paren{1,1-\beta,1-\beta}} \cdot \log W}$. Repeating the same SSSP invocations twice or trice does not add to the overall runtime.

The total runtime is therefore $\tilde O\paren{n^{2+\beta}+n^{2+\gamma}+\paren{ \frac{1}{\varepsilon} }^{O\paren{1}} \cdot\paren{ n^{\omega\paren{1-\beta-\gamma}} +n^{\omega\paren{1,1-\beta,1-\beta}}} \cdot  \log W}$. We \cite{complexityBalancer} select $\beta=\gamma \approx 0.15135313$. Hence the runtime will be: $\tilde O\paren{\paren{\frac{1}{\varepsilon}}^{O\paren{1}}\cdot n^{2.15135313}\cdot \log W}$. 
\end{proof}

This concludes the correcness and runtime analysis of \algref{1+eps,4W2}. 

\section{Multiplicative and additive trade-offs}~\label{tradeoffs}
We recall that \algref{+2Wi}  computed a $+2\msum{W_{i}}{i=1}{k+1}$-APASP and \algref{3ellplus4overellplus2} by Akav and Roditty \cite{AkaRod2021} computed a $\frac{3\ell+4}{\ell+2}$-APASP (See: \appref{3ellplus4overellplus2}). For dense graphs, where $m\in \tilde \Theta\paren{n^{2}}$, the latter is a $\frac{9k+4}{3k+2}$-APASP algorithm with  $\tilde O \paren{n^{2+\frac{1}{3k+2}}}$ runtime, by setting $\ell=3k$.  

\begin{claim2}~\label{clm:tradeoffs}\Copy{clm:tradeoffs}{
    Let $\alpha,\beta\in\mathbb{R}^{+}$.  It is possible to compute a $\paren{\frac{\paren{9k+4}\cdot\alpha+\paren{3k+2}\cdot\beta}{\paren{\alpha+\beta}\cdot\paren{3k+2}},\frac{2\beta}{\alpha+\beta}\msum{W_{i}}{i=1}{k+1}}$-APASP in $\tilde O\paren{n^{2+\frac{1}{3k+2}}}$ time for any $k\in \mathbb{N}$.}
\end{claim2}

The proof is straightforward:

\begin{proof}
    Let $k\in\mathbb{N}$. Invoke both \algref{+2Wi} and \algref{3ellplus4overellplus2}, the latter for $\ell=3k$. The runtime of \algref{+2Wi} is $\tilde O\paren{n^{2+\frac{1}{3k+2}}}$ due to \thmref{+2Wi}. The runtime of \algref{3ellplus4overellplus2} is at most $\tilde O\paren{n^{2+\frac{1}{\ell+2}}}=\tilde O\paren{n^{2+\frac{1}{3k+2}}}$ (for details, see \appref{3ellplus4overellplus2}). Hence, their total runtime remains $\tilde O\paren{n^{2+\frac{1}{3k+2}}}$.

    Invoking both algorithms guarantees us that:
    \begin{enumerate}
        \item $d\bracke{u,v}\leq \delta\paren{u,v}+2\msum{W_{i}}{i=1}{k+1}$,
        \item and $d\bracke{u,v}\leq \frac{9k+4}{3k+2}\cdot\delta\paren{u,v}$.
    \end{enumerate}

    By considering any $\alpha,\beta\in\mathbb{R}^{+}$, we can simply average these upper-bounds:

    \begin{alignat*}{3}
        \paren{\alpha+\beta}\cdot d\bracke{u,v} &\leq \textubrace{ \alpha\cdot\frac{9k+4}{3k+2}\cdot\delta\paren{u,v}+\beta\cdot\delta\paren{u,v} +2\beta\cdot\msum{W_{i}}{i=1}{k+1}}{Summing both upper-bounds} \\
        & \leq \textubrace{\frac{\alpha\cdot\paren{9k+4}+\beta\cdot\paren{3k+2}}{3k+2}\cdot\delta\paren{u,v}+2\beta\cdot\msum{W_{i}}{i=1}{k+1}}{Simply summing} \\
\end{alignat*}

Dividing by $\alpha+\beta$, we conclude that: 
$$ d\bracke{u,v} \leq \frac{\alpha\cdot\paren{9k+4}+\beta\cdot\paren{3k+2}}{\paren{\alpha+\beta}\cdot\paren{3k+2}}\cdot \delta\paren{u,v}+\frac{2\beta}{\alpha+\beta}\cdot \msum{W_{i}}{i=1}{k+1}$$ 
\end{proof}

We observe that \clmref{tradeoffs} partially addresses \queref{tradeoffs}, by presenting a family of APASP algorithms computable in $\tilde{O}(n^{2+\frac{1}{3k+2}})$ time. \clmref{tradeoffs} also partially answers \queref{under2}, as, when $\alpha=\beta=1$, we get a $\paren{\frac{6k+3}{3k+2},\msum{W_{i}}{i=1}{k+1}}$-APASP, whose multiplicative stretch is smaller than $2$. For instance, $k=1$ indicates that a $\paren{\frac{9}{5},W_{1}+W_{2}}$-APASP can be computed in $\tilde O \paren{n^{\frac{11}{5}}}$.

The same idea can be applied as well to \algref{3ellplus4overellplus2+eps} that computed a  $\paren{\frac{3\ell+4}{\ell+2}+\varepsilon}$-APASP and the  $\paren{1+\varepsilon, 2\msum{W_{i}}{i=1}{k+1}}$-APASP algorithm of Saha and Ye \cite{SahYe2023}.
In addition, the $\ell=1$ instance of
\algref{3ellplus4overellplus2+eps} with
\algref{1+eps,4W2} yields a
$\paren{\frac{5}{3}+\varepsilon,
\mmin{W_1,2W_2}}$-APASP computed in $\tilde O\paren{\paren{\frac{1}{\varepsilon}}^{O\paren{1}}\cdot n^{2.15135313}\cdot \log W}$ time.

For completeness, we provide values for the $\paren{\alpha,\beta}$-APASP examples where $\alpha\in\paren{2,\frac{7}{3}}$ that were mentioned in \secref{intro} between \queref{under2} and \queref{tradeoffs}. Consider the following: 

\begin{itemize}
        \item $\alpha=3$, $\beta=1$ and $k=1$ results in $\paren{\frac{11}{5},\frac{W_{1}+W_{2}}{2}}$-APASP in $\tilde O \paren{n^{\frac{11}{5}}}$ time;
        \item $\alpha=2$, $\beta=1$ and $k=2$ results in $\paren{\frac{13}{6},\frac{2\cdot\paren{W_{1}+W_{2}+W_{3}}}{3}}$-APASP in $\tilde O \paren{n^{\frac{17}{8}}}$ time.
\end{itemize}

\newpage
\clearpage
\renewcommand*{\bibfont}{\small}
\begingroup
\phantomsection\printbibliography[heading=bibintoc]
\endgroup

\appendix

\section{Multiplicative $\frac{7}{3}$-APASP}\label{7over3}

\begin{algorithm}[H]
\caption{$\frac{7}{3}$-APASP$\left(G=\left(V,E\right),w\right)$}\label{alg:7over3}
\textbf{Input:} An undirected graph $G=\left(V,E\right)$ and a positive weight function $w:E\rightarrow\mathbb{R}^+$.

\textbf{Output:} A matrix $d$ which is a $\frac{7}{3}$-APASP. 

\medskip

Let $\beta,\gamma \in \paren{0,1}$ to be fixed later

Initialize $d$ as the edges' weights and $\infty$ otherwise



Set $S_{0}\leftarrow V$ and $S_{3}\leftarrow \varnothing$

Sample each vertex of $V$ with probability $n^{-\beta}$ to $S_{1}$

Sample each vertex of $S_{1}$ with probability $n^{-\gamma}$ to $S_{2}$

Compute the distances from any $u\in V$ to its pivot $p_{1}\paren{u}$ (respectively, $p_{2}\paren{u}$)

Construct the set of edges $E_{S_{1}}$ (respectively, $E_{S_{2}}$)

Compute the sets $\ball{u}{S_{1}}{S_{2}}$ and $\ball{u}{S_{1}}$ for every $u\in V$

\For{$\paren{x,y}\in E$}
{
    \For{$v\in\ball{y}{S_{1}}$}
    {      
        \For{$i \in \bracce{0,1,2}$}
        {
            $d\bracke{p_{i}\paren{v},x}\leftarrow \mmin{d\bracke{p_{i}\paren{v},v}+d\bracke{v,y}+d\bracke{x,y}, d\bracke{p_{i}\paren{v},x}}$
        }
    }
}

\For{$i \in \bracce{0,1,2}$}
{
    \For{$s\in  S_{i}$}
    {
        Invoke SSSP from $s$ on $G_{i} \paren{s} = \left( V,E_{ 
 S_{i+1}} \cup \paren{\bracce{s}\times V} \cup H  , d \right)$ and update $d$ accordingly
    }
}

\For{$u,v\in V$}
{
    \For{$s\in\ball{u}{S_{1}}{S_{2}}$}
    {
        $d\bracke{u,v}\leftarrow \mmin{d\bracke{u,s}+d\bracke{s,v},d\bracke{u,v}}$
    }
}

\For{$u,v\in V$} 
{
    \For{$i \in \bracke{1,2}$}
    {
        $d\bracke{u,v}\leftarrow \mmin{d\bracke{u,p_{i}\paren{u}}+d\bracke{p_{i}\paren{u},v},d\bracke{u,v}}$
    }
}

\Return $d$
\end{algorithm}

For completeness, we present a simplified version of the $\frac{7}{3}$-APASP algorithm of Baswana and Kavitha \cite{BasKav2010}. Instead of $\tilde O \paren{nm^{\frac{2}{3}}+n^2}$ time, the runtime for the simplified version is $\tilde O\paren{n^{\frac{7}{3}}}$. The main difference between our versions lies in $\ball{u}{S_{1}}$. While we considered only this set, Baswana and Kavitha considered the $t$ sub-sets leading to it, for $t\in O\paren{\log n}$. These are: $\ball{u}{S_{1,i}}$ for $i\in \bracke{t}$. The runtime then reduces to $\tilde O \paren{nm^{\frac{2}{3}}+n^2}$. Attempting to consider another set of $O\paren{\log n}$ sub-sets $S_{2,1},\ldots,S_{2,t}$ yields no advantage, as other actions that are performed already require a runtime of $\tilde O \paren{n^{2+\gamma}}$. A detailed explanation appears by the end of the proof \thmref{7over3+eps}. 

\newpage
\section{Multiplicative $\frac{3\ell+4}{\ell+2}$-APASP}\label{3ellplus4overellplus2}

\begin{algorithm}[H]
\caption{$\frac{3\ell+4}{\ell+2}$-APASP$\left(G=\left(V,E\right),w\right)$}\label{alg:3ellplus4overellplus2}
\textbf{Input:} An undirected graph $G=\left(V,E\right)$, a positive weight function $w:E\rightarrow\mathbb{R}^+$ and an integer $\ell\in\mathbb{N}$.

\textbf{Output:} A matrix $d$ which is a $\frac{3\ell+4}{\ell+2}$-APASP. 

\medskip

Let $\beta_{1},\ldots,\beta_{\ell+1} \in \paren{0,1}$ to be fixed later

Initialize $d$ as the edges' weights and $\infty$ otherwise

Initialize $S_{0}\leftarrow V$ and set $S_{\ell+2} \leftarrow \varnothing$

\For{$i \in \bracke{\ell+1}$}
{


    \For{$s\in S_{i-1}$}
    {
        Sample $s$ independently into $S_{i}$ with probability $n^{-\beta_{i}}$
    }



    Compute the distances from any $u\in V$ to its pivot $p_{i}\paren{u}$ 

    Construct the set of edges $E_{S_{i}}$

    Compute the set $\ball{u}{S_{i-1}}{S_{i}}$ for all $u\in V$
}

\For{$\paren{x,y}\in E$}
{
    \For{$i \in \bracke{\ell+1}$}
    {
         \For{$v\in\ball{y}{S_{i-1}}{S_{i}}$}
        { 
            $d\bracke{p_{i}\paren{v},x}\leftarrow \mmin{d\bracke{p_{i}\paren{v},v}+d\bracke{v,y}+d\bracke{x,y}, d\bracke{p_{i}\paren{v},x}}$        
        }
    }
}

\For{$i \in \bracce{0}\cup \bracke{\ell+1}$}
{
    \For{$s\in  S_{i}$}
    {
        Invoke SSSP from $s$ on $G_{i} \paren{s} = \paren{ V,E_{ 
 S_{i+1}} \cup \paren{\bracce{s}\times V} \cup H  , d }$ and update $d$ accordingly 
    }
}

\For{$u,v\in V$}
{
    \For{$i \in \bracke{\ell+1}$}
    {
        \For{$s\in\ball{u}{S_{i-1}}{S_{i}}$}
        {
            $d\bracke{u,v}\leftarrow \mmin{d\bracke{u,s}+d\bracke{s,v},d\bracke{u,v}}$
        }
    }
}

\For{$u,v\in V$} 
{           
  $d\bracke{u,v}\leftarrow\mmin{d\bracke{u,p_{i}\paren{u}}+d\bracke{p_{i}\paren{u},v},d\bracke{u,v}}{i=0}{\ell+1}$
}

\Return $d$
\end{algorithm}

For completeness, we provide the $\frac{3\ell+4}{\ell+2}$-APASP algorithm of Akav and Roditty \cite{AkaRod2021} with a simplifying change whose runtime is $\tilde O \paren{n^{2+\frac{1}{\ell+2}}}$ instead of the original $\tilde O \paren{n^{2-\frac{3}{\ell+2}}m^{\frac{2}{\ell+2}}+n^2}$. For the original version, consider $t\in O\paren{\log n}$ sub-sets of $S_{1}$ of the form $S_{1,i}$, for $i\in\bracke{t}$, and their corresponding $\ball{u}{S_{1,i-1}}{S_{1,i}}$. This would result in an improved $\tilde O\paren{n^{2-\frac{3}{\ell+2}}m^{\frac{2}{\ell+2}}+n^2}$ runtime. Similarly to \appref{7over3}, we gain no advantage by performing additional partitioning of other sets $S_{j}$ for $j\geq 2$ into $\tilde O \paren{\log n}$ sets, as is explained both in \thmref{7over3+eps} and in \thmref{3ellplus4overellplus2+eps}. For a step-by-step construction process of this algorithm, we refer the reader to Akav and Roditty \cite{AkaRod2021}.

\begin{algorithm}[H]
\caption{ThorupZwick-Distance Oracle}\label{alg:tz}
\textbf{Input:} An undirected weighted graph $G=\left(V,E,w\right)$, vertices $u,v\in V$, hitting sets $S_0=V\supseteq S_1\supseteq\cdots
\supseteq S_k\supseteq S_{k+1}=\varnothing$ where vertices of $S_{i-1}$ are independently sampled into $S_{i}$  with probability $n^{-\beta_{i}}$ and $\beta_{i}\in\paren{0,1}$,  sets $\ball{x}{S_{i}}{S_{i+1}}$ for all $x\in V$ and $i\in \bracce{0}\cup \bracke{k}$ and an integer $r\in\bracke{k}$.

\textbf{Output:} $d\bracke{u,v}$. 

\medskip

$i\leftarrow r$

\While{$p_{i}\paren{u}\notin \ball{u}{S_{i}}{S_{i+1}}$}
{
    $i\leftarrow i+1$

    Replace between $u,v$
}

\Return $d\bracke{u,p_{i}\paren{u}}+d\bracke{v,p_{i}\paren{u}}$
\end{algorithm}

Akav and Roditty \cite{AkaRod2021} denote by $f=f\paren{u,v,r}$ the value of the index $i$ for which \algref{tz} halts. They then proved the following:

\begin{lemma}[\cite{AkaRod2021}]~\label{lem:tz}
Let $u,v\in V$ and $r\in \bracce{0}\cup \bracke{\ell+1}$. If $f-r$ is even then $\delta\paren{u,p_{f}\paren{u}}\leq \paren{f-r}\cdot \delta\paren{u,v} + \delta\paren{u,p_{r}\paren{u}}$; otherwise: $f-r$ is odd, and then $\delta\paren{v,p_{f}\paren{v}}\leq \paren{f-r}\cdot \delta\paren{u,v} + \delta\paren{u,p_{r}\paren{u}}$.
\end{lemma}

They make use of this within one of their lemmas which proves an upper-bound on $d\bracke{u,v}$, similarly to \lemref{3ellplus4overellplus2+eps_plus}:

\begin{lemma}[\cite{AkaRod2021}]~\label{lem:3ellplus4overellplus2_plus}
$d\bracke{u,v}\leq \delta\paren{u,v}+2\cdot\paren{\ell+1}\cdot\delta\paren{u,a}$ or $d\bracke{u,v}\leq  \delta\paren{u,v}+2\cdot\paren{\ell+1}\cdot\delta\paren{b,v}$
\end{lemma}

Where the above lemma states exactly the samea as does \lemref{3ellplus4overellplus2+eps_plus}, yet refers to \algref{3ellplus4overellplus2}. The vertices $a,b$ are selected in the same way as in \clmref{3ellplus4overellplus2+eps_properties}.

\end{document}